%% file: paper.tex
\documentclass[journal]{IEEEtran}

\usepackage[utf8]{inputenc}
\usepackage{amsmath}
\usepackage{amsthm}
\usepackage{amssymb}
\usepackage[english]{babel}

\newtheorem{theorem}{Theorem}
\newtheorem{lemma}[theorem]{Lemma}
\usepackage{csquotes}
\usepackage[noend]{algorithmic}
\usepackage{algorithm}
\usepackage{url}
\usepackage{graphicx}
\usepackage[caption=false,font=footnotesize]{subfig}
\usepackage{pgfplots}
\usepgfplotslibrary{units}
\usepackage{standalone}
\pgfplotsset{compat=1.14}

\let\existstemp\exists
\let\foralltemp\forall
\renewcommand*{\exists}{\existstemp\mkern2mu}
\renewcommand*{\forall}{\foralltemp\mkern2mu}

\usepackage{calc}
\usepackage{xparse}
\newlength\mytemplena
\newlength\mytemplenb
\DeclareDocumentCommand\myalignalign{sm}
{
  \settowidth{\mytemplena}{$\displaystyle #2$}%
  \setlength\mytemplenb{\widthof{$\displaystyle=$}/2}%
  \hskip-\mytemplena%
  \hskip\IfBooleanTF#1{-\mytemplenb\mkern+4mu}{+\mytemplenb}%
}

\usepackage{xcolor}

\DeclareMathOperator*{\argmin}{arg\,min}
\allowdisplaybreaks
\begin{document}
	
	\title{A Reduced-Complexity Projection Algorithm for ADMM-based LP Decoding}
	
	\author{Florian~Gensheimer, Tobias Dietz, Kira~Kraft,~\IEEEmembership{Student~Member,~IEEE},\\Stefan~Ruzika, and Norbert~Wehn, \IEEEmembership{Senior~Member,~IEEE}%
	    \thanks{This work was supported by the DFG under project-ID WE 2442/9-3
and RU 1524/2-3.}
		\thanks{This paper will be presented in parts at the 10th International Symposium on Turbo Codes and Iterative Information Processing.
		}
		\thanks{F. Gensheimer is with the Mathematical Institute, University of Koblenz-Landau, 56070 Koblenz, Germany (email: gensheimer@uni-koblenz.de).}%
		\thanks{T. Dietz and S. Ruzika are with the Department of Mathematics, Technische Universität Kaiserslautern, 67663 Kaiserslautern, Germany (email: dietz@mathematik.uni-kl.de; ruzika@mathematik.uni-kl.de).}%
		\thanks{K. Kraft and N. Wehn are with the Department of Electrical and Computer Engineering, Technische Universität Kaiserslautern, 67663 Kaiserslautern, Germany (email: kraft@eit.uni-kl.de; wehn@eit.uni-kl.de).}
		}
	
	\maketitle

	\begin{abstract}
		The Alternating Direction Method of Multipliers has recently been adapted for Linear Programming Decoding of Low-Density Parity-Check codes. The computation of the projection onto the parity polytope is the core of this algorithm and usually involves a sorting operation, which is the main effort of the projection.
		
		In this paper, we present an algorithm with low complexity to compute this projection. The algorithm relies on new findings in the recursive structure of the parity polytope and iteratively fixes selected components. It requires up to 37\% less arithmetical operations compared to state-of-the-art projections. Additionally, it does not involve a sorting operation, which is needed in all exact state-of-the-art projection algorithms. These two benefits make it appealing for efficient hard- and software implementations.
	\end{abstract}

	\section{Introduction}

	\IEEEPARstart{L}{inear} Programming (LP) decoding is a rather new decoding approach, that was established in 2003 by Feldman et al.~\cite{Feldman}. The LP decoding problem is a relaxation of the \emph{Maximum-Likelihood} (ML) decoding problem onto a special polytope. Using redundant parity checks~\cite{csa}, the error-correcting performance of LP decoding is close to the ML decoding performance. Therefore, LP decoding has become an interesting area of research for nearly all relevant code classes.
	
	The major advantage of LP decoding is the reduced complexity compared to ML decoding. While ML decoding is NP hard in general~\cite{Berlekamp}, the relaxation onto the special polytope reduces the problem to a linear program, which can be solved in polynomial time. Recently, the \emph{Alternating Direction Method of Multipliers} (ADMM)~\cite{Boyd}, an iterative method from convex optimization, was proposed for solving the LP decoding problem. In this context, the main effort of the ADMM is the projection onto the so-called parity polytope (see Section~\ref{sec:paritypolytopes}). The projection complexity grows with the number of ones in each row of the code's underlying parity-check matrix, thus, ADMM-based LP decoding is mainly performed for \emph{Low-Density Parity-Check} (LDPC) codes. This projection is the key of the ADMM algorithm and requires a sorting operation in all exact state-of-the-art implementations. However, sorting can become a major problem, especially for efficient hardware implementations, where it can heavily impact on latency, area, and power consumption.
	
In this paper, we extend the theory of parity polytopes and reveal their recursive structure. These findings allow us to present a new efficient projection algorithm that does not require sorting operations. The proposed algorithm iteratively fixes selected components of the projection to recursively reduce the problem to a smaller instance. We show that at least one component of the input can be fixed in every step. Therefore, the number of recursions is bounded and the problem size is strictly decreasing in every iteration. Our approach requires up to 37\% less arithmetic operations than state-of-the-art projection algorithms, which directly translates to a reduction in computational complexity. In addition, the sorting operation is circumvented completely.
	
	The outline of the paper is as follows: Section~\ref{sec:relatedwork} describes the related work in the area of the ADMM-based LP decoding. In Section~\ref{sec:ADMMLPDecoding}, preliminaries and the ADMM method for LP decoding are recapitulated. In Section~\ref{sec:paritypolytopes}, we define the considered projection problem, recall some essential properties of the (even) parity polytope and derive the analogous results for its odd counterpart. Section~\ref{sec:geometricidea} shows the geometrical idea of our new projection algorithm in an example. In Section~\ref{sec:fixingcomponents}, we prove the main theorem about this projection. It states that in every iteration, there exists at least one component that can be fixed for the rest of the projection. In Section~\ref{sec:recursion}, we show that the projection can be formulated as a recursive problem and utilize this fact in our efficient projection algorithm. The whole algorithm is then summarized in Section~\ref{sec:algorithm}. Section~\ref{sec:numericalresults} presents numerical results and highlights the benefit of our new projection algorithm. Finally, the paper is concluded in Section~\ref{sec:conclusion}.

	\section{Related Work}\label{sec:relatedwork}
	The first ADMM method for LP decoding was presented by Barman et al. in~\cite{Barman}, where a two-slice representation is used in the projection in order to describe the vectors of the parity polytope. The projection method in~\cite{Barman} needs two sorting operations. A more efficient projection method was presented Zhang and Siegel in~\cite{ADMMSiegel}, which is based on the cut-search algorithm~\cite{csa} of the same authors. Its main effort is a sorting operation on a vector, which is in worst-case as large as the dimension $ d $ of the parity polytope. The projections by Wasson and Draper~\cite{Wasson} and Zhang et al.~\cite{Heusdens} reduce the problem to the projection onto a simplex and use the corresponding algorithms of Duchi et al. from~\cite{projectionl1ball}. The main effort in~\cite{Wasson} is again a sorting operation. In~\cite{Heusdens}, the two main subroutines are partial sorting and a modification of the randomized median finding algorithm from~\cite{Cormen}. An iterative method, which does not require sorting operations, is presented by Wei and Banihashemi in~\cite{Wei}. However, it only outputs an approximate projection. A lookup table is used by Jiao et al. in the projection algorithm in~\cite{ADMMLookup}, where the authors use the symmetric structure of the parity polytope in order to reduce the size of the table. In~\cite{ADMMLookup2}, Jiao et al. further decrease the size by using a non-uniform quantization scheme, which is found by minimizing the mean square error of a sample set.
	
	Apart from the projection onto the parity polytope, many other investigations and improvements of ADMM-based LP decoding are made in the literature. In~\cite{penalizedADMM}, Liu and Draper improve the error correction rate by introducing penalty terms for the objective function that reward binary decision variables. The behavior of ADMM decoding on trapping sets is studied by the same authors in~\cite{trappingsets}. In~\cite{ADMMirregular}, Jiao et al. improve the error-correcting performance of penalized ADMM decoding for irregular LDPC codes by using different penalty parameters for variables with different variable degrees. In~\cite{ReducedComplexityADMM}, Wei et al. reduce the runtime by avoiding projections whenever the change in the input of the projection is sufficiently small. New piecewise penalty terms are introduced by Wang et al. in~\cite{ImprovedPenaltyADMM}. In~\cite{AcceleratedADMM}, Jiao et al. compare two improving techniques of ADMM in the context of LP decoding, namely over-relaxation~\cite{Boyd} and accelerated ADMM~\cite{fastADMM}. A two-step scheme based on ADMM decoding is presented by Jiao and Mu in~\cite{TwoStepADMM}. In order to reduce the error floor, the code structure is changed by eliminating codewords with low weight, and a postprocessing step is added after the ADMM LP decoder. 
	
	In~\cite{WassonHardware}, Wasson et al. propose a hardware architecture for the ADMM LP decoder based on the projection method presented in~\cite{Wasson}. The hardware complexity of ADMM LP decoding is also investigated by Debbabi et al. in~\cite{DebbabiHardware}. A multicore implementation is presented by the same authors in~\cite{MulticoreADMM}. The schedule of the computations in the ADMM LP decoder are changed by Debbabi et al.~\cite{ADMMScheduling} and Jiao et al.~\cite{ADMMScheduling2}. These schedules are combined to a new mixed schedule for ADMM LP decoding of LDPC convolutional codes by Thameur et al. in~\cite{ADMMconvolutional}. In~\cite{ADMMequalization}, Xu et al. propose turbo equalization together with ADMM decoding for communication over the partial response channel.

	\section{ADMM-based LP Decoding}\label{sec:ADMMLPDecoding}
	In this paper, we consider binary linear block codes $ C \subseteq \{0,1\}^n $ with block length $ n $ and a parity-check matrix $ H \subseteq \{0,1 \}^{m\times n} $, where $ J=\{1,\dots,m \} $ denotes the set of check nodes and $ d_j $ denotes the degree of check $ j \in J $, that is, the number of ones in the corresponding row of the parity-check matrix. The set of variable nodes is given by $ I=\{1,\dots,n\} $ and $ N_i\subseteq J $ describes the set of check nodes that include variable node $ i \in I $. The set $ N_j = \{i\in \{1,\dots,n\}:H_{ji}=1 \} \subseteq I$ denotes the set of variable nodes that are considered for check $ j \in \{1,\dots,m \} $. We consider binary-input memoryless channels. In~\cite{Feldman03PhD}, it is shown that, in this case, maximum-likelihood (ML) decoding can be rewritten as the minimization of a linear function. This means that
	\begin{equation*}
	x^{ML} = \underset{x \in C}{\arg\min} \sum_{i=1}^{n}\lambda_ix_i,
	\end{equation*} 
	where
	$\lambda_i=\ln \frac{Pr\left(\widetilde{x}_i|x_i=0\right)}{Pr\left(\widetilde{x}_i|x_i=1\right)}$
	are the so-called log-likelihood ratios (LLR) and $ \widetilde{x} $ is the received vector. The linear programming relaxation of this problem is called linear programming (LP) decoding~\cite{Feldman03PhD}. For ADMM-based LP decoding, the following LP formulation with auxiliary variables $ z_j\in \mathbb{R}^{d_j} $ is used:
	\begin{IEEEeqnarray}{ll}
		\min \quad &\;\lambda^\top x\label{eq:ADMMLP1} \\
		\text{s.\,t.} & T_jx = z_j \quad \forall j \in J\label{eq:ADMMLP2}\\
		& z_j \in \mathcal{P}_{d_j} \quad \forall j \in J.\label{eq:ADMMLP3}
	\end{IEEEeqnarray}
	The matrix $ T_j\in \{0,1\}^{d_j\times n} $ from~\cite{ADMMSiegel} selects the variable nodes $ i \in N_j$ for all $ j \in J $. The ADMM is an iterative method from convex optimization, that combines the strong convergence of the method of multipliers with the decomposability of the dual ascent method~\cite{Boyd}. Mathematically, the ADMM is a gradient method, that solves a special dual problem of (\ref{eq:ADMMLP1})-(\ref{eq:ADMMLP3}) which depends on the augmented Lagrangian. For the LP decoding problem, the augmented Lagrangian with scaled dual variables $ u $ is given by
	\begin{equation*}
	\mathcal{L}_\rho (x,z,u) = \lambda^\top x + \frac{\rho}{2}\sum_{j\in J} \left\lVert T_jx-z_j+u_j\right\rVert_2^2 - \frac{\rho}{2}\sum_{j\in J}\left\lVert u_j \right\rVert_2^2.
	\end{equation*}
	In iteration $k$, the variables are updated as follows:
	\begin{equation}
	x^{k+1}:=\argmin_x \left(\lambda^\top x+\frac{\rho}{2}\sum_{j\in J}\left\lVert T_jx-z_j^k+u_j^k\right\rVert_2^2\right),\label{eq:xupdate}
	\end{equation}
	\begin{equation*}
	z_j^{k+1} := \Pi_{\mathcal{P}_{d_j}}\left(T_jx^{k+1}+u_j^k\right) \quad \forall j \in J,
	\end{equation*}
	\begin{equation*}
	u_j^{k+1}:=u_j^k + T_jx^{k+1}-z_j^{k+1} \quad \forall j \in J.
	\end{equation*}
	The mapping $ \Pi_{\mathcal{P}_{d_j}}(\cdot) $ is defined as the projection onto the parity polytope 
	\begin{equation*}
	\mathcal{P}_{d_j} := \text{conv} \{x\in \{0,1\}^{d_j}:\sum_{i=1}^{d_j}x_i \text{ is even} \}.
	\end{equation*}
	The minimum in  the $ x $-update (\ref{eq:xupdate}) can be computed analytically with the formula
	\begin{equation*}
	x_i = \frac{1}{d_i}\left(\sum_{j\in N_i} \left(\left(u_j^k\right)_i-\left(z_j^k\right)_i\right) - \frac{\lambda}{\rho}\right)  \quad \forall i \in I.
	\end{equation*} The penalty terms used in~\cite{penalizedADMM} and~\cite{ImprovedPenaltyADMM} only change this $ x $-update. The main computational effort are the $ z $-updates which consist of a projection onto the parity polytopes of every parity row $j \in J$.
	
	\section{Even and Odd Parity Polytopes}\label{sec:paritypolytopes}
	In the following, we recall basic properties of \(\mathcal{P}_{d}\) which we call the \emph{even parity polytope} 
	\begin{equation*}
	\mathcal{P}_{d,\text{even}}:= \mathcal{P}_d := \text{conv}\{x\in \{0,1\}^d:\sum_{i=1}^{d}x_i\text{  is even}  \}
	\end{equation*}
	to distinguish it from the \emph{odd parity polytope}
	\begin{equation*}
	\mathcal{P}_{d,\text{odd}} := \text{conv}\{x\in \{0,1\}^d:\sum_{i=1}^{d}x_i\text{  is odd}  \}.
	\end{equation*}
	Projections on the even parity polytope and on the odd parity polytope both play a crucial role in the projection algorithm presented later.   
	
	In~\cite{Feldman03PhD}, Feldman presents the linear programming relaxation
	\begin{IEEEeqnarray}{l}
		0\leq x_i \leq 1 \quad \forall i=1,\dots,n \label{eq:boxconstraints} \\
		0 \leq w_{j,S} \leq 1 \quad \forall S \in E_j \nonumber \\
		\sum_{S\in E_j}w_{j,S} = 1 \label{eq:wjsum} \\
		x_i = \sum_{\substack{S\in E_j \\ S\ni i}}w_{j,S} \quad \forall i \in N_j \label{eq:wjs}
	\end{IEEEeqnarray}
	for the local parity polytope $ C_j=\{x\in \{0,1\}^n: H_jx \equiv 0 \mod 2 \} $, where $ E_j=\{S\subseteq N(j): \lvert S \rvert \text{ even} \} $ is the set of even-sized subsets of $ N_j $. The feasible set of this polyhedron is called $\mathcal{Q}_j$ in~\cite{Feldman03PhD} and $ \dot{\mathcal{Q}_j} = \{x \in \mathbb{R}^n: \exists w: (x,w)^\top \in \mathcal{Q}_j \} $ is the corresponding polyhedron without the auxiliary variables $ w_{j,S} $. 
	
	$ \mathcal{P}_{d,\text{even}} $ can be interpreted as the convex hull of the local parity polytope with parity row $H_1=(1 \dots 1) $. This means that all bit variables $ x_i $ participate in this row. Hence, the sum in (\ref{eq:wjs}) is not empty and it follows that
	\begin{equation*}
	x_i = \sum_{\substack{S\in E_1 \\ S\ni i}}\underbrace{w_{1,S}}_{\geq 0} \geq 0
	\end{equation*}
	and
	\begin{equation*}
	x_i = \sum_{\substack{S\in E_1 \\ S\ni i}}w_{1,S} \leq \sum_{S\in E_1}w_{1,S} \overset{(\ref{eq:wjsum})}{=} 1
	\end{equation*} 
	for all $ i=1,\dots,d $. Thus, the constraints (\ref{eq:boxconstraints}) are redundant and can be removed, so it holds that
	\begin{equation}
	\dot{\mathcal{Q}_1} = \mathcal{P}_{d,\text{even}}=\text{conv} \{x\in \{0,1\}^d:\sum_{i=1}^dx_i \text{ is even} \}, \label{eq:convexhullparitypolytop}
	\end{equation}
	where the variables $ w_{j,S} $ can be interpreted as the coefficients in the convex combinations of the incidence vectors to the even-sized subsets $ S \subseteq \{1,\dots,d \} $. 
	In Theorem 5.15 in~\cite{Feldman03PhD}, Feldman shows that $ \dot{\mathcal{Q}_j} $ can be described by the polyhedron
	\begin{IEEEeqnarray*}{cl}
		0\leq x_i \leq 1 & \quad \forall i=1,\dots,d \\
		\sum_{i \in V}x_i-\sum_{i \in N(j) \setminus V}x_i \leq \lvert V \rvert -1 & \quad \forall V \subseteq N_j \\		
		& \text{ with } \lvert V \rvert \text{ odd.} 
	\end{IEEEeqnarray*}
	Together with (\ref{eq:convexhullparitypolytop}), this shows that $ \mathcal{P}_{d,\text{even}} $ is completely characterized by the box constraints and the forbidden-set inequalities, i.\,e. $ \mathcal{P}_{d,\text{even}} $ is given by
	\begin{IEEEeqnarray}{cl}
		\hspace{60pt} 0\leq x_i \leq 1 \hspace{30pt} & \quad \forall i=1,\dots,d \nonumber  \\
		\myalignalign{\sum_{i \in V}x_i-\sum_{i \in \{1,\dots,d\} \setminus V}x_i \leq |V|-1} &
		\myalignalign*{\sum_{i \in V}x_i-\sum_{i \in \{1,\dots,d\} \setminus V}x_i \leq |V|-1}
		\begin{aligned}
		\sum_{i \in V}x_i-\sum_{i \in \{1,\dots,d\} \setminus V}x_i \leq |V|-1 & \quad \forall V \subseteq \{1,\dots,d\} \\		
		& \text{ with } |V| \text{ odd}.\label{eq:forbiddenset}
		\end{aligned}
	\end{IEEEeqnarray}
	In~\cite{Taghavi}, it is shown, that if for $ x \in [0,1]^d $, one of the forbidden-set inequalities in (\ref{eq:forbiddenset}) defines a \emph{cut}, i.\,e. it is violated, then all other forbidden-set inequalities are fulfilled with strict inequality. In particular this means that at most one forbidden-set inequality of $ \mathcal{P}_{d,\text{even}} $ is violated. In~\cite{csa}, Zhang and Siegel present their so-called cut-search algorithm, that computes this potentially violated forbidden-set inequality. It consists of two steps:
	\begin{enumerate}
		\item[1.] $ \theta_i = \text{sgn}\left(x_i-0.5\right) \quad \forall i=1,\dots,d $
		\item[2.] If $ |\{i:\theta_i=1 \}| $ is even, then determine $ i^*=\argmin_i |0.5-x_i| $ and set $ \theta_{i^*} = -\theta_{i^*} $.
	\end{enumerate}
	The inequality is then given by $ \theta^\top w \leq |V|-1=:p $, where $ |V| = |\{i:\theta_i=1 \}| $. If $ x \notin [0,1]^d $, then the cut-search algorithm for $ \Pi_{[0,1]^d}(x) $ can be used as above with $ x $ in the formulas, because $ x \geq \frac{1}{2} $ holds if and only if $ \Pi_{[0,1]}(x_i) \geq \frac{1}{2} $. For ADMM-based LP decoding, Zhang and Siegel show in~\cite{ADMMSiegel} that if $ \Pi_{[0,1]^d}(x) \notin \mathcal{P}_{d,\text{even}} $, then the projection of $ x $ onto $ \mathcal{P}_{d,\text{even}} $ lies on the face defined by the unique forbidden-set inequality, that is violated by $ \Pi_{[0,1]^d}(x) $:
	\begin{lemma}[\cite{ADMMSiegel}]\label{thm:ADMMfacet}
		Let $ x \in \mathbb{R}^d $, let $ z = \Pi_{[0,1]^d}(x) $. If $ V \subseteq \{1,\dots,d \} $ with $ |V| $ odd is a cutting set of $ z $, i.\,e. $ \theta_V^\top z > |V|-1 $, then $ \mathcal{P}_{d,\text{even}} $ must be on the face of $ \mathcal{P}_{d,\text{even}} $ defined by $ V $, i.\,e. $ \mathcal{P}_{d,\text{even}} \in F_V:= \{w\in[0,1]^d:\theta_V^\top w = |V|-1 \} $.
	\end{lemma}
	The vector $ \theta_V $ denotes the forbidden-set inequality corresponding to $ V $.
	Next, we show that these properties for $ \mathcal{P}_{d,\text{even}} $ are analogously valid for $ \mathcal{P}_{d,\text{odd}} $, where we can use forbidden-set inequalities with even instead of odd subsets:
	\begin{theorem}\label{thm:Podd}
		Let $ d \in  \mathbb{N}_{>0} $. Then it holds:
		\begin{enumerate}
			\item[i)] $\mathcal{P}_{d,\text{odd}} = \{x\in [0,1]^d:\sum_{i \in V}x_i-\sum_{i \in \{1,\dots,d\} \setminus V}x_i \leq |V|-1 \ \forall V \subseteq \{1,\dots,d \}\text{ with } |V| \text{ even} \}$\label{lemma:teil1}
			\item[ii)] If for $ x \in [0,1]^d $, one of the forbidden-set inequalities from i) is a cut, then all other forbidden-set inequalities of i) are fulfilled with strict inequality.
			\item[iii)] The cut from ii) can be found as follows: 
			\begin{enumerate}
				\item[1.] $ \theta_i = \text{sgn}(x_i-0.5) \quad \forall i=1,\dots,d $
				\item[2.] If $ |\{i:\theta_i=1 \}| $ is odd, then determine $ i^*=\argmin_i |0.5-x_i| $ and set $ \theta_{i^*} = -\theta_{i^*} $
			\end{enumerate}
			\item[iv)] 
			Let $ u \in \mathbb{R}^d $, let $ z = \Pi_{[0,1]^d}(x) $. If $ V \subseteq \{1,\dots,d \} $ with $ |V| $ even is a cutting set of $ z $, then $ \mathcal{P}_{d,\text{odd}} $ must be on the face of $ \mathcal{P}_{d,\text{odd}} $ defined by $ V $, i.\,e. $ \mathcal{P}_{d,\text{odd}} \in F_V:= \{w\in[0,1]^d:\theta_V^\top w = |V|-1 \} $.
		\end{enumerate}
	\end{theorem} 
	\begin{IEEEproof}
		\begin{enumerate}
			\item[i)] For $ d=1 $, the parity polytope has the form $ \mathcal{P}_{1,\text{odd}} = \text{conv}\{1\} = \{1\} $ and the right-hand side of i) is given by $ \{x_1\in [0,1]: -x_1\leq -1 \} = \{1\} = \mathcal{P}_{1,\text{odd}} $, because $ V=\emptyset $ defines the only forbidden-set inequality in this case.\\ Next, let us consider the case $ d\geq 2 $.
			 Analogously to the formulation $(\ref{eq:boxconstraints})-(\ref{eq:wjs}) $, we consider the polyhedron $ \mathcal{\widetilde{Q}}_1 $ defined by 
			\begin{IEEEeqnarray*}{l}
				0 \leq w_{j,S} \leq 1 \quad \forall S \in E_1 \\
				\sum_{S\in E_1}w_{1,S} = 1 \label{eq:wjsum1} \\
				x_i = \sum_{\substack{S\in E_1 \\ S\ni i}}w_{1,S} \quad \forall i =1,\dots,d, \label{eq:wjs1}
			\end{IEEEeqnarray*}
			where $ E_1=\{S\subseteq \{1,\dots,d \}: |S| \text{ odd} \} $ is the set of odd-sized subsets of $ \{1,\dots,d \} $. We denote the restriction of $\mathcal{\widetilde{Q}}_1 $ to the variables $ x_i $ by $ \dot{\mathcal{\widetilde{Q}}_1}:= \{x\in \mathbb{R}^d:\exists w: (x,w)^\top \in \mathcal{\widetilde{Q}}_1 \} $.  Since the values $ w_{1,S} $ can be interpreted as the coefficients of a convex combination of the incidence vectors to the sets $ S $, it follows that
			\begin{equation*}
			\dot{\mathcal{\widetilde{Q}}_1} = \mathcal{P}_{d,\text{odd}} = \text{conv}\{x\in \{0,1\}^d:\sum_{i=1}^{d}x_i\text{  is odd}  \}.\label{eq:Podd1}
			\end{equation*}
			We can prove almost exactly as in Theorem 5.15 in~\cite{Feldman03PhD} that
			$ \mathcal{P}_{d,\text{odd}} = \dot{\mathcal{\widetilde{Q}}_1} $ can be described by the polyhedron
			\begin{IEEEeqnarray*}{cl}
				0\leq x_i \leq 1 & \quad \forall i=1,\dots,d \\
				\sum_{i \in V}x_i-\sum_{i \in \{1,\dots,d \} \setminus V}x_i \leq |V|-1 & \quad \forall V \subseteq \{1,\dots,d \} \\		
				& \text{ with } |V| \text{ even.} 
			\end{IEEEeqnarray*}
			In the proof, we only need to replace \enquote{odd} by \enquote{even} and vice versa. Additionally, it is used in the proof of Theorem 5.15 that $\dot{\mathcal{Q}_j}$ is full-dimensional, which is proven in Theorem 2 (c) in~\cite{Jeroslow}. In our proof, we use instead that $ \dot{\mathcal{\widetilde{Q}}_1} $ is full-dimensional, which this is also shown in Theorem 2 (c) of~\cite{Jeroslow}.
			
			\item[ii)] The proof from Theorem 1 in~\cite{Taghavi} can be adopted by replacing every \enquote{odd} by \enquote{even} in the proof. The only statement, that needs to be verified, is that two indicator vectors of two distinct odd subsets have an $\ell_1$-distance of at least 2. This also holds for two distinct even subsets.
			\item[iii)] The forbidden-set inequalities 
			\begin{equation*}
			\sum_{i \in V}x_i - \sum_{i \in \{1,\dots,d\} \setminus V}x_i \leq |V|-1 
			\end{equation*}
			for all $V \subseteq \{1,\dots,d \}$ with $|V|$ even can be rewritten as
			\begin{equation*}
			\sum_{i \in V}(1-x_i) + \sum_{i \in \{1,\dots,d\} \setminus V}x_i \geq 1
			\end{equation*}
			for all $V \subseteq \{1,\dots,d \}$ with $|V|$ even.
			By ii), at most one of these inequalities is violated. If one of these inequalities is violated, it must be the one, where the left-hand side is minimal, since the right-hand side is always $ 1 $. For finding the even-sized set corresponding to this inequality, we define $ V=\{i \in \{1,\dots,d \}:x_i > 0.5\} $. If $ |V| $ is even, then we are done. If $ |V| $ is odd, then we must flip the membership of that index, which increases the sum by the smallest margin. This means that we must find the index $ i^* $, where $ |x_i-0.5| $ is minimized and include it in $ V $ if it was not contained in $ V $ before or vice versa. The cut-search algorithm computes the corresponding coefficient vector to this $ V $. Hence, this cut-search algorithm is correct.
			\item[iv)] The proof in~\cite{ADMMSiegel} can be used word-by-word, only statement ii) is needed.
		\end{enumerate}
		
	\end{IEEEproof}
	As for the polytope $ \mathcal{P}_{d,\text{even}} $, we can use $ x \notin [0,1]^d $ as an input in Theorem~\ref{thm:Podd} iii), when we want to apply this cut-search algorithm to $ \Pi_{[0,1]^d}(x) $.
	
	\section{Geometrical Idea}\label{sec:geometricidea}

	In this section, we want to explain the geometrical idea behind our new projection algorithm in an example. We want to project the point $ \hat{x} = (\frac{1}{2},1,\frac{11}{4})^\top $ onto the parity polytope 
	\begin{equation*}
	\mathcal{P}_{3,\text{even}} := \text{conv} \{x\in \{0,1\}^3:x_1+x_2+x_3 \text{ is even} \},
	\end{equation*}
	i.\,e. we want to compute
	\begin{equation*}
	\begin{pmatrix}
	z_1\\
	z_2\\
	z_3
	\end{pmatrix} := \Pi_{\mathcal{P}_{3,\text{even}}}\begin{pmatrix}
	\frac{1}{2}\\
	1\\
	\frac{11}{4}
	\end{pmatrix}.
	\end{equation*}
	As a first step, we apply the cut-search algorithm of Zhang and Siegel~\cite{csa} to
	\begin{equation*}
	\Pi_{[0,1]^3}\begin{pmatrix}
	\frac{1}{2}\\1\\ \frac{11}{4}
	\end{pmatrix}=
	\begin{pmatrix}
	\frac{1}{2}\\1\\1
	\end{pmatrix},
	\end{equation*} which outputs the forbidden-set inequality $ x_1+x_2+x_3 \leq 2 $.
	Since $ \frac{1}{2} + 1 + 1 = \frac{5}{2} > 2 $, it holds that this inequality is, indeed, violated by $(\frac{1}{2},1,1)^\top$. Hence, it follows that $ (\frac{1}{2},1,1)^\top$ is not in $\mathcal{P}_{3,\text{even}} $ and therefore not the projection of $ \hat{x} $ onto $ \mathcal{P}_{3,\text{even}} $. Hence, it follows from Lemma~\ref{thm:ADMMfacet}, that the projection of $(\frac{1}{2},1,\frac{11}{4})^\top $ onto $ \mathcal{P}_{3,\text{even}} $ lies on the face
	\begin{equation}
	F := \{x\in [0,1]^3:x_1+x_2+x_3=2 \}.\label{eq:face}
	\end{equation}
	The projection onto such a face is a difficult problem, because the face is an intersection of two sets, namely the unit hypercube $ [0,1]^3 $ and the hyperplane $ \{x\in \mathbb{R}^3:x_1+x_2+x_3=2 \} $. However, as an idea in our projection, we use the fact that the projection onto the hypercube $ [0,1]^3 $ and the projection onto a hyperplane can both be computed easily. The projection onto $ [0,1]^3 $ can be computed component-wise by mapping values greater than $ 1 $ to $ 1 $, negative values to $ 0 $ and keeping all other values unchanged. The projection onto the hyperplane can be obtained by subtracting a certain multiple of the normal vector of the hyperplane, which is $ (1,1,1)^\top $ in the example. 
	
\begin{figure}[ht]
		\centering
		\includegraphics[width=\columnwidth]{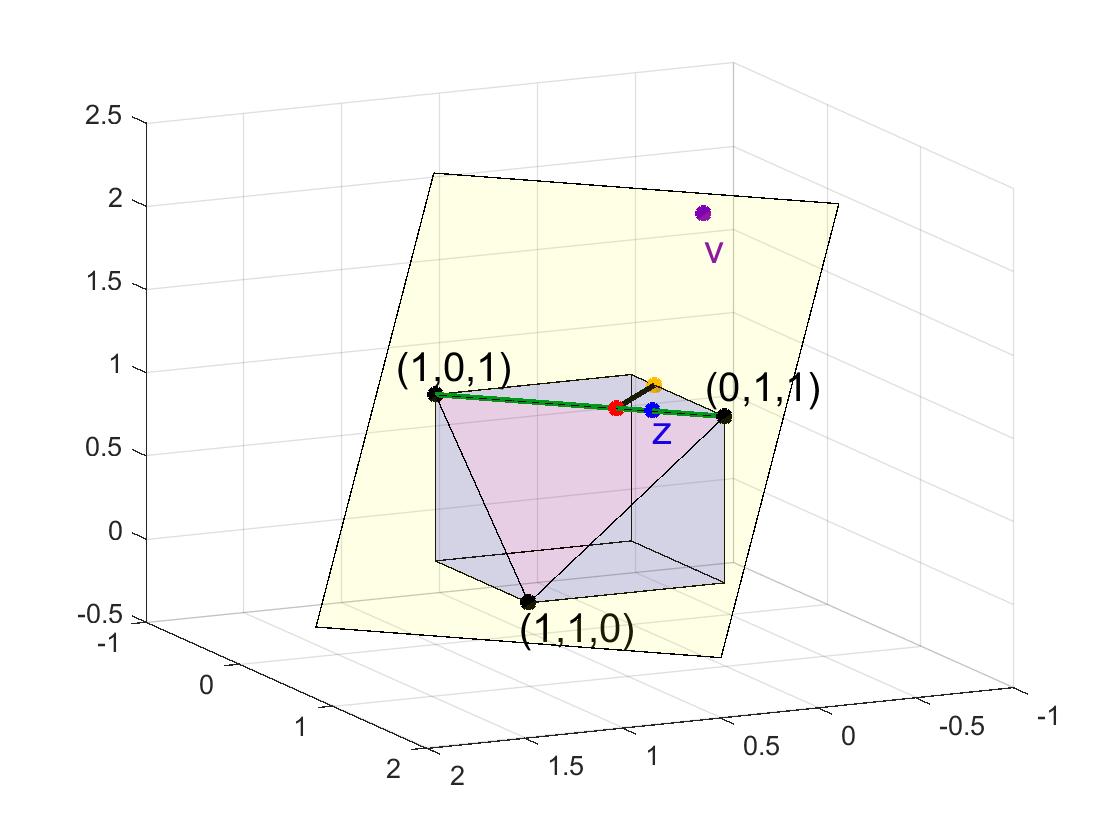}
		\caption{Geometrical Idea}\label{fig:idea}
	\end{figure}

	In our projection, we start with the projection of $ \hat{x} $ onto the hyperplane $ x_1+x_2+x_3=2 $, which leads to the point
	\begin{IEEEeqnarray*}{rcl}
		v&\ =\ &\Pi_{\{x\in \mathbb{R}^3:x_1+x_2+x_3=2\}}\begin{pmatrix}
			\frac{1}{2}\\1\\ \frac{11}{4}
		\end{pmatrix}\\
	&\ =\ &\begin{pmatrix}
		\frac{1}{2}\\1\\ \frac{11}{4}
	\end{pmatrix} - \frac{3}{4}
	\begin{pmatrix}
		1\\1\\1
	\end{pmatrix}=
	\begin{pmatrix}
		-\frac{1}{4}\\ \frac{1}{4}\\ 2
	\end{pmatrix}.
	\end{IEEEeqnarray*}
	The situation is illustrated in Figure~\ref{fig:idea}. 
	In this figure, the yellow area illustrates the hyperplane $ x_1+x_2+x_3=2 $ and the red triangle is the face $F$. The blue volume is the part of the unit hypercube, that fulfills the forbidden-set inequality $ x_1+x_2+x_3\leq 2 $, and
	the blue point $ z $ is the projection of $ \hat{x} $ onto the face, which is - at this moment - still unknown. If the violet point $ v=(-\frac{1}{4},\frac{1}{4},2)^\top $ lies in the unit hypercube $ [0,1]^3 $, then this point will also be the projection of $ \hat{x} $ onto the face $F$. However, this is not the case in this example.
	
	\paragraph{First Attempt}
	As a next idea, we project $ v $ onto $ [0,1]^3 $, which leads to the point
	\begin{equation}
	\Pi_{[0,1]^3}\begin{pmatrix}
	-\frac{1}{4}\\[0.1em] \frac{1}{4}\\[0.1em] 2
	\end{pmatrix}=
	\begin{pmatrix}
	0\\[0.1em] \frac{1}{4}\\[0.1em] 1
	\end{pmatrix}, \label{eq:projcube}
	\end{equation} the orange point in Figure~\ref{fig:idea}. This is no projection onto the face since $0 + \frac{1}{4} + 1 \neq 2$. Even projecting this new point onto the face, which leads to the red point in the figure, is not the desired projection. 
	Hence, projecting all components of $ v $ onto $ [0,1]^3 $ is not a good idea, in general. However, we claim that one component which is not yet contained in $[0,1]$ can be fixed to $0$ or $1$ for the rest of the projection.
	
	\paragraph{Second Attempt}
	As a second attempt, we try to fix the first component $v_1$. Since $v_1 < 0$, we fix $v_1$ to $0$ and obtain the point $(0,\frac{1}{4},2)^\top$.
	However, the only point on the plane $x_1 = 0$, that lies on the face $F$ is the point $(0,1,1)$, which is not the wanted projection.
	
	\paragraph{Third Attempt}
	Since $v_2$ is already contained in $[0,1]$, the next attempt is to fix the third component $v_3$ to $1$, i.\,e. we move from
	\begin{equation*}
	v = 
	\begin{pmatrix}
	-\frac{1}{4}\\[0.1em] \frac{1}{4}\\[0.1em]2
	\end{pmatrix} \text{ to }
	\begin{pmatrix}
	-\frac{1}{4}\\[0.1em] \frac{1}{4}\\[0.1em] 1
	\end{pmatrix}.  
	\end{equation*}
	As can be seen in Figure~\ref{fig:idea}, the green line is the intersection of the plane $v_3 = 1$ with the face $F$. Since $z$ is contained in this green line, fixing $v_3$ to $1$ was correct.
	
	In the second attempt, the point $(0,\frac{1}{4}, 2)^\top$ is not contained in the feasible halfspace of the forbidden-set inequality $x_1 + x_2 + x_3 \leq 2$, in contrast to the point $(-\frac{1}{4}, \frac{1}{4},1)^\top$. It is shown later that this is the criterion for choosing the correct component.
	
	Since we concluded that $ z_3=1 $, we reduce the original problem of projecting $\hat{x}$ onto $\mathcal{P}_{3,\text{even}}$ to the subproblem of projecting $(\hat{x}_1,\hat{x}_2)^\top$ onto
	\begin{equation*}
	    \mathcal{P}_{2,\text{odd}} = \text{conv} \{x\in \{0,1\}^2: x_1+x_2 \text{ is odd}  \},
	\end{equation*}
	which is exactly the green line $\{x \in F: x_1 + x_2 = 1\}$ in Figure~\ref{fig:idea}. This subproblem has one dimension less, and the \enquote{type} of the parity polytope changed from even to odd. We can now apply the same approach again. The cut-search algorithm applied to $(\frac{1}{2},1)^\top$ leads to the forbidden-set inequality $x_1 + x_2 \leq 1$, which is the same as inserting $x_3 = 1$ to the original forbidden-set inequality $x_1 + x_2 + x_3 \leq 2$. Later it is proven that this fast update procedure is always correct, such that the cut-search algorithm does not have to be applied again from scratch.

\begin{figure}[ht]
		\centering
		\includegraphics[width=\columnwidth]{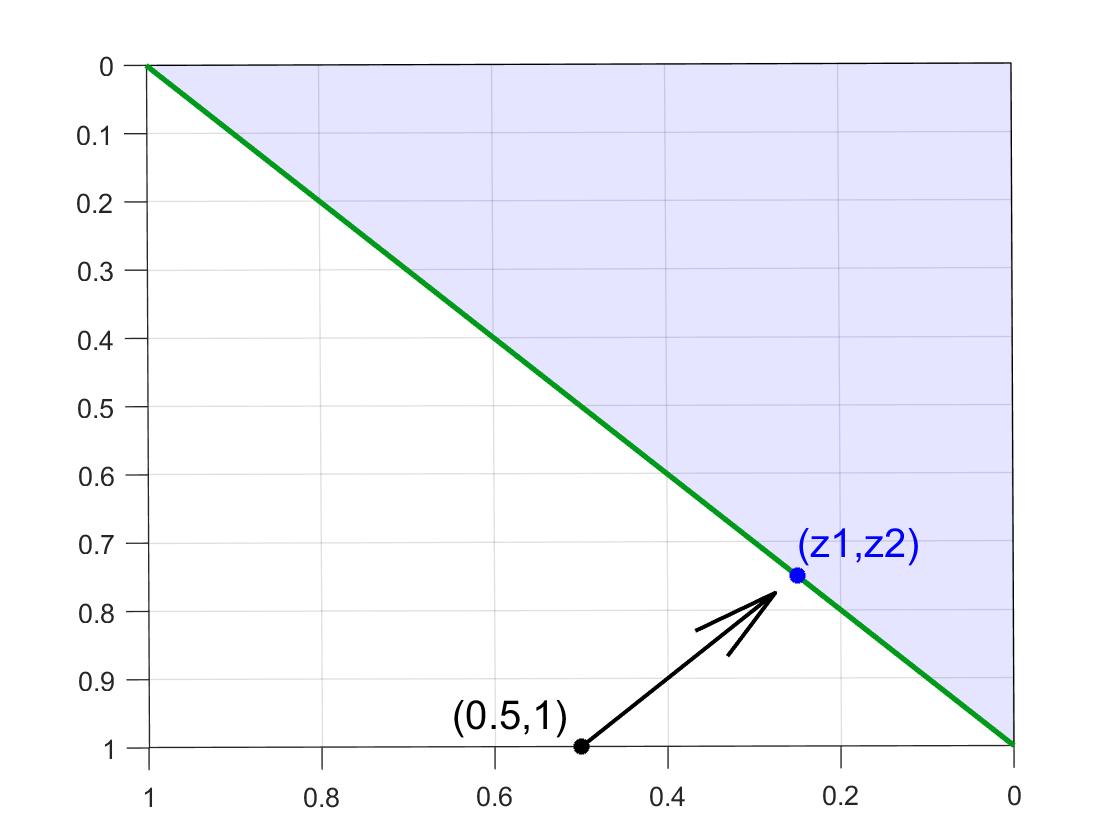}
		\caption{Recursive Projection}\label{fig:idea3}
	\end{figure}

Next, we project $ (\frac{1}{2},1)^\top $ onto the hyperplane $ x_1+x_2=1 $. This situation is illustrated in Figure~\ref{fig:idea3}.
	The resulting point on the hyperplane is
	\begin{equation*}
	\widetilde{v}=\Pi_{\{x\in \mathbb{R}^2:x_1+x_2=1\}}\begin{pmatrix}
	\frac{1}{2}\\1
	\end{pmatrix} = \begin{pmatrix}
	\frac{1}{2}\\1
	\end{pmatrix}-\frac{1}{4}
	\begin{pmatrix}
	1\\1
	\end{pmatrix}=
	\begin{pmatrix}
	\frac{1}{4}\\[0.1em] \frac{3}{4}
	\end{pmatrix}.
	\end{equation*}
	Since $ \widetilde{v} \in [0,1]^2 $, it holds that
	\begin{equation*}
	\Pi_{\mathcal{P}_{2,\text{odd}}}\begin{pmatrix}
	\frac{1}{2}\\1
	\end{pmatrix}=
	\begin{pmatrix}
	\frac{1}{4}\\[0.1em]
	\frac{3}{4}
	\end{pmatrix}.
	\end{equation*}
	Hence, the solution of the original projection problem is
	\begin{equation*}
	z = \Pi_{\mathcal{P}_{3,\text{even}}}\begin{pmatrix}
	\frac{1}{2}\\[0.1em] 1 \\[0.1em] \frac{11}{4}
	\end{pmatrix}=
	\begin{pmatrix}
	\frac{1}{4}\\[0.1em]
	\frac{3}{4}\\[0.1em]
	1
	\end{pmatrix}.
	\end{equation*}
	
	\section{Fixing Components of the Projection}\label{sec:fixingcomponents}
	In this section, we generalize the previous example and present the main theorem of this paper, that states how components in our projection can be fixed. Its proof is inspired by the geometric idea.
	
	The goal of the projection is to compute $ z = \Pi_{\mathcal{P}_{d,\text{even}}}(x) $, 
	i.\,e. the projection of some given point $ x \in \mathbb{R}^d $ onto the even parity polytope $ \mathcal{P}_{d,\text{even}} $. As in other projection algorithms (see e.\,g.~\cite{ADMMSiegel}), we start with the application of the cut-search algorithm of Zhang and Siegel~\cite{csa} to $ x \in \mathbb{R}^d $ to obtain some forbidden-set inequality $ \theta^\top w \leq p $. 
	If $ \theta^\top \Pi_{[0,1]^d}(x) \leq p $, then we know that all other forbidden-set inequalities are also fulfilled, 
	because $ \theta^\top w \leq p $ is, as the output of the cut-search algorithm, the only forbidden-set inequality of $ \mathcal{P}_{d,\text{even}} $, that is potentially violated. Hence, in this case it holds that $ \Pi_{[0,1]^d}(x) \in \mathcal{P}_{d,\text{even}} $ and that $ \Pi_{\mathcal{P}_{d,\text{even}}}(x) = \Pi_{[0,1]^d}(x) $. If $ \theta^\top \Pi_{[0,1]^d}(x) > p $, we know from Lemma~\ref{thm:ADMMfacet} that $ z $ lies on the face $ \{w\in [0,1]^d: \theta^\top w =p\} $. In this case, we compute the orthogonal projection of $ x $ onto the hyperplane $ \theta^\top w = p $. The following lemma states, that the orthogonal projection onto the hyperplane moves the point $x$ in the correct direction, namely in the direction of the desired projection $ z = \Pi_{\mathcal{P}_{d,\text{even}}}(x) $.
	
	\begin{lemma}\label{thm:proj}
		Let $ U:=\{w\in \mathbb{R}^d: \theta^\top w = p  \} $ be a hyperplane and $\emptyset\neq F:= U \cap [0,1]^d $ its intersection with the unit hypercube. Let $ x \in \mathbb{R}^d $. Then it holds that 
		$ \Pi_F(x) \in \argmin_{y\in F} \left\lVert \Pi_U(x) -y \right\rVert_2 $. This means, that the projection of $ x $ onto $ F $ is the point on $ F $, which has the smallest distance to the projection of $ x $ onto the hyperplane.
	\end{lemma}
	\begin{IEEEproof}
		Let $ z=\Pi_F(x) $ be the projection of $ x $ onto $ F $ and let $ v=\Pi_U(x) $ be the projection of $ x $ onto the hyperplane $ U $. Let $ y \in F $. The projection $ v $ can be written as $ x -\lambda \theta $ for some $ \lambda \in \mathbb{R} $, because $ \theta $ is a normal vector of the hyperplane $ \theta^\top w= p $. Hence, it follows that
		\begin{IEEEeqnarray*}{rcl}
			\left\lVert y-x\right\rVert_2^2 &\ =\ & \left\lVert y-v+v-x\right\rVert_2^2\\
			&\ =\ &\left\lVert y-v\right\rVert_2^2 + \left\lVert v-x\right\rVert_2^2 + \langle y-v,-\lambda \theta \rangle\\
			&\ =\ & \left\lVert y-v\right\rVert_2^2 + \left\lVert v-x\right\rVert_2^2 - \lambda (\theta^\top y - \theta^\top v)\\
			&\ =\ & \left\lVert y-v\right\rVert_2^2 + \left\lVert v-x\right\rVert_2^2 - \lambda (p-p)\\
			&\ =\ & \left\lVert y-v\right\rVert_2^2 + \left\lVert v-x\right\rVert_2^2.
		\end{IEEEeqnarray*}
		We can conclude that 
		\begin{equation}\label{eq:lemmaproj}
		\left\lVert y-v\right\rVert_2^2 = \left\lVert y-x\right\rVert_2^2 - \left\lVert v-x\right\rVert_2^2.
		\end{equation}
		By replacing $ y $ with $ z $, one obtains that
		\begin{equation}\label{eq:lemmaproj2}
		\left\lVert z-v\right\rVert_2^2 = \left\lVert z-x\right\rVert_2^2 - \left\lVert v-x\right\rVert_2^2.
		\end{equation}
		Hence, it follows that 
		\begin{IEEEeqnarray*}{rl}
			\left\lVert z-v\right\rVert_2^2 & \overset{(\ref{eq:lemmaproj2})}{=} \left\lVert z-x\right\rVert_2^2-\left\lVert v-x\right\rVert_2^2\\
			& \overset{\text{def. of }z}{\leq}\left\lVert y-x\right\rVert_2^2-\left\lVert v-x\right\rVert_2^2 \overset{(\ref{eq:lemmaproj})}{=}\left\lVert y-v\right\rVert_2^2.
		\end{IEEEeqnarray*}
	\end{IEEEproof}
	
	To perform the projection onto the hyperplane, we need to subtract a multiple of the normal vector of the hyperplane $ \theta^\top w = p $ from $ x $, i.\,e. we want to find the step length $\lambda$ such that $x-\lambda \theta$ lies on the hyperplane.
	Therefore, the equation
	\begin{equation*}
	p=\theta^\top (x-\lambda \theta) = \theta^\top x - \lambda \sum_{i=1}^{d}(\underbrace{\theta_i}_{\in \{\pm 1 \}})^2 = \theta^\top x - \lambda d.
	\end{equation*}
	needs to be fulfilled. Hence, the projection of $ x $ onto $ \theta^\top w = p $ is given by
	\begin{equation*}
	v = x - \frac{\theta^\top x - p}{d}\theta.
	\end{equation*}
	If $ v $ lies in $ [0,1]^d $, then $ v \in \{w\in [0,1]^d: \theta^\top w = p \} $ holds. By Lemma~\ref{thm:ADMMfacet}, this means that $ v $ is the wanted projection onto $ \mathcal{P}_{d,\text{even}} $ in this case. If $ v \notin [0,1]^d $, we claim that we can fix at least one component $ z_i $ with $ v_i \notin [0,1] $ to $ 0 $ or $ 1 $. We claim that we can fix those components $ z_i $, where the projection of $ v_i $ onto $ [0,1] $ would move the point $ v $ into the feasible halfspace $ \theta^\top w \leq p $ of the violated forbidden-set inequality. If $ v_i>1 $, this would mean that we move into the direction \(-e_i\), i.\,e. $ \theta^\top(v-e_i) \leq p $ shall be fulfilled where \(e_i\)~denotes the \(i\)-th unit vector. With $ \theta^\top v = p $ and $ \theta \in \{\pm 1 \} $, we can conclude that
	\begin{equation*}
	\theta^\top (v-e_i) \leq p \Leftrightarrow -\theta^\top e_i \leq 0 \Leftrightarrow \theta_i \geq 0 \Leftrightarrow \theta_i = 1.
	\end{equation*}
	For the case $ v_i < 0 $, projecting onto $ [0,1] $ means to move into the direction $ e_i $, i.\,e. $ \theta^\top (v+e_i) \leq p $ shall be fulfilled. In the same way, we get that
	\begin{equation*}
	\theta^\top (v+e_i) \leq p \Leftrightarrow \theta^\top e_i \leq 0 \Leftrightarrow \theta_i \leq 0 \Leftrightarrow \theta_i = -1.
	\end{equation*}
	This means that, if $ v_i>1 $ and $ \theta_i = 1 $, we claim that $ z_i=1 $. If $ v_i<0 $ and $ \theta_i=-1 $, we claim that $ z_i=0 $. This claim is formalized and proven in the following main theorem of the paper:
	\begin{theorem}\label{thm:fixcomponents} Let $ x \in \mathbb{R}^d $ with $ d \geq 2 $. Let 
		\begin{equation*}
		\theta^\top w = \sum_{j \in V}w_j - \sum_{j \in \{1,\dots,d\} \setminus V}w_j\leq |V|-1
		\end{equation*} with $ V \subseteq \{1,\dots,d \} $ ($ |V| $ even or odd) be a forbidden-set inequality. Let $ v=\Pi_{\{w\in \mathbb{R}^d:\theta^\top w = |V|-1\}}(x) $ be the projection of $ x $ onto the hyperplane $ \theta^\top w = |V|-1 $ and let $ z = \Pi_{F}(x) $ be the projection of $ x $ onto the face $F:= \{w\in [0,1]^d: \theta^\top w = |V|-1 \} $. Let $ i \in \{1,\dots,d \} $. Then it holds:
		\begin{enumerate}
			\item If $ v_i>1 $ and $ \theta_i=1 $, then $ z_i=1 $.
			\item If $ v_i<0 $ and $ \theta_i=-1 $, then $ z_i=0 $.
		\end{enumerate}
	\end{theorem}
	\begin{IEEEproof}
		to 1.: It is proven by contradiction. Assume that $ z_i <1 $:
		
		\begin{figure}[ht]
			\centering
			\includegraphics[width=\columnwidth]{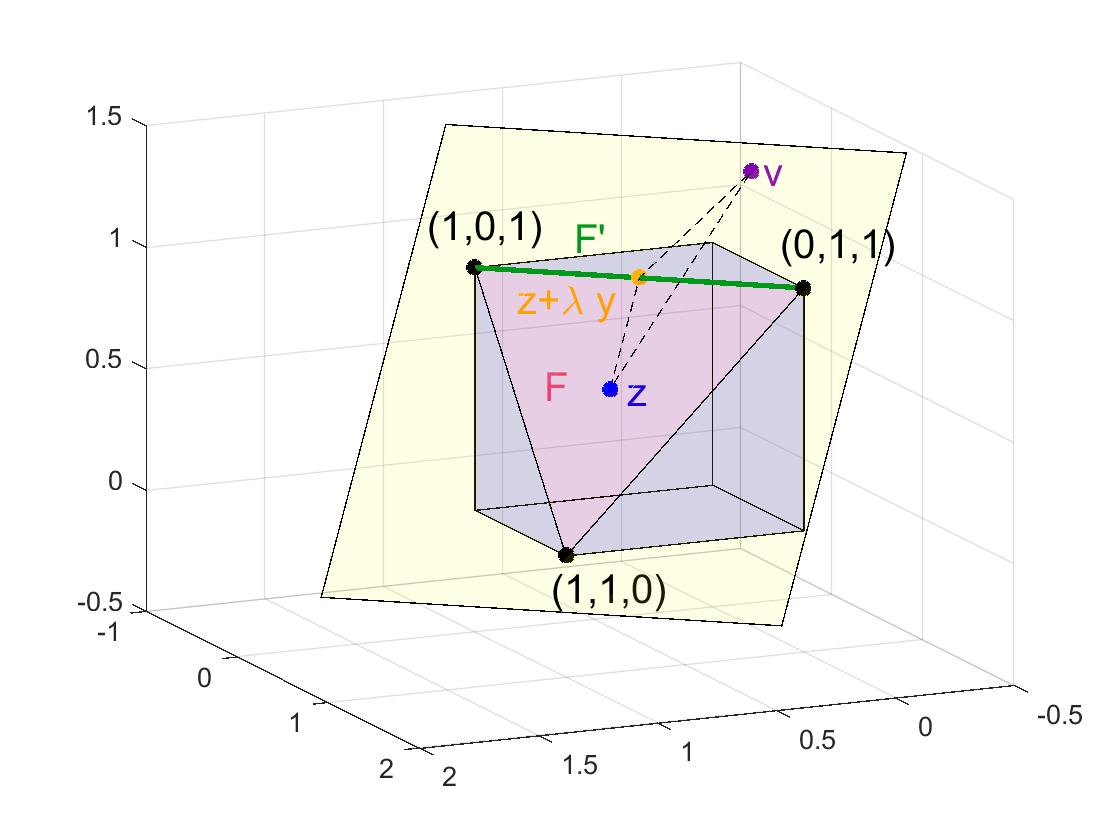}
			\caption{Geometric idea of this proof}\label{fig:proof}
		\end{figure}
		The idea of this proof is to construct another point $ z+\lambda y $ on the face $ F $, which has a shorter distance to $ v $ than $ z $. This results in a contradiction to Lemma~\ref{thm:proj}. In order to construct $ z+\lambda y $, we start in $ z $ and move along the hyperplane perpendicular to the face $ F':=\{w\in F: w_i=1 \} $ until we intersect it. The intersection point is then the wanted point in $ F $ with the shorter distance to $ v $. The situation is illustrated in Figure~\ref{fig:proof}.  
		\paragraph{Finding direction of improvement $ y $}
		For finding the improving direction $ y $, we increase the component $ z_i $ along the hyperplane $ \theta^\top w = |V|-1 $. This means that the direction $ y $ is the orthogonal projection of $ e_i $ onto $ \theta^\top w = 0 $. Since $ \theta $ is a normal vector of $ \theta^\top w = 0 $, we get that
		\begin{equation*}
		y=e_i - \frac{\langle e_i,\theta \rangle}{\langle \theta,\theta \rangle}\theta = e_i - \frac{\theta_i}{\sum_{j=1}^{d}\underbrace{\theta_j^2}_{=1}}\theta 
		\overset{\theta_i=1}{=} e_i - \frac{\theta}{d}.
		\end{equation*}
		For $ j \in \{1,\dots,d \} \setminus \{i\} $, this leads to
		\begin{equation*}
		y_j = (e_i)_j - \frac{\theta_j}{d} = -\frac{\theta_j}{d}.
		\end{equation*}
		For $ j=i $, it leads to
		\begin{equation*}
		y_i = (e_i)_i - \frac{\theta_i}{d} \overset{\theta_i=1}{=} 1 - \frac{1}{d} = \frac{d-1}{d}.
		\end{equation*}
		By multiplying with the denominator $ d $, we get that
		\begin{equation*}
		y_j = \begin{cases}
		-\theta_j& \text{ if } j\neq i\\
		d-1& \text{ if } j=i
		\end{cases}\quad \forall j=1,\dots,d.
		\end{equation*} 
		\paragraph{Finding the intersection point}
		Next, we determine the step length $ \lambda $, such that $ z+\lambda y \in F' $.  Hence, we set 
		\begin{equation*}
		1 \overset{!}{=} (z+\lambda y)_i = z_i + \lambda y_i = z_i + \lambda (d-1).
		\end{equation*}
		It follows that $ \lambda=\frac{1-z_i}{d-1} $. Next, we show that $ z+\lambda y $ lies in $F$. Since $ z \in F $ and $ \theta^\top y = 0 $, it follows by 
		\begin{equation}\label{eq:pointonplane}
		\theta^\top(z+\lambda y) = \theta^\top z +\lambda \theta^\top y = |V|-1+\lambda \cdot 0 = |V|-1
		\end{equation}
		that $z+\lambda y$ is contained in the hyperplane.
		Therefore, it is left to show that $ 0\leq z_j+\lambda y_j \leq 1 $ for all $ j \in \{1,\dots,d\} \setminus \{i\} $. Let $ j \in \{1,\dots,d\}\setminus \{i\} $.
		
		Case 1: $ \theta_j = 1 $\\
		Since $ z_j\leq 1 $, $ z_i < 1 $ and $ d > 1$, it follows that $ (z+\lambda y)_j = z_j + \frac{1-z_i}{d-1}\cdot (-\theta_j) = z_j - \frac{1-z_i}{d-1} \leq 1 $. In the following, we use that $ z $ lies on the hyperplane, i.\,e.
		\begin{equation}
		\sum_{l=1}^{d}\theta_lz_l = |V|-1 \label{eq:plane}.
		\end{equation}
		Since 
		\begin{equation}
		\theta_l=1  \Leftrightarrow  l \in V \quad \forall l=1,\dots,d \label{eq:thetaundV} 
		\end{equation}
		and since $ \theta_i=\theta_j=1 $, it follows that 
		\begin{equation}
		|\{l\in\{1,\dots,d\}\setminus \{j,i\}:l\in V  \}| = |V|-2. \label{eq:card}
		\end{equation}
		Together with $ d\geq 2$ and $ 0 \leq z_j \leq 1 $, we can conclude that
		\begin{IEEEeqnarray*}{l}
			\sum_{\substack{l=1\\ l\notin \{i,j\} }}^{d}\theta_lz_l - \underbrace{z_j}_{\geq 0}\underbrace{(d-2)}_{\geq 0} \leq \sum_{\substack{l=1\\ l\notin \{i,j\} }}^d\theta_lz_j\\
			= \sum_{\substack{l=1\\ l\notin \{i,j\}\\ l \in V }}^d\underbrace{z_j}_{\leq 1} - \sum_{\substack{l=1\\ l\notin \{i,j\}\\ l \notin V }}^d\underbrace{z_j}_{\geq 0} \leq \sum_{\substack{l=1\\ l\notin \{i,j\}\\ l \in V }}^d 1 \overset{(\ref{eq:card})}{=} |V|-2.
		\end{IEEEeqnarray*}
		By using some algebra, it follows that
		\begin{IEEEeqnarray*}{cccc}
			&\sum_{\substack{l=1\\ l\notin \{i,j\} }}^d \theta_lz_l - z_j(d-2) &\ \leq \ & |V|-2\\
			\overset{\theta_j=1}{\Leftrightarrow} & \sum_{\substack{l=1\\l\neq i}}^d \theta_lz_l - z_j(d-1) &\ \leq \ & |V|-2\\
			\overset{(\ref{eq:plane})}{\Leftrightarrow}& |V|-1-\theta_iz_i - z_j(d-1) &\ \leq \ &|V|-2\\
			\overset{-(|V|-2)}{\Leftrightarrow} & 1 -\theta_iz_i -z_j(d-1) &\ \leq \ &0\\
			\overset{\theta_i=1=\theta_j}{\Leftrightarrow}& (1-z_i)\theta_j - z_j(d-1) &\ \leq \ & 0\\
			\overset{:-(d-1)}{\Leftrightarrow}& z_j + \underbrace{\frac{1-z_i}{d-1}}_{=\lambda}\underbrace{(-\theta_j)}_{=y_j}&\ \geq \ &0 
		\end{IEEEeqnarray*}
		Hence, it follows that $ (z-\lambda y)_j \geq 0 $.
		
		Case 2: $ \theta_j = -1 $\\
		This case follows the same idea as the first case. With $ z_j \geq 0$, $ z_i < 1 $ and $ d>1 $, we get
		that
		\begin{equation*}
		(z+\lambda y)_j = z_j + \frac{1-z_i}{d-1}(-\theta_j) = z_j + \frac{1-z_i}{d-1} \geq 0.
		\end{equation*}\\
		Since $ \theta_i = 1 $ and $ \theta_j = -1 $, it follows with (\ref{eq:thetaundV}) that
		\begin{equation}
		|\{l\in\{1,\dots,d \}\setminus \{j,i\}:l\in V  \}| = |V|-1. \label{eq:cardinalitycase2}
		\end{equation}
		With $ d\geq 2 $ and $0 \leq z_j \leq 1 $, we can conclude that
		\begin{IEEEeqnarray*}{l}
			-\underbrace{(d-2)}_{\geq 0}\underbrace{(1-z_j)}_{\geq 0} + \sum_{\substack{l=1\\ l\notin \{i,j\} }}^d\theta_lz_l \leq \sum_{\substack{l=1\\ l\notin \{i,j\} }}^d\theta_lz_l\\ = \sum_{\substack{l=1\\ l\notin \{i,j\}\\ l \in V }}^d\underbrace
			{z_j}_{\leq 1} - \sum_{\substack{l=1\\ l\notin \{i,j\}\\ l \notin V }}^d\underbrace{z_j}_{\geq 0} \leq \sum_{\substack{l=1\\ l\notin \{i,j\} }}^d 1 \overset{(\ref{eq:cardinalitycase2})}{=} |V|-1.
		\end{IEEEeqnarray*}
		By using some algebra, $ \theta_j=-1 $ and $ \theta_i=1 $, we get that
		{\allowdisplaybreaks
			\begin{IEEEeqnarray*}{clcl}
				&-(d-2)(1-z_j) + \sum_{\substack{l=1\\ l\notin \{i,j\} }}^d\theta_lz_l &\ \leq \ & |V|-1\\
				\Leftrightarrow& -(d-1)(1-z_j) + 1 - z_j + \sum_{\substack{l=1\\ l\notin \{i,j\} }}^d\theta_lz_l &\ \leq \ &|V|-1\\
				\Leftrightarrow& -(d-1)(1-z_j) + 1 + \sum_{\substack{l=1\\l\neq i}}^d\theta_lz_l &\ \leq \ & |V|-1\\
				\Leftrightarrow& -(d-1)(1-z_j) + 1 + \sum_{\substack{j=1\\j\neq i}}^d\theta_lz_l - (|V|-1) &\ \leq \ & 0\\
				{\Leftrightarrow}& -(d-1)(1-z_j) + 1 - \theta_iz_i &\ \leq \ & 0\\
				\Leftrightarrow& -(d-1)(1-z_j) + (1-z_i)(-\theta_j) &\ \leq \ & 0\\
				\Leftrightarrow& (d-1)z_j + (1-z_i)(-\theta_j) &\ \leq \ & d-1\\
				\Leftrightarrow& z_j + \underbrace{\frac{1-z_i}{d-1}}_{=\lambda}\underbrace{(-\theta_j)}_{=y_j} &\ \leq \ & 1
		\end{IEEEeqnarray*}}
		Hence, it follows that $ (z-\lambda y)_j \leq 1 $.
		
		\paragraph{Distances to $ v $}
		We have shown that $ z+\lambda y \in F'\subset F $. Next, we show that $ z+\lambda y $ is a point on the face, that is closer to $ v $ than $ z $.
		
		In the following, we use that
		\begin{IEEEeqnarray}{ll}
		\begin{aligned}
			y& = \begin{pmatrix}
				-\theta_1\\ \vdots \\ -\theta_{i-1}\\d-1\\-\theta_{i+1}\\ \vdots \\ -\theta_d
			\end{pmatrix} =
			\begin{pmatrix}
				-\theta_1\\ \vdots \\ -\theta_{i-1}\\-\theta_i + \theta_i +d-1\\-\theta_{i+1}\\ \vdots \\ -\theta_d
			\end{pmatrix} \\
			& = -\theta + (\theta_i + d-1)e_i \overset{\theta_i=1}{=} -\theta + de_i.\label{eq:y}
			\end{aligned}
		\end{IEEEeqnarray}
		Calculating the distance between $v$ and $z$, we obtain
		\begin{IEEEeqnarray}{cl}
		\begin{aligned}
			&\left\lVert v-z\right\rVert_2^2= \left\lVert v-(z+\lambda y) + \lambda y\right\rVert_2^2\\
			=& \left\lVert v-(z+\lambda y)\right\rVert_2^2 + \lambda^2\left\lVert y\right\rVert_2^2 + 2\langle v-(z+\lambda y), \lambda y \rangle. \label{eq:algebra}
			\end{aligned}
		\end{IEEEeqnarray}
		The last two terms can be transformed to
		\begin{IEEEeqnarray*}{cl}
			& \lambda^2\left\lVert y\right\rVert_2^2 + 2\langle v-(z+\lambda y), \lambda y \rangle\\
			=& \lambda^2\left\lVert y\right\rVert_2^2 + 2\lambda \langle v-z,y \rangle - 2\lambda^2\left\lVert y\right\rVert_2^2\\
			\overset{(\ref{eq:y})}{=}&  - \lambda^2\left\lVert y\right\rVert_2^2 + 2\lambda \langle v-z,-\theta +de_i  \rangle\\
			=&  \lambda \left( -\lambda \left\lVert y\right\rVert_2^2 + 2 \langle v-z,-\theta +de_i  \rangle \right)\\
			=&  \lambda\left(-\lambda\left(\sum_{\substack{j=1\\j\neq i}}^d \underbrace{\left(-\theta_j\right)^2}_{=1} + \left(d-1\right)^2\right) + 2 \langle v-z,-\theta +de_i  \rangle \right)\\
			=&  \lambda\left(\frac{z_i-1}{d-1}\left(\left(d-1\right) + \left(d-1\right)^2\right) + 2 \langle v-z,-\theta +de_i  \rangle\right)\\
			=&  \lambda \left(\left(z_i-1\right)d -2\theta^\top v + 2\theta^\top z + 2d\langle v-z,e_i \rangle\right)\\
			=&  \lambda\left(\left(z_i-1\right)d - 2\left(|V|-1\right)+2\left(|V|-1\right) + 2d\left(v_i-z_i\right)\right)\\
			=&  \lambda d( z_i-1+2v_i-2z_i )\\
			=&  \underbrace{\frac{1-z_i}{d-1}}_{>0}\underbrace{d}_{>0}(\underbrace{v_i-1}_{>0}+\underbrace{v_i-z_i}_{>0})>0.
		\end{IEEEeqnarray*}
		Hence, it follows with (\ref{eq:algebra}) that
		\begin{equation*}
		\left\lVert v-z\right\rVert_2^2 > \left\lVert v-(z+\lambda y)\right\rVert_2^2.
		\end{equation*}
		However, because of Lemma~\ref{thm:proj} and $ z+\lambda y\in F $, it holds that $ \left\lVert v-z\right\rVert_2^2 > \left\lVert v-(z+\lambda y)\right\rVert_2^2 $ is a contradiction to $ z $ being the projection onto the face $ F $. Hence, the assumption $ z_i < 1 $ was wrong and it must hold that $ z_i=1 $.
		
		to 2.: The proof is similar to the first part. To obtain a contradiction, we assume that $ z_i > 0 $.
		
		We want to decrease the component of $ z_i $ along the hyperplane $ \theta^\top w = |V|-1 $. Hence, we choose the direction vector $ y $ as the orthogonal projection of $ -e_i $ onto $ \theta^\top w = 0 $. We get that
		\begin{equation*}
		y=-e_i-\frac{\langle-e_i,\theta \rangle}{\langle \theta,\theta \rangle}\theta = -e_i + \frac{\theta_i}{d}\theta \overset{\theta_i=-1}{=}-e_i-\frac{\theta}{d}.
		\end{equation*}
		For $ j \in \{1,\dots,d \} \setminus \{i \} $, this leads to 
		\begin{equation*}
		y_j = -(e_i)_j - \frac{\theta_j}{d} = -\frac{\theta_j}{d}.
		\end{equation*}
		For $ j=i $, we obtain
		\begin{equation*}
		y_i = -(e_i)_i - \frac{\theta_i}{d} \overset{\theta_i=-1}{=}-1+\frac{1}{d} = -\frac{d-1}{d}.
		\end{equation*}
		By multiplying with $ d $, we obtain the direction vector
		\begin{equation*}
		y_j = \begin{cases}
		-\theta_j& \text{ if } j\neq i\\
		-(d-1)& \text{ if } j=i
		\end{cases}\quad \forall j=1,\dots,d.
		\end{equation*} 
		For determining the step length $ \lambda $, $ z+\lambda y $ shall be the intersection point with the face $ F'=\{w\in F:w_i=0 \} $. Hence, $ z+\lambda y $ shall fulfill the equation 
		\begin{equation*}
		0 \overset{!}{=} (z+\lambda y)_i = z_i -\lambda (d-1).
		\end{equation*}
		Hence, we get that $ \lambda = \frac{z_i}{d-1} $.
		
		Again, we want to show that $ z+\lambda y \in F' \subset F $. As in (\ref{eq:pointonplane}), it follows that $ \theta^\top(z+\lambda y)=|V|-1 $. It remains to show that $0 \leq z_j+\lambda y_j \leq 1$ for all $ j \in \{1,\dots,d \} \setminus \{i\} $. For this purpose, we distinguish two cases:
		
		Case 1: $ \theta_j=1 $\\
		We get that
		\begin{equation*}
		    (z+\lambda y)_j = z_j + \frac{z_i}{d-1}(-\theta_j) = \underbrace{z_j}_{\leq 1} - \underbrace{\frac{z_i}{d-1}}_{>0} \leq 1.
		\end{equation*}
		With $ \theta_j=1=-\theta_i $ and (\ref{eq:thetaundV}) we can conclude that 
		\begin{equation}\label{eq:thetaundV2}
		|\{l\in \{1,\dots,d\}\setminus \{i,j\}:l\in V\} | = |V|-1 
		\end{equation}
		and that
		\begin{IEEEeqnarray*}{l}
			-\underbrace{(d-2)}_{\geq 0}\underbrace{z_j}_{\geq 0} + \sum_{\substack{l=1\\ l\notin \{i,j\} }}^d\theta_lz_l \leq \sum_{\substack{l=1\\ l\notin \{i,j\} }}^d\theta_lz_l\\ = \sum_{\substack{l=1\\ l\notin \{i,j\}\\ l \in V }}^d\underbrace{z_l}_{\leq 1} - \sum_{\substack{l=1\\ l\notin \{i,j\}\\ l \notin V }}^d\underbrace{z_l}_{\geq 0} \leq \sum_{\substack{l=1\\ l\notin \{i,j\}\\ l \in V }}^d1 \overset{(\ref{eq:thetaundV2})}{=} |V|-1.
		\end{IEEEeqnarray*}
		By using some algebra, we get that
		\begin{IEEEeqnarray*}{cccc}
			&-(d-2)z_j + \sum_{\substack{l=1\\ l\notin \{i,j\} }}^d\theta_lz_l &\ \leq \ & |V|-1\\
			\overset{\cdot (-1)}{\Leftrightarrow}& (d-2)z_j - \sum_{\substack{l=1\\ l\notin \{i,j\} }}^d\theta_lz_l &\ \geq \ & -(|V|-1)\\
			\overset{\theta_j=1}{\Leftrightarrow}& (d-1)z_j - \sum_{\substack{l=1\\l\neq i}}^d\theta_lz_l &\ \geq \ & -(|V|-1)\\
			\overset{+(|V|-1)}{\Leftrightarrow}& (d-1)z_j - \sum_{\substack{l=1\\l\neq i}}^d\theta_lz_l + (|V|-1) &\ \geq \ & 0\\
			\overset{(\ref{eq:plane})}{\Leftrightarrow}& (d-1)z_j + \theta_iz_i &\ \geq \ & 0\\
			\overset{\theta_i=-1=-\theta_j}{\Leftrightarrow}& (d-1)z_j + z_i(-\theta_j) &\ \geq \ & 0\\
			\overset{:(d-1)}{\Leftrightarrow}& z_j + \underbrace{\frac{z_i}{d-1}}_{=\lambda}\underbrace{(-\theta_j)}_{=y_j} &\ \geq \ & 0. 
		\end{IEEEeqnarray*}
		Hence, it holds that $ z_j + \lambda y_j \geq 0 $.
		
		Case 2: $ \theta_j=-1 $\\
		It holds that
		\begin{equation*}
		(z+\lambda y)_j = z_j + \frac{z_i}{d-1}(-\theta_j)=\underbrace{z_j}_{\geq 0} + \underbrace{\frac{z_i}{d-1}}_{>0} \geq 0.
		\end{equation*}
		In this case, we have $ \theta_i=\theta_j = -1 $. Together with (\ref{eq:thetaundV}), it follows that
		\begin{equation*}
		|\{l\in \{1,\dots,d\}\setminus \{i,j\}:l\in V \}| = |V|
		\end{equation*}
		and
		\begin{IEEEeqnarray*}{l}
			-\underbrace{(1-z_j)}_{\geq 0}\underbrace{(d-2)}_{\geq 0} + \sum_{\substack{l=1\\ l\notin \{i,j\} }}^d\theta_lz_l \leq \sum_{\substack{l=1\\ l\notin \{i,j\} }}^d\theta_lz_l\\ = \sum_{\substack{l=1\\ l\notin \{i,j\}\\ l \in V }}^d\underbrace{z_j}_{\leq 1} - \sum_{\substack{l=1\\ l\notin \{i,j\}\\ l \notin V }}^d\underbrace{z_j}_{\geq 0} \leq \sum_{\substack{l=1\\ l\notin \{i,j\}\\ l \in V }}^d1 \overset{(\ref{eq:thetaundV})}{=} |V|.
		\end{IEEEeqnarray*}
		We get that
		\begin{IEEEeqnarray*}{cccc}
			&-(1-z_j)(d-2) + \sum_{\substack{l=1\\ l\notin \{i,j\} }}^d\theta_lz_l&\ \leq \ &|V|\\
			\Leftrightarrow& -(1-z_j)(d-1) + 1-z_j + \sum_{\substack{l=1\\ l\notin \{i,j\} }}^d\theta_lz_l &\ \leq \ & |V|\\
			\overset{\theta_j=-1}{\Leftrightarrow}& -(1-z_j)(d-1) + 1 + \sum_{\substack{l=1\\l\neq i}}^d\theta_lz_l&\ \leq \ &|V|\\
			\overset{-|V|}{\Leftrightarrow}& -(1-z_j)(d-1) - (|V|-1) + \sum_{\substack{l=1\\l\neq i}}^d\theta_lz_l &\ \leq \ &0\\
			\overset{(\ref{eq:plane})}{\Leftrightarrow}& -(1-z_j)(d-1) - \theta_iz_i &\ \leq \ & 0\\
			\overset{-\theta_i=1=-\theta_j}{\Leftrightarrow}& -(1-z_j)(d-1) + z_i(-\theta_j) &\ \leq \ &0\\
			\overset{:(d-1)}{\Leftrightarrow}& -1+z_j + \frac{z_i}{d-1}(-\theta_j) &\ \leq \ &0\\
			\overset{+1}{\Leftrightarrow}& z_j + \underbrace{\frac{z_i}{d-1}}_{=\lambda}\underbrace{(-\theta_j)}_{=y_j} &\ \leq \ & 1
		\end{IEEEeqnarray*}
		Hence, it follows that $ (z+ \lambda y)_j \leq 1$ and $ z+\lambda y \in F' \subset F $.\\
		For the distance comparison, we use 
		\begin{IEEEeqnarray}{lcl}
		\begin{aligned}
			y &=& \begin{pmatrix}
				-\theta_1\\ \vdots \\ -\theta_{i-1}\\-(d-1)\\-\theta_{i+1}\\ \vdots \\ -\theta_d
			\end{pmatrix} =
			\begin{pmatrix}
				-\theta_1\\ \vdots \\ -\theta_{i-1}\\-\theta_i + \theta_i -(d-1)\\-\theta_{i+1}\\ \vdots \\ -\theta_d
			\end{pmatrix}\\
			&=& -\theta + (\theta_i - d-1)e_i \overset{\theta_i=-1}{=} -\theta - de_i.\label{eq:y1}
			\end{aligned}
		\end{IEEEeqnarray}
		As in case 1, it holds that
		\begin{equation*}
		\left\lVert v-z\right\rVert_2^2=\left\lVert v-(z+\lambda y)\right\rVert_2^2 + \lambda^2\left\lVert y\right\rVert_2^2 + 2\langle v-(z+\lambda y), \lambda y \rangle.
		\end{equation*}
		For the last two terms, we get that
		{\allowdisplaybreaks
			\begin{IEEEeqnarray*}{cl}
				&\lambda^2\left\lVert y\right\rVert_2^2 + 2\langle v-(z+\lambda y), \lambda y \rangle\\
				=& \lambda^2\left\lVert y\right\rVert_2^2 + 2\lambda \langle v-z,y \rangle - 2\lambda^2\left\lVert y\right\rVert_2^2\\
				\overset{(\ref{eq:y1})}{=}&  \lambda \left( -\lambda \left\lVert y\right\rVert_2^2 + 2 \langle v-z,-\theta -de_i  \rangle \right)\\
				=&  \lambda\left(-\lambda\left(\sum_{\substack{j=1\\j\neq i}}^d \underbrace{\left(-\theta_j\right)^2}_{=1} + \left(d-1\right)^2\right) + 2 \langle v-z,-\theta -de_i  \rangle \right)\\
				=&  \lambda\left(\frac{-z_i}{d-1}\left(\left(d-1\right) + \left(d-1\right)^2\right) + 2 \langle v-z,-\theta -de_i  \rangle\right)\\
				=&  \lambda \left(-z_id -2\theta^\top v + 2\theta^\top z - 2d\langle v-z,e_i \rangle\right)\\
				=&  \lambda\left(-z_id - 2\left(|V|-1\right)+2\left(|V|-1\right) - 2d\left(v_i-z_i\right)\right)\\
				=&  \lambda d(-z_i-2v_i+2z_i )\\
				=&  \underbrace{\frac{z_i}{d-1}}_{>0}\underbrace{d}_{>0}(\underbrace{z_i}_{>0}-2\underbrace{v_i}_{<0})>0.
		\end{IEEEeqnarray*}}
		Hence, it follows that $\left\lVert v-z\right\rVert_2^2>\left\lVert v-(z+\lambda y)\right\rVert_2^2$. Again, Lemma~\ref{thm:proj}, $ z+\lambda y\in F $ and $ \left\lVert v-z\right\rVert_2^2 > \left\lVert v-(z+\lambda y)\right\rVert_2^2 $ are a contradiction to the fact, that $ z $ is the projection of $ x $ onto the face $ F $. Hence, the assumption $ z_i>0 $ was wrong and it follows that $ z_i=0 $.
	\end{IEEEproof}
	The next theorem shows that the conditions of the last theorem are fulfilled for at least one component $ v_i $ if $ v \notin [0,1]^d $:
	\begin{theorem}\label{thm:existencedfixing} Let $ x \in \mathbb{R}^d $ 
		and let
		\begin{equation*}
		\theta^\top w = \sum_{i \in V} w_i - \sum_{i \in \{1,\dots,d\} \setminus V}w_i \leq |V|-1
		\end{equation*}
		with $ V \subseteq \{1,\dots,d \} $ ($ |V| $ odd or even) a forbidden-set inequality. Let $ v = \Pi_{\{w\in \mathbb{R}^d:\theta^\top w = |V|-1\}}(x) $ be the projection of $ x $ onto the hyperplane $ \theta^\top w = |V|-1 $ with $ v\notin [0,1]^d $. Then there exists at least one $ i \in \{1,\dots,d \} $ such that
		\begin{equation*}
		v_i > 1 \text{ and } \theta_i = 1
		\end{equation*}
		or 
		\begin{equation*}
		v_i < 0 \text{ and } \theta_i = -1.
		\end{equation*}
	\end{theorem}
	\begin{IEEEproof}
		The proof is again by contradiction:\\
		Let us assume that for all $ v_i \notin [0,1] $, it holds that
		\begin{equation}
		v_i < 0 \text{ and } \theta_i = 1 \label{eq:existencefixing1}
		\end{equation}
		or 
		\begin{equation}
		v_i > 1 \text{ and } \theta_i = -1. \label{eq:existencefixing2}
		\end{equation}
		In particular, it holds that $ v_i \leq 1 $ for all $ i=1,\dots,d $ with $ \theta_i=1 $ and that $ v_i\geq 0 $ for all $ i=1,\dots,d $ with $ \theta_i=-1 $.
		Since $ v\notin [0,1]^d $, there exists at least one component $ v_j $ with $ j \in \{1,\dots,d \} $ that fulfills (\ref{eq:existencefixing1}) or (\ref{eq:existencefixing2}). If $ v_j < 0 $ and $ \theta_j=1 $, it follows that
		\begin{IEEEeqnarray*}{lcl}
			|V|-1 &\ =\ & \theta^\top v = \sum_{\substack{i \in V\\ i \neq j}}\underbrace{v_i}_{\leq 1} + \underbrace{v_j}_{<0} - \sum_{i \in \{1,\dots,d\} \setminus V}\underbrace{v_i}_{\geq 0}\\
			&<& \sum_{\substack{i \in V\\ i \neq j}}1 = |V|-1.
		\end{IEEEeqnarray*}
		If $ v_j > 1 $ and $ \theta_j = -1 $, it holds that
		\begin{IEEEeqnarray*}{lcl}
			|V|-1&\ =\ &\theta^\top v = \sum_{i \in V}\underbrace{v_i}_{\leq 1} - \underbrace{v_j}_{>1} - \sum_{\substack{i \in \{1,\dots,d \}\setminus V \\ i \neq j}} \underbrace{v_j}_{\geq 0}\\
			&<& \sum_{i \in V}1 - 1 = |V|-1.
		\end{IEEEeqnarray*}
		Both cases lead to a contradiction. Hence, the assumption was wrong and the claim follows.
	\end{IEEEproof} 	 
	\section{Recursive Structure of the Projection}\label{sec:recursion} 
	Up to now, we established the following procedure for computing the projection $ z = \Pi_{\mathcal{P}_{d,\text{even}}}(x) $: First, we compute the potentially violated forbidden-set inequality $ \theta^\top w \leq p$ of $ \Pi_{[0,1]^d}(x) $ and check whether $ \theta^\top \Pi_{[0,1]^d}(x) \leq p$. If this is true, then $ z = \Pi_{[0,1]^d}(x) $. Otherwise, we know from Lemma~\ref{thm:ADMMfacet}, that $ z \in \{w\in [0,1]^d: \theta^\top w= p \} $ and compute the projection $ v = x - \frac{\theta^\top x - p}{d}\theta  $ of $ x $ onto the hyperplane. If $ v \in [0,1]^d $, then $ z=v $. Otherwise, Theorem~\ref{thm:existencedfixing} tells us that we can compute at least one component of $ z $ by using Theorem~\ref{thm:fixcomponents}. Next, we show that the remaining components of $ z $ are the solution of a smaller-dimensional projection problem onto $ \mathcal{P}_{\widetilde{d},\text{even}} $ or $ \mathcal{P}_{\widetilde{d},\text{odd}} $, where $ \widetilde{d}< d $. First, we consider the case that only one component of $ z $ was fixed with Theorem~\ref{thm:fixcomponents}. For this purpose, we show that the points in the parity polytopes exhibit the following recursive structure:
	
	\begin{theorem}\label{thm:Pevenodd}
		Let $ d \geq 2 $ and $ i \in \{1,\dots,d \} $. Then it holds:
		\begin{IEEEeqnarray*}{cl}
			1)\ &  \{x\in \mathcal{P}_{d,\text{even}}:x_i=1 \} =\\ & \{\widetilde{x}_1,\dots,\widetilde{x}_{i-1},1,\widetilde{x}_i,\dots,\widetilde{x}_{d-1})^\top \in \mathbb{R}^d: \widetilde{x} \in \mathcal{P}_{d-1,\text{odd}} \} \\
			2)\ & \{x\in \mathcal{P}_{d,\text{even}}:x_i=0 \} =\\ & \{(\widetilde{x}_1,\dots,\widetilde{x}_{i-1},0,\widetilde{x}_i,\dots,\widetilde{x}_{d-1})^\top \in \mathbb{R}^d: \widetilde{x} \in \mathcal{P}_{d-1,\text{even}} \} \\
			3)\ & \{x\in \mathcal{P}_{d,\text{odd}}:x_i=1 \} = \\ & \{(\widetilde{x}_1,\dots,\widetilde{x}_{i-1},1,\widetilde{x}_i,\dots,\widetilde{x}_{d-1})^\top \in \mathbb{R}^d: \widetilde{x} \in \mathcal{P}_{d-1,\text{even}} \} \\
			4)\ & \{x\in \mathcal{P}_{d,\text{odd}}:x_i=0 \} = \\ & \{(\widetilde{x}_1,\dots,\widetilde{x}_{i-1},0,\widetilde{x}_i,\dots,\widetilde{x}_{d-1})^\top \in \mathbb{R}^d: \widetilde{x} \in \mathcal{P}_{d-1,\text{odd}} \} 
		\end{IEEEeqnarray*}
	\end{theorem}
	\begin{IEEEproof}
		\text{For 1.) }: Since 
		$$ \mathcal{P}_{d,\text{even}} = \text{conv} \{ x\in \{0,1\}^d: \sum_{j=1}^{d} \text{ is even} \} $$ 
		is a convex hull of finitely many points, it follows that it is a polyhedron. Since $ \mathcal{P}_{d,\text{even}} \subseteq [0,1]^d $, it follows that $ x_i\leq 1 $ is a valid inequality and $ \{x\in \mathcal{P}_{d,\text{even}}: x_i=1 \} $ is a face of $ \mathcal{P}_{d,\text{even}} $. Hence, all extreme points of $ \{x\in \mathcal{P}_{d,\text{even}}:x_i=1 \} $ are also extreme points of $ \mathcal{P}_{d,\text{even}} $. Together with the fact that all extreme points of $ \mathcal{P}_{d,\text{even}} $ are binary vectors, it follows that $ \{x\in \mathcal{P}_{d,\text{even}}:x_i=1 \} $ does also only have binary vectors as extreme points. Since $ \{x\in \mathcal{P}_{d,\text{even}}:x_i=1 \} \subseteq [0,1]^d $, it even follows that its binary solutions are exactly its extreme points. For any binary vector $ x \in \{0,1\}^d $, it holds that
		\begin{IEEEeqnarray}{ll}
			&x\in \mathcal{P}_{d,\text{even}} \text{ and } x_i=1\label{eq:Pevenodd0} \\
			\Leftrightarrow &\ x_i=1 \text{ and } \sum_{j=1}^{d}x_j \text{ is even}\\
			\Leftrightarrow &\ x_i=1 \text{ and } \sum_{\substack{j=1\\j\neq i}}^dx_j \text{ is odd}\label{eq:Pevenodd}
		\end{IEEEeqnarray}
		Since $ \{x\in \mathcal{P}_{d,\text{even}}:x_i=1 \} \subseteq [0,1]^d $ is a polytope, it follows from Minkowski's Theorem that $ \{x\in \mathcal{P}_{d,\text{even}}:x_i=1 \} $ is the convex hull of its extreme points, i.\,e.
		{\allowdisplaybreaks
			\begin{IEEEeqnarray*}{ll}
				&\{x\in \mathcal{P}_{d,\text{even}}:x_i=1 \}\\
				\overset{(\ref{eq:Pevenodd})}{=}& \text{conv}\left\{		
				(\widetilde{x}_1,\dots,\widetilde{x}_{i-1},1,\widetilde{x}_i,\dots,\widetilde{x}_{d-1})^\top \in \{0,1\}^d:\phantom{\sum_{j=1}^{d-1}} \right.\\
				&\left. \hspace*{10mm} \sum_{j=1}^{d-1}\widetilde{x}_j \text{ is odd} \right\}\\
				\overset{(*)}{=}&
				\left\{ \begin{pmatrix}
					\sum_{l=1}^L\lambda_l \begin{pmatrix}
						\widetilde{x}_1^l\\ \vdots \\ \widetilde{x}_{i-1}^l
					\end{pmatrix}\\
					1\\
					\sum_{l=1}^L\lambda_l \begin{pmatrix}
						\widetilde{x}_{i}^l\\ \vdots \\ \widetilde{x}_{d-1}^l
					\end{pmatrix}
				\end{pmatrix}| \sum_{l=1}^L\lambda_l=1; \widetilde{x}^l\in \{0,1\}^{d-1};\right.\\
				&\left. \hspace*{5mm}  \sum_{j=1}^{d-1}\widetilde{x}_j^l \text{ is odd}; 0\leq \lambda_l \leq 1 \ \forall l \in \{1,\ldots,L\}, L\in \mathbb{N}_{>0} \right\} \\
				=& \{(\widetilde{x}_1,\dots,\widetilde{x}_{i-1},1,\widetilde{x}_i,\dots,\widetilde{x}_{d-1})^\top \in \mathbb{R}^d: \widetilde{x}\in \mathcal{P}_{d-1,\text{odd}} \}.
		\end{IEEEeqnarray*}} 
		In $(*)$, we use that $ \sum_{l=1}^{L}\lambda_l\cdot 1=1 $.
		The proofs for 2) to 4) are completely analogous to 1). In 2) and 4), we need to replace $ 1 $ by $ 0 $ in the corresponding $ i $-th components and use $ \sum_{l=1}^L \lambda_l\cdot 0 = 0 $ instead of $ \sum_{l=1}^L\lambda_l=1 $. The equivalences in (\ref{eq:Pevenodd0}) - (\ref{eq:Pevenodd}) are replaced by:
		
		\begin{enumerate}
		    \item[2)] 
		    $\begin{aligned}[t]
			& x\in \mathcal{P}_{d,\text{even}} \ \text{and}\  x_i=0\\  \Leftrightarrow \quad & x_i=0\ \text{and} \ \sum_{j=1}^{d}x_j \text{ is even} \\  \Leftrightarrow \quad & x_i=0 \ \text{and} \sum_{\substack{j=1\\j\neq i}}^dx_j \text{ is even.}
			\end{aligned}$
					    \item[3)] 
		    $\begin{aligned}[t]
			& x\in \mathcal{P}_{d,\text{odd}} \ \text{and}\  x_i=1\\  \Leftrightarrow \quad & x_i=1\ \text{and} \ \sum_{j=1}^{d}x_j \text{ is odd} \\  \Leftrightarrow \quad & x_i=1 \ \text{and} \sum_{\substack{j=1\\j\neq i}}^dx_j \text{ is even.}
			\end{aligned}$
					    \item[4)] 
		    $\begin{aligned}[t]
			& x\in \mathcal{P}_{d,\text{odd}} \ \text{and}\  x_i=0\\  \Leftrightarrow \quad & x_i=0\ \text{and} \ \sum_{j=1}^{d}x_j \text{ is odd} \\  \Leftrightarrow \quad & x_i=0 \ \text{and} \sum_{\substack{j=1\\j\neq i}}^dx_j \text{ is odd.}
			\end{aligned}$
		\end{enumerate}
	\end{IEEEproof} 
	Next, we show that after fixing one component of $ z_i $, the remaining entries of $ z $ are again the solution of a projection problem onto a parity polytope. For this purpose, we show that fixing one component reduces the problem to a projection of the remaining components onto a smaller-dimensional parity polytope.
	\begin{theorem}\label{thm:recursion}
		\begin{enumerate}
			\item Let $ z=\Pi_{\mathcal{P}_{d,\text{even}}}(x) $ with $ z_i=1 $ for some $ i \in \{1,\dots,d \} $. Then it holds: 
			\begin{IEEEeqnarray*}{l}
				(z_1,\dots,z_{i-1},z_{i+1},\dots,z_d) =\\
				\Pi_{\mathcal{P}_{d-1},\text{odd}}(x_1,\dots,x_{i-1},x_{i+1},\dots,x_d)^\top
			\end{IEEEeqnarray*} 
			\item Let $ z=\Pi_{\mathcal{P}_{d,\text{even}}}(x) $ with $ z_i=0 $ for some $ i \in \{1,\dots,d \} $. Then it holds:
			\begin{IEEEeqnarray*}{l} (z_1,\dots,z_{i-1},z_{i+1},\dots,z_d) =\\ \Pi_{\mathcal{P}_{d-1},\text{even}}(x_1,\dots,x_{i-1},x_{i+1},\dots,x_d)^\top 
			\end{IEEEeqnarray*}
			\item Let $ z=\Pi_{\mathcal{P}_{d,\text{odd}}}(x) $ with $ z_i=1 $ for some $ i \in \{1,\dots,d \} $. Then it holds:
			\begin{IEEEeqnarray*}{l}
				(z_1,\dots,z_{i-1},z_{i+1},\dots,z_d) =\\ \Pi_{\mathcal{P}_{d-1},\text{even}}(x_1,\dots,x_{i-1},x_{i+1},\dots,x_d)^\top 
			\end{IEEEeqnarray*}
			\item Let $ z=\Pi_{\mathcal{P}_{d,\text{odd}}}(x) $ with $ z_i=0 $ for some $ i \in \{1,\dots,d \} $. Then it holds:
			\begin{IEEEeqnarray*}{l}
				(z_1,\dots,z_{i-1},z_{i+1},\dots,z_d) =\\ \Pi_{\mathcal{P}_{d-1},\text{odd}}(x_1,\dots,x_{i-1},x_{i+1},\dots,x_d)^\top 
			\end{IEEEeqnarray*}
		\end{enumerate}
	\end{theorem}
	\begin{IEEEproof}
		It follows directly from Theorem~\ref{thm:Pevenodd} and 
		\begin{IEEEeqnarray*}{rcl}
			\left\lVert z-x\right\rVert_2^2&=&\sum_{j=1}^{d}(z_j-x_j)^2 = \sum_{\substack{j=1\\j\neq i}}^{d}(z_j-x_j)^2 + |z_i-x_i|^2\\
			&=&\lVert (z_1,\dots,z_{i-1},z_{i+1},\dots,z_d)-\\
			&&(x_1,\dots,x_{i-1},x_{i+1},\dots,x_d)\rVert_2^2 + |z_i-x_i|^2.
		\end{IEEEeqnarray*}
	\end{IEEEproof}
	If more than one component of $ z $ was fixed with Theorem~\ref{thm:fixcomponents}, then Theorem~\ref{thm:recursion} can be applied several times inductively so that the remaining components of $ z $ are again the solution of a projection problem onto a parity polytope.
	
	Hence, the idea of our algorithm is to repeat our mentioned steps (see the beginning of this section) recursively in order to solve the corresponding smaller-dimensional projection problem onto some parity polytope $ \mathcal{P}_{\widetilde{d},\text{even}} $ or $ \mathcal{P}_{\widetilde{d},\text{odd}} $ with $ \widetilde{d} < d $. Since we showed in Theorem~\ref{thm:Podd} that the required properties for the parity polytope $ \mathcal{P}_{\widetilde{d},\text{even}} $ from the literature are also true in the analogous way for $ \mathcal{P}_{\widetilde{d},\text{odd}} $, it is not relevant in our problem, on which type of parity polytope we need to project. In the following, we will just talk about (csa), when we mean the cut-search algorithm of Zhang and Siegel~\cite{csa} for $ \mathcal{P}_{\widetilde{d},\text{even}} $ or the analog cut-search algorithm from Theorem~\ref{thm:Podd} iii) for $ \mathcal{P}_{\widetilde{d},\text{even}} $, respectively.
	
	The first step in our proposed projection algorithm was the application of (csa) to $ \Pi_{[0,1]^d}(x)$. Let us assume we are now in the recursive part of the algorithm with $ \widetilde{d} < d$, i.\,e. we  already fixed some components of $ z $ and want to project the remaining components $ \widetilde{x} \in \mathbb{R}^{\widetilde{d}} $ of $ x $ to the corresponding (even or odd) smaller-dimensional parity polytope. Next, we will show that the (csa) applied to $ \Pi_{[0,1]^{\widetilde{d}}}(\widetilde{x}) $ can be computed very efficiently. We do not need to apply (csa) again from the scratch. Instead, we can use the output $ \theta^\top w \leq p $ with $ \theta \in \{\pm 1\}^d $ of (csa) applied to $ \Pi_{[0,1]^{d}}(x) $, which was computed in the previous step. The output of (csa) for $ \Pi_{[0,1]^{\widetilde{d}}}(\widetilde{x}) $ is then given by the components of $ \theta $, where $ z_i $ is not yet fixed. The right-hand side $ \widetilde{p} $ is the right-hand side of the previous step minus the number of components of $ z $ that were fixed to $ 1 $ in the previous step. The next theorem shows this statement for the case that one component of $ z $ was fixed in the previous step.

	\begin{theorem}\label{thm:csarecursion}
		Let $ x=(\widetilde{x}_1,\dots,\widetilde{x}_{i-1},y,\widetilde{x}_i,\dots,\widetilde{x}_{d-1})^\top \in \mathbb{R}^d $ with $ d \geq 2 $.
		\begin{enumerate}
			\item Let $ (\theta,p)\in \mathbb{R}^{d+1} $ be the forbidden-set inequality returned by (csa) applied to $x$ and $ \mathcal{P}_{d,\text{even}} $. Let $ \theta_i=1 $. Then $ ((\theta_1,\dots,\theta_{i-1},\theta_{i+1},\dots,\theta_d)^\top,p-1) $ is the output of (csa) applied to $ \widetilde{x} $ and $ \mathcal{P}_{d-1,\text{odd}} $.
			\item Let $ (\theta,p)\in \mathbb{R}^{d+1} $ be the forbidden-set inequality returned by (csa) applied to $ x $ and $ \mathcal{P}_{d,\text{odd}} $. Let $ \theta_i=1 $. Then $ ((\theta_1,\dots,\theta_{i-1},\theta_{i+1},\dots,\theta_d)^\top,p-1) $ is the output of (csa) applied to $ \widetilde{x} $ and $ \mathcal{P}_{d-1,\text{even}} $.
			\item Let $ (\theta,p)\in \mathbb{R}^{d+1} $ be the forbidden-set inequality returned by (csa) applied to $ x $ and $ \mathcal{P}_{d,\text{even}} $. Let $ \theta_i=-1 $. Then $ ((\theta_1,\dots,\theta_{i-1},\theta_{i+1},\dots,\theta_d)^\top,p) $ is the output of (csa) applied to $ \widetilde{x} $ and $ \mathcal{P}_{d-1,\text{even}} $.
			\item Let $ (\theta,p)\in \mathbb{R}^{d+1} $ be the forbidden-set inequality returned by (csa) applied to $ x$ and $ \mathcal{P}_{d,\text{odd}} $. Let $ \theta_i=-1 $. Then $ ((\theta_1,\dots,\theta_{i-1},\theta_{i+1},\dots,\theta_d)^\top,p) $ is the output of (csa) applied to $ \widetilde{x} $ and $ \mathcal{P}_{d-1,\text{odd}} $.
		\end{enumerate}
		
	\end{theorem}
	
	\begin{IEEEproof}
	We use that the right-hand side of any forbidden-set inequality $ \theta^\top w \leq p $ is given by $ p = |\{i\in \{1,\dots,\}: \theta_i=1 \}|-1 $.
	
		to 1): We consider two cases:
		
		Case 1: $ |\{j\in \{1,\dots,d\}:x_j>0.5 \}| $ is odd\\
		In this case, the (csa) for $ x $ and $ \mathcal{P}_{d,\text{even}} $ stops after step 1 and it holds that $ \theta_j= 1$ if and only if $ {x}_j > 0.5 $ for all $ j=1,\dots,d $. Since $ \theta_i=1 $, it holds that $ y>0.5 $ and that
		\begin{IEEEeqnarray*}{l}
			|\{j\in \{1,\dots,d \}\setminus \{i\}:x_j > 0.5 \} | =\\ |\{j\in \{1,\dots,d\}:x_j>0.5 \} |-1
		\end{IEEEeqnarray*}
		is even. This means that the (csa) for $ \widetilde{x} $ and $ \mathcal{P}_{d-1,\text{odd}} $ also stops after step 1 and computes\\ $ ((\theta_1,\dots,\theta_{i-1},\theta_{i+1},\dots,\theta_d)^\top,p-1) $.
		
		Case 2: $ |\{j\in \{1,\dots,d\}:x_j>0.5 \}| $ is even\\
		In this case, the (csa) applied to $ x $ and $ \mathcal{P}_{d,\text{even}} $ would need to do step 2, i.\,e. it would compute $ i^* $ and set $ \theta_{i^*} := - \theta_{i^*} $. We consider two cases:
		
		Case 2a: $ i^*=i $\\
		Because of $ \theta_i=1 $ and $ i^*=i $, it holds in this case that $ y\leq 0.5 $. Then, the (csa) for $ \widetilde{x} $ and $ \mathcal{P}_{d-1,\text{odd}} $ would compute the vector $ (\theta_1,\dots,\theta_{i-1},\theta_{i+1},\dots,\theta_d)^\top $ in step 1 and stop because
		\begin{IEEEeqnarray*}{l}
			|\{j\in \{1,\dots,d \}\setminus \{i\}: x_j > 0.5 \}| =\\ |\{j\in \{1,\dots,d \}:x_j > 0.5  \} | 
		\end{IEEEeqnarray*}
		is even. The right-hand side is $ p-1=|\{j\in \{1,\dots,d \}\setminus \{i\}: x_j > 0.5 \}|-1 $, because no component of $ \theta $ is flipped and $ p-1 $, the right-hand side after step 1, is not increased to $ p $ in step 2, as it is done in the (csa) applied to $ x $ and $ \mathcal{P}_{d,\text{even}} $.
		
		Case 2b: $ i^* \neq i $\\
		Because of $ \theta_i=1 $ and $ i^*\neq i $, it holds that $ y>0.5 $. Hence, it follows that 
		\begin{IEEEeqnarray}{l}
		\begin{aligned}
			|\{j\in \{1,\dots,d\}\setminus \{i\}:x_j > 0.5 \} | =\\ |\{j\in \{1,\dots,d\}:x_j > 0.5 \} |-1 \label{eq:proofcsa}
			\end{aligned}
		\end{IEEEeqnarray}
		is odd. Since, $ i^*\neq i $, the same entry of $ \theta $ is flipped by (csa) for $ x $ and $\widetilde{x} $. Hence, the (csa) applied to $\widetilde{x} $ and $ \mathcal{P}_{d-1,\text{odd}} $ would output the coefficient vector $ (\theta_1,\dots,\theta_{i-1},\theta_{i+1},\dots,\theta_d)^\top $. From (\ref{eq:proofcsa}) and the flipping of $ \theta_{i^*} $ in the (csa) for $ x $ and $ \widetilde{x} $, it follows that the right-hand side for $ \widetilde{x} $ is one less than the one for $ x$ after step 1 and step 2 of (csa). Hence, the right-hand side $ p-1 $ is outputted by the (csa) applied to $ \widetilde{x} $ and $ \mathcal{P}_{d-1,\text{odd}} $.\\
		
		to 2): the proof of 1) can be used, where every \enquote{odd} is replaced by \enquote{even} and vice versa.
		
		to 3): We consider two cases:
		
		Case 1: $ |\{j\in \{1,\dots,d\}: x_j > 0.5 \} | $ is odd\\
		In this case, the (csa) applied to $x $ and $ \mathcal{P}_{d,\text{even}} $ stops after step 1 and $ x_j > 0.5 $ is equivalent to $ \theta_j=1 $ for all $ j=1,\dots,d $. Hence, it follows from $ \theta_i=-1 $ that $ y \leq 0.5 $ and that
		\begin{IEEEeqnarray*}{l}
			|\{j\in \{1,\dots,d \}\setminus \{i\}: {x}_j > 0.5 \} | =\\ |\{j\in\{1,\dots,d\}: {x}_j > 0.5  \} |
		\end{IEEEeqnarray*}
		is odd. Hence, the (csa) applied to $ \widetilde{x}$ also stops after step 1 and outputs 
		$ ((\theta_1,\dots,\theta_{i-1},\theta_{i+1},\dots,\theta_d)^\top,p) $.
		
		Case 2: $ |\{j\in \{1,\dots,d \}: x_j > 0.5 \} | $ is even\\
		In this case, (csa) applied to $x$ and $ \mathcal{P}_{d,\text{even}} $ is not finished after step 1 and has to flip $ \theta_{i^*} $. There can occur two cases:
		
		Case 2a: $ i^*=i $.\\
		Since $ \theta_i=-1 $, it must hold that $ y > 0.5 $ and that
		\begin{IEEEeqnarray*}{l}
			|\{j\in \{1,\dots,d \}\setminus \{i\}:x_j > 0.5 \}| =\\ |\{j\in \{1,\dots,d \}:x_j>0.5 \}|-1
		\end{IEEEeqnarray*}
		is odd. Hence, the (csa) applied to $ \widetilde{x}$ and $ \mathcal{P}_{d-1,\text{even}} $ stops after step 1 and returns the coefficient vector $ (\theta_1,\dots,\theta_{i-1},\theta_{i+1},\dots,\theta_d)^\top $. After step 1 of (csa), the right-hand side to $ x $ is one unit larger than the one of $ \widetilde{x} $. However, since the (csa) applied to $ x $ flips $ \theta_i $ from $ 1 $ to $ -1 $, the computed right-hand sides to $ x $ and $ \widetilde{x} $ are the same after the termination of (csa). Hence the (csa) applied to $ \widetilde{x} $ and $ \mathcal{P}_{d-1,\text{even}} $ returns $ ((\theta_1,\dots,\theta_{i-1},\theta_{i+1},\dots,\theta_d)^\top,p) $.
		
		Case 2b: $ i^*\neq i $\\
		In this case, $ \theta_i $ is always $ -1 $ during the whole (csa) applied to $ x$ and $ \mathcal{P}_{d,\text{even}} $. Hence, it holds that $ y\leq 0.5 $ and that
		\begin{IEEEeqnarray*}{l}
			|\{j\in \{1,\dots,d \}\setminus \{i\}:x_j>0.5 \}| =\\ |\{j\in \{1,\dots,d \}: x_j > 0.5 \}|
		\end{IEEEeqnarray*}
		is even. Hence, the (csa) for $ x $ and $ \widetilde{x} $ (with $ \mathcal{P}_{d,\text{even}} $ and $ \mathcal{P}_{d-1,\text{even}} $) will compute the same right-hand side in step 1 and flip the same entry of $ \theta $. Hence, the (csa) applied to $ \widetilde{x}$ and $ \mathcal{P}_{d-1,\text{even}} $ outputs\\ $ ((\theta_1,\dots,\theta_{i-1},\theta_{i+1},\dots,\theta_d)^\top,p) $.
		
		to 4): the proof of 3) can be used by replacing every \enquote{even} by \enquote{odd} and vice versa. 
	\end{IEEEproof}
	If more than one component of $ z $ was fixed in the last iteration, then the last theorem can be applied several times inductively. The next step after computing the potentially violated forbidden-set inequality $ \widetilde{\theta}^\top \widetilde{w} \leq \widetilde{p} $ of $ \Pi_{[0,1]^{\widetilde{d}}}(\widetilde{x}) $ would be the check whether $ \widetilde{\theta}^\top \Pi_{[0,1]^{\widetilde{d}}}(\widetilde{x}) \leq \widetilde{p} $. If the check is fulfilled, then $\Pi_{[0,1]^{\widetilde{d}}}(\widetilde{x})$ would be the projection of $ \widetilde{x} $ onto $ \mathcal{P}_{\widetilde{d},\text{even}} $ or $ \mathcal{P}_{\widetilde{d},\text{odd}} $, respectively. However, we show that this check is never fulfilled and therefore redundant in the recursive part of our projection algorithm. If the check is true, then the corresponding very first check in the beginning of the algorithm would have also been true and the algorithm would have terminated in the beginning with $ \Pi_{[0,1]^{d}}(x) $. This is shown in the next theorem:     
	\begin{theorem}\label{thm:checkcube}
		The check $ \theta^\top \Pi_{[0,1]^d}(x) \leq p $ is only necessary in the beginning of the proposed projection algorithm, i.\,e. it is redundant in the recursive calls of the algorithm. 
	\end{theorem}
	\begin{IEEEproof}
		Let $ x \in \mathbb{R}^d $ and $ (\theta,p) $ the input of a recursive call of the described projection algorithm. Since it is a recursive call, at least one component of the wanted projection $ z $ was fixed before in the algorithm by using Theorem~\ref{thm:fixcomponents}. With Theorem~\ref{thm:csarecursion}, it follows that $ \theta^\top w\leq p $ is the output of the corresponding cut-search algorithm applied to $ \Pi_{[0,1]^d}(x) $ and $ \mathcal{P}_{d,\text{even}} $ (or $ \mathcal{P}_{d,\text{odd}} $). If we undo the last fixing of a component $ z_j $ in the algorithm, we obtain a larger vector $ \begin{pmatrix}
		x\\ \widetilde{x}_{d+1}
		\end{pmatrix} \in \mathbb{R}^{d+1}$. (Without loss of generality, the last component of $ z $ was fixed). There are two possible cases when we fix components of $ z $ according to Theorem~\ref{thm:fixcomponents}:
		
		Case 1: $ \theta_{d+1} = 1 $ and $ v_{d+1}> 1 $
		
		From the way, how components of $ z $ are fixed in Theorem~\ref{thm:fixcomponents}, and from Theorem~\ref{thm:csarecursion}, it follows that
		\begin{equation}
		\begin{pmatrix}
		\theta \\ 1
		\end{pmatrix}^\top \widetilde{w} \leq p+1 \label{eq:largercheck1}
		\end{equation}
		is the result of the cut-search algorithm applied to $ \Pi_{[0,1]^{d+1}}\begin{pmatrix}
		x\\ \widetilde{x}_{d+1}
		\end{pmatrix} $ with $ \mathcal{P}_{d+1,\text{odd}} $ (or $ \mathcal{P}_{d+1,\text{even}} $).\\ If $ \theta^\top \Pi_{[0,1]^d}(x) \leq p $, then it holds that 
		\begin{IEEEeqnarray*}{l}
			\begin{pmatrix}
				\theta \\ 1
			\end{pmatrix}^\top \Pi_{[0,1]^{d+1}}\begin{pmatrix}
				x\\ \widetilde{x}_{d+1}
			\end{pmatrix} =\\ \underbrace{\theta^\top \Pi_{[0,1]^d}(x)}_{\leq p} + \underbrace{\Pi_{[0,1]}(\widetilde{x}_{d+1})}_{\leq 1} \leq p+1.
		\end{IEEEeqnarray*}
		
		Case 2: $ \theta_{d+1}=-1 $ and $ v_{d+1}<0 $
		
		From Theorem~\ref{thm:csarecursion} and the way, how components of $ z $ are fixed in Theorem~\ref{thm:fixcomponents}, it follows that 
		\begin{equation}
		\begin{pmatrix}
		\theta \\ -1
		\end{pmatrix}^\top \widetilde{w} \leq p \label{eq:largercheck2}
		\end{equation}
		is the result of the cut-search algorithm applied to $ \Pi_{[0,1]^{d+1}}\begin{pmatrix}
		x\\ \widetilde{x}_{d+1}
		\end{pmatrix} $ with $ \mathcal{P}_{d+1,\text{even}} $ (or $ \mathcal{P}_{d+1,\text{odd}} $).\\ If $ \theta^\top \Pi_{[0,1]^d}(x) \leq p $, then it holds that
		\begin{IEEEeqnarray*}{l}
			\begin{pmatrix}
				\theta \\ -1
			\end{pmatrix}^\top \Pi_{[0,1]^{d+1}}\begin{pmatrix}
				x\\ \widetilde{x}_{d+1}
			\end{pmatrix} =\\ \underbrace{\theta^\top \Pi_{[0,1]^d}(x)}_{\leq p} - \underbrace{\Pi_{[0,1]}(\widetilde{x}_{d+1})}_{\geq 0} \leq p.
		\end{IEEEeqnarray*}
		Summarizing both cases, we can conclude that if the check in Theorem~\ref{thm:checkcube} is fulfilled during a recursive call of the proposed projection algorithm, then the corresponding vector $ \begin{pmatrix}
		x\\ \widetilde{x}_{d+1}
		\end{pmatrix} $ fulfills the corresponding higher-dimensional check (\ref{eq:largercheck1}) or (\ref{eq:largercheck2}). Inductively, it follows that if the check in Theorem~\ref{thm:checkcube} is fulfilled, then the very first check of the algorithm, that tests whether the projection onto the unit hypercube lies in the parity polytope, was also fulfilled. But in this case, the algorithm terminates with the projection onto the unit hypercube without starting the recursion. Hence, if our projection algorithm is currently in a recursive call, then the check from Theorem~\ref{thm:checkcube} is never fulfilled, i.\,e. we can omit the check during the recursive part of the proposed algorithm.
	\end{IEEEproof}
	After the projection onto the hyperplane, there are three possibilites: A component $ v_i $ could lie outside of $ [0,1] $ and fulfill one of the two fixing conditions from Theorem~\ref{thm:fixcomponents}, such that we can compute $ z_i $, or $ v_i $ lies outside of $ [0,1] $ and cannot be fixed, or $ v_i \in [0,1] $. If all components of $ v $ lie in $ [0,1] $, the algorithm terminates and $ z = v $. Next, we want to investigate the question, how the components of $ v $ can switch between these three cases in the subsequent recursive steps.
	
	For this purpose, let us assume first that the projection $ v $ of $ x $ onto the current hyperplane $ \theta^\top w = p $ is not in $ [0,1]^d $ (otherwise, $ z=v $ and we are finished) and that we could only fix one component of $ z $, i.\,e. there exists exactly one $ i \in \{1,\dots,d \} $ with 
	\begin{equation*}
	v_i > 1 \text{ and } \theta_i=1 
	\end{equation*} 
	or 
	\begin{equation*}
	v_i<0 \text{ and } \theta_i=-1. 
	\end{equation*}
	Let $ \widetilde{x}=(x_1,\dots,x_{i-1},x_{i+1},\dots,x_d)^\top $ be the remaining components of $ x $ in the next recursive step and let $ \widetilde{\theta} = (\theta_1,\dots,\theta_{i-1},\theta_{i+1},\dots,\theta_{d})^\top $ and
	\begin{equation*}
	\widetilde{p}= \begin{cases}
	p & \text{ if } v_i < 0\\
	p-1 & \text{ if } v_i > 1
	\end{cases}
	\end{equation*}
	be the violated forbidden-set inequality in the next recursive step.
	Let $ (\widetilde{v}_1,\dots,\widetilde{v}_{i-1},\widetilde{v}_{i+1},\dots,\widetilde{v}_d)^\top = \widetilde{x} - \frac{\widetilde{\theta}^\top\widetilde{x}-\widetilde{p} }{d-1}\widetilde{\theta} $ be the projection of $ \widetilde{x} \in \mathbb{R}^{\widetilde{d}} = \mathbb{R}^{d-1} $ onto the hyperplane $ \widetilde{\theta}^\top \widetilde{w} = \widetilde{p} $ in the next recursive step. For simplifying the notation, we assume without loss of generality, that $ j < i $. Then, the following holds for any component $ \widetilde{v}_j $:
	\begin{theorem} \label{thm:movement}
		\begin{enumerate}
			\item[i)] If $ \theta_j=1 $, then $ \widetilde{v}_j > v_j $.
			\item[ii)] If $ \theta_j=-1 $, then $ \widetilde{v}_j < v_j $.
		\end{enumerate}
	\end{theorem}
	\begin{IEEEproof}
		We make a case distinction:\\
		In both cases, we use that 
		\begin{equation}
		-\frac{1}{d-1}=-\frac{1}{d}-\frac{1}{(d-1)d}.\label{eq:d}
		\end{equation}
		Case 1: $ v_i > 1 $, $ \theta_i = 1 $:
		\begin{IEEEeqnarray*}{lcl}
			\widetilde{v}_j & = &x_j - \frac{\sum\limits_{\substack{l =1\\ l\neq i}}^d\theta_lx_l-(p-1)}{d-1}\theta_j\\
			& = &x_j - \frac{\sum\limits_{l =1}^d\theta_lx_l-p}{d-1}\theta_j + \frac{\theta_ix_i\theta_j}{d-1} - \frac{1}{d-1}\theta_j\\
			& \overset{(\ref{eq:d})}{=} &\underbrace{x_j -  \frac{\sum\limits_{l =1}^d\theta_lx_l-p}{d}\theta_j}_{=v_j} -  \frac{\sum\limits_{l =1}^d\theta_lx_l-p}{(d-1)d}\theta_j\\
			& &+ \frac{\theta_ix_i\theta_j}{d-1} - \frac{1}{d-1}\theta_j\\
			& = &v_j + \frac{\theta_j}{d-1}\left(\theta_ix_i - \frac{\sum\limits_{l =1}^d\theta_lx_l-p}{d} - 1 \right)\\
			& \overset{\theta_i=1}{=} &v_j + \frac{\theta_j}{d-1}\left(\underbrace{x_i-\frac{\theta^\top x-p}{d}\theta_i}_{=v_i}-1\right)\\
			& = &v_j + \frac{\theta_j}{d-1}(v_i-1). 
		\end{IEEEeqnarray*}
		If $ \theta_j=1 $, then it holds that $ \widetilde{v}_j = v_j + \underbrace{\frac{1}{d-1}}_{>0}\underbrace{(v_i-1)}_{>0} > v_j $.\\
		If $ \theta_j=-1 $, then it holds that $ \widetilde{v}_j = v_j - \underbrace{\frac{1}{d-1}}_{>0}\underbrace{(v_i-1)}_{>0} < v_j $.
		
		Case 2: $ v_i < 0 $, $ \theta_i = -1 $:
		\begin{IEEEeqnarray*}{lcl}\
			\widetilde{v}_j & = &x_j - \frac{\sum\limits_{\substack{l =1\\ l\neq i}}^d\theta_lx_l-p}{d-1}\theta_j\\
			& = &x_j - \frac{\sum\limits_{l =1}^d\theta_lx_l-p}{d-1}\theta_j + \frac{\theta_ix_i\theta_j}{d-1}\\
			& \overset{(\ref{eq:d})}{=} &\underbrace{x_j -  \frac{\sum\limits_{l =1}^d\theta_lx_l-p}{d}\theta_j}_{=v_j} -  \frac{\sum\limits_{l =1}^d\theta_lx_l-p}{(d-1)d}\theta_j + \frac{\theta_ix_i\theta_j}{d-1}\\
			& = &v_j - \frac{\theta_j}{d-1}\left(-\theta_ix_i + \frac{\sum\limits_{l =1}^d\theta_lx_l-p}{d} \right)\\
			& \overset{\theta_i=-1}{=} &v_j - \frac{\theta_j}{d-1}\left(\underbrace{x_i-\frac{\theta^\top x-p}{d}\theta_i}_{=v_i}\right) = v_j - \frac{\theta_j}{d-1}v_i.
		\end{IEEEeqnarray*}
		If $ \theta_j=1 $, then it holds that $ \widetilde{v}_j = v_j - \underbrace{\frac{1}{d-1}}_{>0}\underbrace{v_i}_{<0} > v_j $.\\
		If $ \theta_j=-1 $, then it holds that $ \widetilde{v}_j = v_j + \underbrace{\frac{1}{d-1}}_{>0}\underbrace{v_i}_{<0} < v_j $.	  
	\end{IEEEproof}
	If more than one component was fixed, the analogous result follows from applying the last theorem inductively. Hence, it follows that components $ v_j $ of the projection onto the hyperplane are strictly monotonically increasing in the case of $ \theta_j=1 $ and strictly monotonically decreasing in the case of $ \theta_j=-1 $. Since we can fix components of $ z $ in the cases $ (v_j>1, \theta_j=1) $ and $ (v_j<0, \theta_j=-1) $, this means that the $ v_j's $ move into the direction where Theorem~\ref{thm:fixcomponents} is applicable. Additionally, this means if $ v_j \in [0,1] $, then there are two possibilities:
	\begin{enumerate}
		\item $ v_j $ stays in $ [0,1] $ in every following recursion.
		\item $ v_j $ is fixed in the first recursion, where it is not in $ [0,1] $ anymore.
	\end{enumerate} 
	\section{Projection Algorithm}\label{sec:algorithm}
	
	Our projection method is summarized in Algorithm~\ref{alg:exactprojidea}.
	
	\begin{figure}[t!]
	\begin{algorithm}[H]
		\caption{Projection Algorithm}
		\label{alg:exactprojidea}
		\begin{algorithmic}[1]{}
			\REQUIRE $ x \in \mathbb{R}^d $, Output: $ z=\Pi_{\mathcal{P}_{d},\text{even}}(x) $
			\STATE $ \theta_i: = \text{sgn}(x_i-0.5) \quad \forall i=1,\dots,d$
			\IF{$ |\{ i:\theta_i=1\}| $ is even}
			\STATE $ i^*:=\argmin_i |x_i-0.5| $
			\STATE $ \theta_{i^*} =-\theta_{i^*} $
			\ENDIF
			\STATE $ p = |\{ i:\theta_i=1\}| - 1$
			\STATE $ u=\Pi_{[0,1]^d}(x) $
			\IF{$ \theta^\top u \leq p $}
			\RETURN u
			\ENDIF
			\STATE $ l = d $, $ f = 1 $, $ fold = 1 $
			\STATE $ Q=[1,\dots,d] $
			\WHILE{True}
			\IF{$ l=1 $}
			\IF{$ \theta_f=1 $}
			\STATE $ z_f=0 $
			\ELSE
			\STATE $ z_f=1 $
			\ENDIF
			\STATE \textbf{break}
			\ENDIF
			\STATE $ v_{f,\dots,d} = x_{f,\dots,d} - \frac{\langle\theta_{f,\dots,d}, x_{f,\dots,d}\rangle-p}{l}\theta_{f,\dots,d} $
			\FOR{$ i=fold,\dots,d $}
			\IF{$ v_i > 1 $}
			\IF{$ \theta_i=1 $}
			\STATE swap $ Q_f $ and $ Q_i $
			\STATE $ x_i=x_f $
			\STATE $ \theta_i=\theta_f $
			\STATE $ z_f=1 $
			\STATE $ f=f+1 $
			\STATE $ p=p-1 $
			\ENDIF
			\ELSIF{$ v_i < 0  $}
			\IF{$ \theta_i=-1 $}
			\STATE swap $ Q_f $ and $ Q_i $
			\STATE $ x_i=x_f $
			\STATE $ \theta_i=\theta_f $
			\STATE $ z_f=0 $
			\STATE $ f=f+1 $
			\ENDIF
			\ENDIF
			\ENDFOR
			\IF{$ fold=f $}
			\STATE$ z_{fold,\dots,d}=v_{fold,\dots,d} $
			\STATE \textbf{break}
			\ENDIF
			\STATE $ l=d-f+1 $, $ fold=f $
			\ENDWHILE
			\RETURN $ z[Q] $		
		\end{algorithmic}
		\end{algorithm}
	\end{figure}
	
	In the lines $ 1-5 $, we apply the cut-search algorithm to $ x $, which leads to the same result as applying it to $ \Pi_{[0,1]^{d}}(x) $. In the lines $ 6-8 $, it is checked whether $ \Pi_{[0,1]^{d}}(x) $ is lying in the parity polytope. If this is true, then $ \Pi_{[0,1]^{d}}(x) $ is the projection of $ x $ onto $ \mathcal{P}_{d,\text{even}} $ and therefore returned. If $ \Pi_{[0,1]^{d}} \notin \Pi_{\mathcal{P}_{d,\text{even}}} $, we enter the while loop, the recursive part of the projection. Before entering the loop, we initialize $ l $  by $ d $. This variable tracks the number of components of $ z $, that are not yet computed. For the implementation, we swap the computed components of $ z $ to the beginning of the vector and continue with the remaining vector. The variable $ f $ tracks the index of the first uncomputed component of $ z $, whereas $ fold $ stores the value of $ f $ from the beginning of the current while loop iteration. The index vector $ Q $ tracks all swaps, that were made. The lines $ 12-17 $ describe one of the two stopping criteria. If the current dimension $ l $ of the recursive subproblem is $ 1 $, then there are the following two possibilities for the corresponding parity polytope: 
	The first case is
	\begin{equation*}
	\mathcal{P}_{1,\text{even}} = \text{conv} \{x\in \{0,1\}:x_1 \text{ is even} \} = \text{conv} \{0 \} = \{0\}.
	\end{equation*}
	In this case, $ \mathcal{P}_{1,\text{even}} $ can be described by the box constraints $ 0\leq x \leq 1 $ and the only forbidden-set inequality $ x\leq 0 $. This means that $ \theta_f=1 $ and $ p= 0 $ in this case.
	The second possibility is the odd parity polytope
	\begin{equation*}
	\mathcal{P}_{1,\text{odd}} = \text{conv} \{x\in \{0,1\}:x_1 \text{ is odd} \} = \text{conv} \{1 \} = \{1\}.
	\end{equation*}
    \begin{figure}
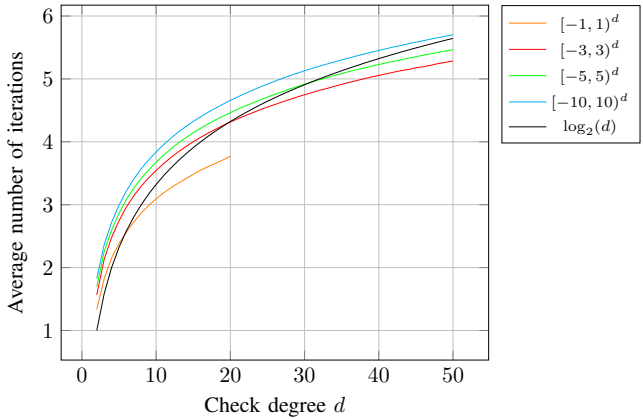

    \centering
    	\includestandalone[height=0.31\textwidth]{loops}
    		\caption{Number of while loop iterations in the case $ \Pi_{[0,1]^{d}}(x) \notin \mathcal{P}_{d,\text{even}} $}\label{fig:iterations}
    \end{figure}
	\begin{figure*}
	    \centering
	    \subfloat[Low-complexity operations]{		
	    \includestandalone[height=0.31\textwidth]{low_complexity_1}
		    \label{fig:Simpleops1}} \hspace{10pt}
	    \subfloat[Multiplications and divisions]{
    	\includestandalone[height=0.31\textwidth]{mult_div_1}
        \label{fig:multdivops1}}
    	\caption{Comparison of arithmetical operations for different projection methods with    random input from $ [-1,1)^d $}
    	\label{fig:comparison1}
    \end{figure*}
    \begin{figure*}
    	\centering
    	\subfloat[Low-complexity operations]{		
    	\includestandalone[height=0.31\textwidth]{low_complexity_3}
    		\label{fig:Simpleops3}} \hspace{10pt}
    	\subfloat[Multiplications and divisions]{
    	\includestandalone[height=0.31\textwidth]{mult_div_3}
        \label{fig:multdivops3}}
    	\caption{Comparison of arithmetical operations for different projection methods with random input from $ [-3,3)^d $}
    	\label{fig:comparison3}
    \end{figure*}

	In this case, $ \mathcal{P}_{1,\text{odd}} $ can be described by $ 0\leq x \leq 1 $ and the only forbidden-set inequality $ -x\leq -1 $. This means that $ \theta_f=-1 $ and $ p= -1 $ in this case. Hence, we can make the check $ \theta_f=1 $ to distinguish both cases.
	If this stopping criterion is not fulfilled, we continue and compute the projection of the remaining components of $ x $ onto the hyperplane defined by the current forbidden-set inequality in line 18. In the lines 19-34, we go through all components of $ v_{f,\dots,d} $ and check to which components of $ z $ Theorem~\ref{thm:fixcomponents} can be applied, i.\,e. which components of $ z $ can be fixed in this iteration. For the swaps in $ x $ and $ \theta $ in the lines 23-24 and 31-32, it is sufficient to update $ x_i $ and $\theta_i $, because $ x_f $ and $ \theta_f $ are not needed anymore. Theorem~\ref{thm:existencedfixing} says that at least one component can be fixed in the case of $ v_{f,\dots,d}\notin [0,1]^l  $. Hence, if no component was fixed, i.\,e. the check in line 35 is true, then we are in the case $ v_{f,\dots,d}\in [0,1]^l $ and we can stop with $ z_{f,\dots,d}= v_{f,\dots,d}$. Otherwise, we update the current problem size $ l $, update $ fold $ and consider the corresponding smaller-dimensional projection problem in the next while loop iteration. In the algorithm, we do not check whether $ l=0 $, which could happen theoretically. However, the next theorem shows that this situation cannot occur:
	\begin{theorem}
		In Algorithm~\ref{alg:exactprojidea}, the variable $ l $ cannot become zero.
	\end{theorem}
	
	\begin{IEEEproof}
		Assume the claim is wrong, i.\,e. $ l $ is set to zero during an iteration of the while loop in Algorithm~\ref{alg:exactprojidea}. Let $ v \in \mathbb{R}^{\widetilde{d}} $ be the current projection of $ x \in \mathbb{R}^{\widetilde{d}} $ onto the current hyperplane $ \theta^\top w = p = |\{i:\theta_i=1 \}|-1 $ in the iteration of the while loop, where $ l $ is set to zero. Since $ l $ is set to zero in line 38, this means that all remaining components of the projection on $ x $ onto the face $ \{w\in [0,1]^{\widetilde{d}}:\theta^\top w = p \} $ were fixed with Theorem~\ref{thm:fixcomponents}. This means that for all $ i=1,\dots,\widetilde{d} $, it must hold that
		\begin{equation*}
		v_i > 1 \text{ and } \theta_i = 1
		\end{equation*}
		or 
		\begin{equation*}
		v_i < 0 \text{ and } \theta_i = -1.
		\end{equation*}
		Since $ v $ lies on the hyperplane $ \theta^\top v=p $, it follows that
		\begin{IEEEeqnarray*}{l}
			p=\theta^\top v = \sum_{\substack{i =1\\ \theta_i=1}}^{\widetilde{d}}
			\underbrace{v_i}_{>1} - \sum_{\substack{i =1\\ \theta_i=-1}}^{\widetilde{d}}\underbrace{v_i}_{<0} > \sum_{\substack{i =1\\ \theta_i=1}}^{\widetilde{d}}1 =\\ |\{i:\theta_i=1 \} | = p+1,	
		\end{IEEEeqnarray*} 
		which is a contradiction. Hence, the assumption was wrong and the claim follows.
	\end{IEEEproof}
	Since we fix, by Theorem~\ref{thm:existencedfixing}, at least one component of $ z $ in every iteration of the while loop in Algorithm~\ref{alg:exactprojidea}, it follows that the worst-case complexity of Algorithm~\ref{alg:exactprojidea} is $ \mathcal{O}(d) + d\cdot\mathcal{O}(d) = \mathcal{O}(d^2) $. Figure~\ref{fig:iterations} displays the actual average number of iterations in the case $\Pi_{[0,1]^d}(x) \notin \mathcal{P}_{d,\text{even}}$ that we measured in our numerical results for different ranges of input vectors. In the Figure, one can see that the actual number is close to $\log_2(d)$. Hence, we make the conjecture that half of the components are fixed in average. A mathematical intuition behind this conjecture is that when we compute the projection $ v $ onto the hyperplane $ \theta^\top x = p $, then every component $v_j$ can be in $[0,1]$ or not. If $v_j\in [0,1]$, Theorem~\ref{thm:movement} states that $v_j$ stays in $[0,1]$ for the whole algorithm or is fixed in the first iteration where it leaves $[0,1]$. Staying in $[0,1]$ would be a good situation, because if $v\in [0,1]^d $, then the algorithm terminates and $v$ is the wanted projection onto the parity polytope (see line 35 of Algorithm~\ref{alg:exactprojidea}). In the more difficult situation $v_j \notin [0,1]$, there are four possibilities: $(v_j<0,\theta_j=1)$, $(v_j<0,\theta_j=-1)$, $(v_j>1,\theta_j=1)$ and $(v_j>1,\theta_j=-1)$. In two of four, i.e. in half of the cases, we can fix the component $z_j$ of the projection by using Theorem~\ref{thm:fixcomponents}. Under the conjecture that all four cases occur with equal probability, we would obtain an average complexity of $\mathcal{O}(d) + \mathcal{O}(\frac{d}{2})+\mathcal{O}(\frac{d}{4})+\dots=\mathcal{O}(2d)=\mathcal{O}(d)$, i.\,e. we would have a quadratic worst-case and a linear average-case complexity, similar to the algorithm in~\cite{Heusdens}.
	
    \section{Numerical Results}\label{sec:numericalresults}
    In this section, we compare our new projection algorithm with the projection algorithms from the literature, namely with the algorithm of Zhang and Siegel~\cite{ADMMSiegel}, Wasson and Draper~\cite{Wasson}, Zhang et al.~\cite{Heusdens}, and Wei and Banihashemi~\cite{Wei}. To have an estimation of the potential complexity for efficient hard- and software implementations, we count the number of arithmetic operations. For this purpose, we count the number of divisions, the number of multiplications and the number of all other low-complexity operations, as e.\,g. comparisons, additions, substractions or negations. We avoided unnecessary divisions by replacing the comparisons $ \frac{\delta}{\zeta}>t_i $ in Algorithm 3 of~\cite{ADMMSiegel} by $ \delta>\zeta t_i $ and by storing $ is_i $ instead of $s_i$ for the computation of $\rho$ in Algorithm 2 of~\cite{Wasson}. Comparisons with $ 0 $, $ \max(x,0) $, the floor operation and assignments to $ 1 $, $ 0 $, $ -1 $ or other assignments without arithmetic operations were not counted to the low-complexity operations due to their negligible effort in hardware. Since the coefficient vector $ \theta\in \{\pm1\}^d $ (or the vector $ f \in \{0,1\}^d $  in~\cite{Wasson}) could be stored as a boolean array, comparisons of the form $ \theta_i=1 $ or $ \theta_i=-1 $ are also negligible and were not counted. Operations of the form $ a+\theta_i\cdot b $, as they appear in several algorithms, were therefore counted as one low-complexity operation, because one can check whether $ \theta_i=1 $ and then make an addition or substraction to compute $ a + b $ or $ a - b $, respectively. The sortings from~\cite{ADMMSiegel} and~\cite{Wasson} were implemented with the quicksort algorithm with the last element as pivot element. For simplification, we also chose the last element as pivot element in the partial sorting and the projection onto the simplex in~\cite{Heusdens}.
    
\begin{figure*}
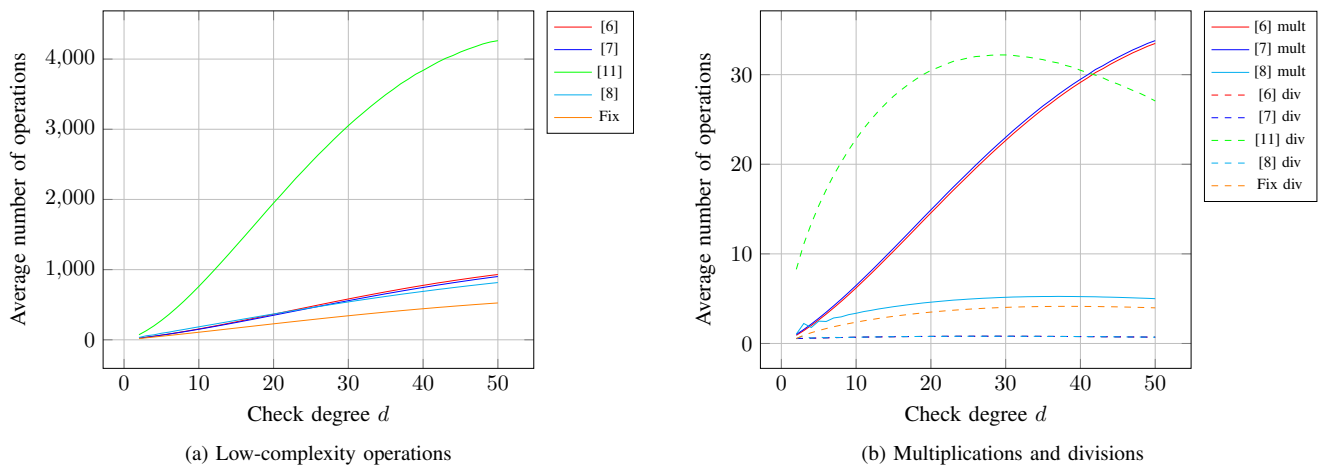

    	\centering
    	\subfloat[Low-complexity operations]{		
    	\includestandalone[height=0.31\textwidth]{low_complexity_10}
    		\label{fig:Simpleops10}} \hspace{10pt}
    	\subfloat[Multiplications and divisions]{
    	\includestandalone[height=0.31\textwidth]{mult_div_10}
        \label{fig:multdivops10}}
    	\caption{Comparison of arithmetical operations for different projection methods with random input from $ [-10,10)^d $}
    	\label{fig:comparison10}
    \end{figure*}

\begin{figure}[h]
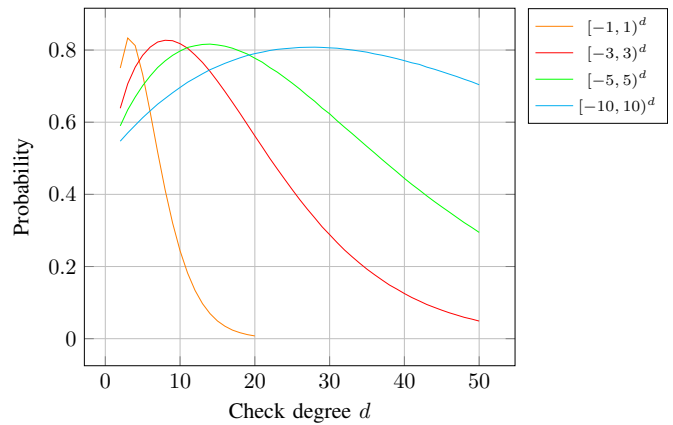

    \centering
    	\includestandalone[height=0.31\textwidth]{probability}
    		\caption{Probability of the case $ \Pi_{[0,1]^{d}}(x) \notin \mathcal{P}_{d,\text{even}} $}\label{fig:prob}
    \end{figure}

    We simulated projections onto $ \mathcal{P}_{d,\text{even}} $
    for input vectors from $ [-1,1)^d $, $ [-3,3)^d $, $ [-5,5)^d $, and $ [-10,10)^d $. For $ [-1,1)^d $, the dimensions $d=2,\dots,20$ were simulated. For the other three simulations, the dimensions $ d=2,\dots,50 $ were investigated. For every $ d $, one million randomly generated vectors were projected. The results are shown in Figures~\ref{fig:comparison1} to~\ref{fig:comparison10}. Our proposed algorithm is denoted by \enquote{Fix}. Figures~\ref{fig:Simpleops1} to~\ref{fig:Simpleops10} show the average number of low-complexity operations and Figure~\ref{fig:multdivops1} to~\ref{fig:multdivops10} the average number of multiplications and divisions for different check degrees. After our mentioned simplifications, our algorithm and the algorithm from~\cite{Wei} do not need any multiplications. Hence, the corresponding two plots are left out in the Figures~\ref{fig:multdivops1} to~\ref{fig:multdivops10}. The algorithms from~\cite{ADMMSiegel} and~\cite{Wasson} need both exactly one division in the difficult case of $\Pi_{[0,1]^d}(x) \notin \mathcal{P}_{d,\text{even}} $ and zero in the simple case $\Pi_{[0,1]^d}(x) \in \mathcal{P}_{d,\text{even}} $. The number of divisions of~\cite{Heusdens} are very similar and form basically the same lines in the Figures~\ref{fig:multdivops1} to~\ref{fig:multdivops10}. In these Figures, the number of multiplications and divisions is monotonically decreasing for all five algorithms after a certain value of $ d $, although the problem size increases for growing $ d $. The same situation can be observed for the low-complexity operations of~\cite{Wei}. This behavior can be explained by Figure~\ref{fig:prob}. It shows the probabilities, measured through our simulations, that the difficult case $ \Pi_{[0,1]^d}(x)\notin \mathcal{P}_{d,\text{even}} $ occurs for some random input from our considered ranges. For all four ranges, we can see that the probabilities decrease after a certain certain check degree. In our proposed and in the two algorithms from~\cite{ADMMSiegel,Wasson} it is checked whether $ \Pi_{[0,1]^d}(x)\in \mathcal{P}_{d,\text{even}} $. In~\cite{Heusdens}, this check is done in the main case of their algorithm. The approximative algorithm from~\cite{Wei} stops after one iteration 
    in that case. Hence, the number of multiplications and divisions (and low-complexity operations for~\cite{Wei}) decrease for large $ d $. Additionally, this means that for $[-1,1)^d$ and $d>20$, the projection is very simple, because the hard case $ \Pi_{[0,1]^d}(x)\notin \mathcal{P}_{d,\text{even}} $ is happening only very rarely. Hence, we only considered the degrees $d\leq 20$ for this range.
    \begin{table}[h]
    \renewcommand{\arraystretch}{1.3}
    \caption{Maximum Gain}
    \label{table:gain}
    \centering
    	\begin{tabular}{l|c|c}
    		\hline
    		input & low-complexity operations & all arith. op.\\
    		\hline
    		\hline
    		$ [-1,1)^d $ & -13\% & -14\% \\
    		\hline
    		$ [-3,3)^d $ & -31\% & -32\% \\
    		\hline
    		$ [-5,5)^d $ & -36\% & -37\% \\
    		\hline
    		$ [-10,10)^d $ & -37\% & -37\% \\
    		\hline
    	\end{tabular}
    \end{table}
        \begin{figure*}
    	\centering
    	\subfloat[Low-complexity operations]{		
    	\includestandalone[height=0.31\textwidth]{low_complexity_5}
    		\label{fig:Simpleops5}} \hspace{10pt}
    	\subfloat[Multiplications and divisions]{
    	\includestandalone[height=0.31\textwidth]{mult_div_5}
        \label{fig:multdivops5}}
    	\caption{Comparison of arithmetical operations for different projection methods with random input from $ [-5,5)^d $}
    	\label{fig:comparison5}
    \end{figure*}
    
    As can be seen in the Figures~\ref{fig:comparison1} to~\ref{fig:comparison10}, our implementation  
    needs less low-complexity operations than all other implementations for every dimension $d$ and every range the random input is created from. The Figures also show that low-complexity operations form the large majority of operations in all considered algorithms. Regarding high-complexity operations, our algorithm needs slightly more divisions than the implementations from~\cite{ADMMSiegel},~\cite{Wasson} and~\cite{Heusdens}. However, they are less than the sum of divisions and multiplications of every other algorithm in all considered cases. Table~\ref{table:gain} shows the maximum gain in arithmetical operations for different input intervals. The values were obtained by comparing our proposed algorithm with the best result from the other four considered algorithms. Overall, we need up to $ 37\% $ less arithmetical operations, resulting in lower implementation complexity and making LP decoding more attractive for efficient hardware implementation.

    \section{Conclusion}\label{sec:conclusion}
    In this paper, we presented a new reduced-complexity projection algorithm for ADMM-based LP decoding. By establishing the theory of the odd parity polytope similar to that of the even parity polytope, the projection algorithm can be regarded as a recursive problem, where the projections are varying between projections on the even or odd parity polytope. As some components of the input are fixed in every iteration, the problem size is constantly decreasing.
    In contrast to other exact state-of-the-art projections, the proposed algorithm needs up to 37\% less arithmetical operations and additionally requires no sorting operation. 
    These properties make it a very good choice for future hardware implementations.
    
    \section*{Acknowledgement}
    We gratefully acknowledge financial support by the DFG (project-ID: WE 2442/9-3
    and RU 1524/2-3).
    \bibliographystyle{IEEEtran}
    \bibliography{paper}
\end{document}

%% file: loops.tex
\begin{tikzpicture}[scale=0.85]
		\begin{axis}[
		xlabel=Check degree $d$,
		ylabel=Average number of iterations,
		grid=both,
		legend style={legend pos=outer north east,font=\scriptsize}]
		
		\addplot[color=orange] plot coordinates {
			( 2 , 1.33339421289 )
			( 3 , 1.8243537237 )
			( 4 , 2.15488573166 )
			( 5 , 2.38197350451 )
			( 6 , 2.56546327674 )
			( 7 , 2.72566510263 )
			( 8 , 2.86857243982 )
			( 9 , 2.9883186238 )
			( 10 , 3.09466148933 )
			( 11 , 3.18771158774 )
			( 12 , 3.27097395616 )
			( 13 , 3.3434697093 )
			( 14 , 3.41389471868 )
			( 15 , 3.48272210796 )
			( 16 , 3.55027496382 )
			( 17 , 3.60549187555 )
			( 18 , 3.65783306292 )
			( 19 , 3.71208089409 )
			( 20 , 3.7706855792 )

		};
		\addlegendentry{$[-1,1)^d$}
		
%
%
%
%
%
		
		\addplot[color=red] plot coordinates {
			( 2 , 1.56483700919 )
			( 3 , 2.13201105235 )
			( 4 , 2.47882785927 )
			( 5 , 2.74557085299 )
			( 6 , 2.9648763501 )
			( 7 , 3.14488454625 )
			( 8 , 3.29695131977 )
			( 9 , 3.42836680034 )
			( 10 , 3.54535792993 )
			( 11 , 3.65289759136 )
			( 12 , 3.75135572941 )
			( 13 , 3.84091633404 )
			( 14 , 3.92395360293 )
			( 15 , 4.0007285556 )
			( 16 , 4.07086568201 )
			( 17 , 4.13760608239 )
			( 18 , 4.19979357358 )
			( 19 , 4.25876726618 )
			( 20 , 4.31479402382 )
			( 21 , 4.36567356611 )
			( 22 , 4.4156818732 )
			( 23 , 4.4623085245 )
			( 24 , 4.50936232068 )
			( 25 , 4.55270589773 )
			( 26 , 4.59313778884 )
			( 27 , 4.63474452555 )
			( 28 , 4.67326682443 )
			( 29 , 4.71146196604 )
			( 30 , 4.74857412306 )
			( 31 , 4.78202933802 )
			( 32 , 4.8183697975 )
			( 33 , 4.84785823043 )
			( 34 , 4.8819862435 )
			( 35 , 4.91352594793 )
			( 36 , 4.94433610263 )
			( 37 , 4.97287045335 )
			( 38 , 5.00217762424 )
			( 39 , 5.03010929638 )
			( 40 , 5.05531666613 )
			( 41 , 5.08193314157 )
			( 42 , 5.10677242805 )
			( 43 , 5.13121918777 )
			( 44 , 5.14989860817 )
			( 45 , 5.17744507442 )
			( 46 , 5.19401426104 )
			( 47 , 5.21661734647 )
			( 48 , 5.24454583861 )
			( 49 , 5.26264373103 )
			( 50 , 5.28617086663 )

		};
		\addlegendentry{$[-3,3)^d$}
		
		\addplot[color=green] plot coordinates {
			( 2 , 1.69488939223 )
			( 3 , 2.25586894798 )
			( 4 , 2.59509048558 )
			( 5 , 2.86262999018 )
			( 6 , 3.08434612046 )
			( 7 , 3.26699727362 )
			( 8 , 3.42086020248 )
			( 9 , 3.55424765858 )
			( 10 , 3.67564530572 )
			( 11 , 3.7878510821 )
			( 12 , 3.8880220618 )
			( 13 , 3.98278126368 )
			( 14 , 4.06656451974 )
			( 15 , 4.14454896845 )
			( 16 , 4.21604195885 )
			( 17 , 4.28412872102 )
			( 18 , 4.34771297253 )
			( 19 , 4.40663741044 )
			( 20 , 4.4638371556 )
			( 21 , 4.51854244229 )
			( 22 , 4.56986246681 )
			( 23 , 4.61854507753 )
			( 24 , 4.66685552408 )
			( 25 , 4.71209917682 )
			( 26 , 4.75550392167 )
			( 27 , 4.79819924838 )
			( 28 , 4.83851728262 )
			( 29 , 4.8778612099 )
			( 30 , 4.91548159177 )
			( 31 , 4.95236431786 )
			( 32 , 4.98834802345 )
			( 33 , 5.02007754596 )
			( 34 , 5.05435339332 )
			( 35 , 5.08431074306 )
			( 36 , 5.11698822506 )
			( 37 , 5.14608211546 )
			( 38 , 5.1729045554 )
			( 39 , 5.20186913866 )
			( 40 , 5.22840050775 )
			( 41 , 5.255935497 )
			( 42 , 5.2812124595 )
			( 43 , 5.30433222484 )
			( 44 , 5.32947209256 )
			( 45 , 5.3535038993 )
			( 46 , 5.37697516287 )
			( 47 , 5.40010050759 )
			( 48 , 5.4220116754 )
			( 49 , 5.44242716905 )
			( 50 , 5.46522237988 )

		};
		\addlegendentry{$[-5,5)^d$}
		
		\addplot[color=cyan] plot coordinates {
			( 2 , 1.82684993369 )
			( 3 , 2.36534881193 )
			( 4 , 2.71828752464 )
			( 5 , 2.99370597859 )
			( 6 , 3.22067976017 )
			( 7 , 3.41108051141 )
			( 8 , 3.57209904467 )
			( 9 , 3.71293870138 )
			( 10 , 3.83892922025 )
			( 11 , 3.95502776906 )
			( 12 , 4.05947414029 )
			( 13 , 4.15556384864 )
			( 14 , 4.24265076793 )
			( 15 , 4.32492095163 )
			( 16 , 4.39899264531 )
			( 17 , 4.46927140222 )
			( 18 , 4.53509356872 )
			( 19 , 4.59789497553 )
			( 20 , 4.65911020944 )
			( 21 , 4.71461506886 )
			( 22 , 4.76927025044 )
			( 23 , 4.81988849415 )
			( 24 , 4.87035442818 )
			( 25 , 4.9184757847 )
			( 26 , 4.96376474961 )
			( 27 , 5.00776932953 )
			( 28 , 5.0501732759 )
			( 29 , 5.09038467494 )
			( 30 , 5.12975942954 )
			( 31 , 5.16776094339 )
			( 32 , 5.20213106604 )
			( 33 , 5.23657659617 )
			( 34 , 5.26986189682 )
			( 35 , 5.30208659975 )
			( 36 , 5.33370537552 )
			( 37 , 5.36541910356 )
			( 38 , 5.39485420372 )
			( 39 , 5.42353251941 )
			( 40 , 5.45290043616 )
			( 41 , 5.48029294448 )
			( 42 , 5.50776062548 )
			( 43 , 5.53258150192 )
			( 44 , 5.55922236504 )
			( 45 , 5.58375159932 )
			( 46 , 5.60817673531 )
			( 47 , 5.63269976838 )
			( 48 , 5.65626903574 )
			( 49 , 5.67917978016 )
			( 50 , 5.70232801316 )

		};
		\addlegendentry{$[-10,10)^d$}
		
		\addplot[domain=2:50,samples=49] {ln(x)/ln(2)};
		\addlegendentry{$ \log_2(d) $}
		
		\end{axis}
		\end{tikzpicture}

%% file: low_complexity_1.tex
\begin{tikzpicture}[scale=0.8]
		\begin{axis}[
		xlabel=Check degree $d$,
		ylabel=Average number of operations,
		grid=both,
		legend style={legend pos=outer north east,font=\scriptsize}]
		\addplot[color=red] plot coordinates {
			( 2 , 21.00096 )
			( 3 , 35.389077 )
			( 4 , 48.693957 )
			( 5 , 59.249796 )
			( 6 , 66.784593 )
			( 7 , 71.527264 )
			( 8 , 74.245635 )
			( 9 , 75.589155 )
			( 10 , 76.111031 )
			( 11 , 76.512681 )
			( 12 , 77.107027 )
			( 13 , 78.106857 )
			( 14 , 79.569953 )
			( 15 , 81.526918 )
			( 16 , 84.102277 )
			( 17 , 87.01588 )
			( 18 , 90.486048 )
			( 19 , 94.140723 )
			( 20 , 98.096166 )

		};
		\addlegendentry{\cite{ADMMSiegel}}
		
		\addplot[color=blue] plot coordinates {
			( 2 , 26.001281 )
			( 3 , 41.047399 )
			( 4 , 54.826475 )
			( 5 , 66.01304 )
			( 6 , 74.389424 )
			( 7 , 80.052474 )
			( 8 , 83.743863 )
			( 9 , 86.17121 )
			( 10 , 87.849235 )
			( 11 , 89.448197 )
			( 12 , 91.278401 )
			( 13 , 93.547487 )
			( 14 , 96.276634 )
			( 15 , 99.525315 )
			( 16 , 103.389401 )
			( 17 , 107.596187 )
			( 18 , 112.346767 )
			( 19 , 117.284792 )
			( 20 , 122.511479 )

		};
		\addlegendentry{\cite{Wasson}}
		
		\addplot[color=green] plot coordinates {
			( 2 , 36.789112 )
			( 3 , 67.276948 )
			( 4 , 92.435176 )
			( 5 , 109.502035 )
			( 6 , 118.915281 )
			( 7 , 122.018306 )
			( 8 , 120.89143 )
			( 9 , 117.368335 )
			( 10 , 112.994284 )
			( 11 , 108.937773 )
			( 12 , 105.757904 )
			( 13 , 103.804538 )
			( 14 , 103.113547 )
			( 15 , 103.640835 )
			( 16 , 105.376441 )
			( 17 , 108.032569 )
			( 18 , 111.531451 )
			( 19 , 115.489922 )
			( 20 , 120.003695 )

		};
		\addlegendentry{\cite{Wei}}

		\addplot[color=cyan] plot coordinates {
			( 2 , 38.757381 )
			( 3 , 60.695703 )
			( 4 , 80.225422 )
			( 5 , 95.248185 )
			( 6 , 105.575094 )
			( 7 , 112.067529 )
			( 8 , 116.324004 )
			( 9 , 119.247662 )
			( 10 , 121.810498 )
			( 11 , 124.537147 )
			( 12 , 127.957945 )
			( 13 , 132.026748 )
			( 14 , 136.955323 )
			( 15 , 142.48425 )
			( 16 , 148.769183 )
			( 17 , 155.573553 )
			( 18 , 162.85543 )
			( 19 , 170.510869 )
			( 20 , 178.246418 )

		};
		\addlegendentry{\cite{Heusdens}}
		
		\addplot[color=orange] plot coordinates {
			( 2 , 18.750959 )
			( 3 , 32.43268 )
			( 4 , 44.058137 )
			( 5 , 52.47325 )
			( 6 , 58.293877 )
			( 7 , 62.048593 )
			( 8 , 64.48811 )
			( 9 , 66.103431 )
			( 10 , 67.376607 )
			( 11 , 68.744935 )
			( 12 , 70.43833 )
			( 13 , 72.530224 )
			( 14 , 75.024331 )
			( 15 , 77.914665 )
			( 16 , 81.280821 )
			( 17 , 84.860445 )
			( 18 , 88.834679 )
			( 19 , 92.91482 )
			( 20 , 97.198521 )

		};
		\addlegendentry{Fix}
		
		\end{axis}
		\end{tikzpicture}

%% file: mult_div_1.tex
\begin{tikzpicture}[scale=0.8]
		\begin{axis}[
		xlabel=Check degree $d$,
		ylabel=Average number of operations,
		grid=both,
		legend style={legend pos=outer north east,font=\scriptsize}]
		\addplot[color=red] plot coordinates {
			( 2 , 0.500017 )
			( 3 , 1.061905 )
			( 4 , 1.567347 )
			( 5 , 1.920213 )
			( 6 , 2.098581 )
			( 7 , 2.109237 )
			( 8 , 1.995356 )
			( 9 , 1.798891 )
			( 10 , 1.556151 )
			( 11 , 1.308269 )
			( 12 , 1.068262 )
			( 13 , 0.85312 )
			( 14 , 0.667805 )
			( 15 , 0.511941 )
			( 16 , 0.388714 )
			( 17 , 0.287742 )
			( 18 , 0.214627 )
			( 19 , 0.155066 )
			( 20 , 0.111227 )

		};
		\addlegendentry{\cite{ADMMSiegel} mult}
		
		\addplot[color=blue] plot coordinates {
			( 2 , 1.000199 )
			( 3 , 1.582753 )
			( 4 , 2.069212 )
			( 5 , 2.371892 )
			( 6 , 2.486868 )
			( 7 , 2.429894 )
			( 8 , 2.251641 )
			( 9 , 1.998801 )
			( 10 , 1.708666 )
			( 11 , 1.422822 )
			( 12 , 1.152669 )
			( 13 , 0.914221 )
			( 14 , 0.711456 )
			( 15 , 0.543148 )
			( 16 , 0.410666 )
			( 17 , 0.302848 )
			( 18 , 0.225186 )
			( 19 , 0.162387 )
			( 20 , 0.116062 )

		};
		\addlegendentry{\cite{Wasson} mult}
		
		\addplot[color=cyan] plot coordinates {
			( 2 , 1.125636 )
			( 3 , 1.943698 )
			( 4 , 2.420689 )
			( 5 , 2.641112 )
			( 6 , 2.562352 )
			( 7 , 2.310727 )
			( 8 , 1.97475 )
			( 9 , 1.612918 )
			( 10 , 1.276888 )
			( 11 , 0.985217 )
			( 12 , 0.745673 )
			( 13 , 0.552683 )
			( 14 , 0.405164 )
			( 15 , 0.290701 )
			( 16 , 0.20821 )
			( 17 , 0.145837 )
			( 18 , 0.103201 )
			( 19 , 0.070955 )
			( 20 , 0.048458 )

		};
		\addlegendentry{\cite{Heusdens} mult}
		
		\addplot[color=red,dashed] plot coordinates {
			( 2 , 0.750115 )
			( 3 , 0.833351 )
			( 4 , 0.812386 )
			( 5 , 0.7322 )
			( 6 , 0.628253 )
			( 7 , 0.517251 )
			( 8 , 0.412501 )
			( 9 , 0.320767 )
			( 10 , 0.243626 )
			( 11 , 0.182253 )
			( 12 , 0.133928 )
			( 13 , 0.097109 )
			( 14 , 0.069566 )
			( 15 , 0.049109 )
			( 16 , 0.03455 )
			( 17 , 0.023817 )
			( 18 , 0.016641 )
			( 19 , 0.011274 )
			( 20 , 0.007614 )

		};
		\addlegendentry{\cite{ADMMSiegel} div}
		
		\addplot[color=blue,dashed] plot coordinates {
			( 2 , 0.750115 )
			( 3 , 0.833351 )
			( 4 , 0.812386 )
			( 5 , 0.7322 )
			( 6 , 0.628253 )
			( 7 , 0.517251 )
			( 8 , 0.412501 )
			( 9 , 0.320767 )
			( 10 , 0.243626 )
			( 11 , 0.182253 )
			( 12 , 0.133928 )
			( 13 , 0.097109 )
			( 14 , 0.069566 )
			( 15 , 0.049109 )
			( 16 , 0.03455 )
			( 17 , 0.023817 )
			( 18 , 0.016641 )
			( 19 , 0.011274 )
			( 20 , 0.007614 )

		};
		\addlegendentry{\cite{Wasson} div}
		
		\addplot[color=green,dashed] plot coordinates {
			( 2 , 3.81743 )
			( 3 , 5.31507 )
			( 4 , 5.790658 )
			( 5 , 5.617599 )
			( 6 , 5.113294 )
			( 7 , 4.465604 )
			( 8 , 3.801553 )
			( 9 , 3.191989 )
			( 10 , 2.670857 )
			( 11 , 2.248647 )
			( 12 , 1.914383 )
			( 13 , 1.65897 )
			( 14 , 1.468189 )
			( 15 , 1.327703 )
			( 16 , 1.227123 )
			( 17 , 1.156373 )
			( 18 , 1.107148 )
			( 19 , 1.071518 )
			( 20 , 1.047975 )

		};
		\addlegendentry{\cite{Wei} div}

		\addplot[color=cyan,dashed] plot coordinates {
			( 2 , 0.750115 )
			( 3 , 0.833351 )
			( 4 , 0.812386 )
			( 5 , 0.7322 )
			( 6 , 0.628253 )
			( 7 , 0.517251 )
			( 8 , 0.412501 )
			( 9 , 0.320767 )
			( 10 , 0.243626 )
			( 11 , 0.182253 )
			( 12 , 0.133928 )
			( 13 , 0.097109 )
			( 14 , 0.069566 )
			( 15 , 0.049109 )
			( 16 , 0.03455 )
			( 17 , 0.023817 )
			( 18 , 0.016641 )
			( 19 , 0.011274 )
			( 20 , 0.007614 )

		};
		\addlegendentry{\cite{Heusdens} div}
		
		\addplot[color=orange,dashed] plot coordinates {
			( 2 , 0.750115 )
			( 3 , 1.395593 )
			( 4 , 1.688414 )
			( 5 , 1.712735 )
			( 6 , 1.596134 )
			( 7 , 1.402175 )
			( 8 , 1.179433 )
			( 9 , 0.956554 )
			( 10 , 0.752978 )
			( 11 , 0.580493 )
			( 12 , 0.437817 )
			( 13 , 0.324583 )
			( 14 , 0.237429 )
			( 15 , 0.171006 )
			( 16 , 0.122638 )
			( 17 , 0.085866 )
			( 18 , 0.060868 )
			( 19 , 0.041849 )
			( 20 , 0.028709 )

		};
		\addlegendentry{Fix div}
		
		\end{axis}
		\end{tikzpicture}

%% file: low_complexity_3.tex
\begin{tikzpicture}[scale=0.8]
		\begin{axis}[
		xlabel=Check degree $d$,
		ylabel=Average number of operations,
		grid=both,
		legend style={legend pos=outer north east,font=\scriptsize}]
		\addplot[color=red] plot coordinates {
			( 2 , 23.413445 )
			( 3 , 38.631998 )
			( 4 , 55.028302 )
			( 5 , 72.396133 )
			( 6 , 90.384193 )
			( 7 , 108.730079 )
			( 8 , 126.870465 )
			( 9 , 144.77665 )
			( 10 , 161.994267 )
			( 11 , 178.548952 )
			( 12 , 193.903768 )
			( 13 , 208.371561 )
			( 14 , 221.582162 )
			( 15 , 233.566085 )
			( 16 , 244.329573 )
			( 17 , 253.902256 )
			( 18 , 262.374352 )
			( 19 , 269.591405 )
			( 20 , 275.667921 )
			( 21 , 280.921954 )
			( 22 , 284.953997 )
			( 23 , 288.511237 )
			( 24 , 291.171355 )
			( 25 , 293.048361 )
			( 26 , 294.18547 )
			( 27 , 295.021821 )
			( 28 , 295.581096 )
			( 29 , 294.919488 )
			( 30 , 295.139656 )
			( 31 , 294.223347 )
			( 32 , 293.143001 )
			( 33 , 292.533653 )
			( 34 , 291.658001 )
			( 35 , 290.524528 )
			( 36 , 289.848451 )
			( 37 , 288.975933 )
			( 38 , 287.929696 )
			( 39 , 287.994611 )
			( 40 , 286.920172 )
			( 41 , 286.57517 )
			( 42 , 286.540039 )
			( 43 , 286.104786 )
			( 44 , 286.63597 )
			( 45 , 287.152428 )
			( 46 , 287.624437 )
			( 47 , 288.69639 )
			( 48 , 289.458922 )
			( 49 , 291.104749 )
			( 50 , 292.162122 )

		};
		\addlegendentry{\cite{ADMMSiegel}}
		
		\addplot[color=blue] plot coordinates {
			( 2 , 24.383864 )
			( 3 , 39.067296 )
			( 4 , 55.227057 )
			( 5 , 72.37912 )
			( 6 , 90.123304 )
			( 7 , 108.17176 )
			( 8 , 126.082989 )
			( 9 , 143.796384 )
			( 10 , 160.917811 )
			( 11 , 177.436505 )
			( 12 , 192.914843 )
			( 13 , 207.657392 )
			( 14 , 221.260316 )
			( 15 , 233.782427 )
			( 16 , 245.286375 )
			( 17 , 255.747709 )
			( 18 , 265.19131 )
			( 19 , 273.559016 )
			( 20 , 280.936412 )
			( 21 , 287.596273 )
			( 22 , 293.267118 )
			( 23 , 298.397662 )
			( 24 , 302.813623 )
			( 25 , 306.519368 )
			( 26 , 309.522837 )
			( 27 , 312.351366 )
			( 28 , 314.94195 )
			( 29 , 316.388022 )
			( 30 , 318.614249 )
			( 31 , 319.847751 )
			( 32 , 321.004882 )
			( 33 , 322.509762 )
			( 34 , 323.806958 )
			( 35 , 324.88226 )
			( 36 , 326.290791 )
			( 37 , 327.588438 )
			( 38 , 328.737842 )
			( 39 , 330.809207 )
			( 40 , 331.882544 )
			( 41 , 333.62838 )
			( 42 , 335.638148 )
			( 43 , 337.308362 )
			( 44 , 339.79525 )
			( 45 , 342.308667 )
			( 46 , 344.744074 )
			( 47 , 347.722976 )
			( 48 , 350.426308 )
			( 49 , 353.893138 )
			( 50 , 356.8393 )

		};
		\addlegendentry{\cite{Wasson}}
		
		\addplot[color=green] plot coordinates {
			( 2 , 48.521514 )
			( 3 , 93.231496 )
			( 4 , 143.001999 )
			( 5 , 196.541752 )
			( 6 , 251.99552 )
			( 7 , 307.843039 )
			( 8 , 361.77018 )
			( 9 , 414.141106 )
			( 10 , 462.289067 )
			( 11 , 507.761512 )
			( 12 , 546.858843 )
			( 13 , 582.00086 )
			( 14 , 612.573856 )
			( 15 , 637.559532 )
			( 16 , 656.951324 )
			( 17 , 672.152242 )
			( 18 , 683.83894 )
			( 19 , 691.349837 )
			( 20 , 693.826148 )
			( 21 , 694.693426 )
			( 22 , 689.989858 )
			( 23 , 685.240394 )
			( 24 , 677.175354 )
			( 25 , 667.360612 )
			( 26 , 655.104356 )
			( 27 , 642.927703 )
			( 28 , 628.872664 )
			( 29 , 614.315533 )
			( 30 , 600.591248 )
			( 31 , 584.420822 )
			( 32 , 567.615496 )
			( 33 , 554.512028 )
			( 34 , 538.878657 )
			( 35 , 525.74856 )
			( 36 , 512.646795 )
			( 37 , 498.384684 )
			( 38 , 484.963076 )
			( 39 , 475.400514 )
			( 40 , 463.183714 )
			( 41 , 452.389313 )
			( 42 , 443.413654 )
			( 43 , 434.454205 )
			( 44 , 426.961677 )
			( 45 , 420.271794 )
			( 46 , 415.058695 )
			( 47 , 409.156309 )
			( 48 , 404.535101 )
			( 49 , 401.637111 )
			( 50 , 397.970041 )

		};
		\addlegendentry{\cite{Wei}}

		\addplot[color=cyan] plot coordinates {
			( 2 , 36.184841 )
			( 3 , 59.821918 )
			( 4 , 79.767022 )
			( 5 , 102.869638 )
			( 6 , 124.293345 )
			( 7 , 145.404252 )
			( 8 , 164.923103 )
			( 9 , 183.689541 )
			( 10 , 200.715361 )
			( 11 , 216.88338 )
			( 12 , 231.429331 )
			( 13 , 245.342818 )
			( 14 , 257.762174 )
			( 15 , 269.327826 )
			( 16 , 279.673807 )
			( 17 , 289.489342 )
			( 18 , 298.294995 )
			( 19 , 306.398723 )
			( 20 , 313.54468 )
			( 21 , 320.535022 )
			( 22 , 326.619107 )
			( 23 , 332.625988 )
			( 24 , 338.066758 )
			( 25 , 343.113766 )
			( 26 , 347.757086 )
			( 27 , 352.29722 )
			( 28 , 356.941979 )
			( 29 , 360.823979 )
			( 30 , 365.529857 )
			( 31 , 369.509178 )
			( 32 , 373.506087 )
			( 33 , 377.958378 )
			( 34 , 382.309784 )
			( 35 , 386.619663 )
			( 36 , 391.161983 )
			( 37 , 395.827832 )
			( 38 , 400.365378 )
			( 39 , 405.496366 )
			( 40 , 410.382767 )
			( 41 , 415.462472 )
			( 42 , 420.91716 )
			( 43 , 426.15635 )
			( 44 , 432.130006 )
			( 45 , 438.082372 )
			( 46 , 443.98619 )
			( 47 , 450.088644 )
			( 48 , 456.275592 )
			( 49 , 463.219785 )
			( 50 , 469.196164 )

		};
		\addlegendentry{\cite{Heusdens}}
		
		\addplot[color=orange] plot coordinates {
			( 2 , 17.856039 )
			( 3 , 31.164937 )
			( 4 , 44.260431 )
			( 5 , 57.302304 )
			( 6 , 70.214262 )
			( 7 , 82.853528 )
			( 8 , 94.974187 )
			( 9 , 106.587925 )
			( 10 , 117.552646 )
			( 11 , 127.957173 )
			( 12 , 137.556145 )
			( 13 , 146.518343 )
			( 14 , 154.750317 )
			( 15 , 162.224871 )
			( 16 , 169.067352 )
			( 17 , 175.25749 )
			( 18 , 180.88013 )
			( 19 , 185.8651 )
			( 20 , 190.305242 )
			( 21 , 194.371369 )
			( 22 , 197.981545 )
			( 23 , 201.275879 )
			( 24 , 204.279139 )
			( 25 , 206.916231 )
			( 26 , 209.247262 )
			( 27 , 211.541122 )
			( 28 , 213.812951 )
			( 29 , 215.478648 )
			( 30 , 217.637426 )
			( 31 , 219.342909 )
			( 32 , 221.044968 )
			( 33 , 222.959075 )
			( 34 , 224.897904 )
			( 35 , 226.69108 )
			( 36 , 228.739605 )
			( 37 , 230.732581 )
			( 38 , 232.789023 )
			( 39 , 235.283793 )
			( 40 , 237.249628 )
			( 41 , 239.750937 )
			( 42 , 242.243057 )
			( 43 , 244.663221 )
			( 44 , 247.493893 )
			( 45 , 250.436581 )
			( 46 , 253.238209 )
			( 47 , 256.396825 )
			( 48 , 259.480695 )
			( 49 , 262.824234 )
			( 50 , 266.077276 )

		};
		\addlegendentry{Fix}
		
		\end{axis}
		\end{tikzpicture}

%% file: mult_div_3.tex
\begin{tikzpicture}[scale=0.8]
		\begin{axis}[
		xlabel=Check degree $d$,
		ylabel=Average number of operations,
		grid=both,
		legend style={legend pos=outer north east,font=\scriptsize}]
		\addplot[color=red] plot coordinates {
			( 2 , 0.749829 )
			( 3 , 1.346609 )
			( 4 , 2.045302 )
			( 5 , 2.800246 )
			( 6 , 3.578558 )
			( 7 , 4.362149 )
			( 8 , 5.117247 )
			( 9 , 5.847173 )
			( 10 , 6.521409 )
			( 11 , 7.14888 )
			( 12 , 7.694305 )
			( 13 , 8.186369 )
			( 14 , 8.600171 )
			( 15 , 8.9375 )
			( 16 , 9.207838 )
			( 17 , 9.40936 )
			( 18 , 9.550155 )
			( 19 , 9.620362 )
			( 20 , 9.633222 )
			( 21 , 9.600212 )
			( 22 , 9.498063 )
			( 23 , 9.378359 )
			( 24 , 9.215957 )
			( 25 , 9.009383 )
			( 26 , 8.768067 )
			( 27 , 8.516089 )
			( 28 , 8.248436 )
			( 29 , 7.923048 )
			( 30 , 7.644 )
			( 31 , 7.311719 )
			( 32 , 6.973451 )
			( 33 , 6.665175 )
			( 34 , 6.341976 )
			( 35 , 6.012636 )
			( 36 , 5.707567 )
			( 37 , 5.394737 )
			( 38 , 5.068361 )
			( 39 , 4.808293 )
			( 40 , 4.499186 )
			( 41 , 4.209203 )
			( 42 , 3.956825 )
			( 43 , 3.676636 )
			( 44 , 3.450576 )
			( 45 , 3.217639 )
			( 46 , 2.991896 )
			( 47 , 2.794721 )
			( 48 , 2.579163 )
			( 49 , 2.420614 )
			( 50 , 2.220238 )

		};
		\addlegendentry{\cite{ADMMSiegel} mult}
		
		\addplot[color=blue] plot coordinates {
			( 2 , 0.999488 )
			( 3 , 1.640944 )
			( 4 , 2.366593 )
			( 5 , 3.138118 )
			( 6 , 3.927635 )
			( 7 , 4.717645 )
			( 8 , 5.478213 )
			( 9 , 6.20854 )
			( 10 , 6.880749 )
			( 11 , 7.504363 )
			( 12 , 8.043293 )
			( 13 , 8.52786 )
			( 14 , 8.931666 )
			( 15 , 9.258119 )
			( 16 , 9.516114 )
			( 17 , 9.70606 )
			( 18 , 9.833589 )
			( 19 , 9.889648 )
			( 20 , 9.888942 )
			( 21 , 9.842145 )
			( 22 , 9.726953 )
			( 23 , 9.594094 )
			( 24 , 9.418867 )
			( 25 , 9.199999 )
			( 26 , 8.946019 )
			( 27 , 8.682091 )
			( 28 , 8.404025 )
			( 29 , 8.06625 )
			( 30 , 7.77757 )
			( 31 , 7.4353 )
			( 32 , 7.087671 )
			( 33 , 6.770462 )
			( 34 , 6.440003 )
			( 35 , 6.102292 )
			( 36 , 5.790271 )
			( 37 , 5.47081 )
			( 38 , 5.137722 )
			( 39 , 4.872356 )
			( 40 , 4.557425 )
			( 41 , 4.262617 )
			( 42 , 4.005499 )
			( 43 , 3.720996 )
			( 44 , 3.491473 )
			( 45 , 3.2549 )
			( 46 , 3.025585 )
			( 47 , 2.825675 )
			( 48 , 2.607008 )
			( 49 , 2.446115 )
			( 50 , 2.243162 )

		};
		\addlegendentry{\cite{Wasson} mult}
		
		\addplot[color=cyan] plot coordinates {
			( 2 , 1.124784 )
			( 3 , 2.176923 )
			( 4 , 2.300087 )
			( 5 , 2.907395 )
			( 6 , 3.21563 )
			( 7 , 3.560068 )
			( 8 , 3.7668 )
			( 9 , 3.94414 )
			( 10 , 4.031812 )
			( 11 , 4.095664 )
			( 12 , 4.093892 )
			( 13 , 4.080643 )
			( 14 , 4.020839 )
			( 15 , 3.947383 )
			( 16 , 3.845778 )
			( 17 , 3.73901 )
			( 18 , 3.614836 )
			( 19 , 3.481767 )
			( 20 , 3.334517 )
			( 21 , 3.193874 )
			( 22 , 3.036497 )
			( 23 , 2.891438 )
			( 24 , 2.737701 )
			( 25 , 2.587799 )
			( 26 , 2.438013 )
			( 27 , 2.292863 )
			( 28 , 2.153964 )
			( 29 , 2.010164 )
			( 30 , 1.88429 )
			( 31 , 1.753131 )
			( 32 , 1.629219 )
			( 33 , 1.516806 )
			( 34 , 1.408481 )
			( 35 , 1.303929 )
			( 36 , 1.20672 )
			( 37 , 1.116077 )
			( 38 , 1.02544 )
			( 39 , 0.951531 )
			( 40 , 0.870798 )
			( 41 , 0.799026 )
			( 42 , 0.735349 )
			( 43 , 0.669971 )
			( 44 , 0.616924 )
			( 45 , 0.564961 )
			( 46 , 0.51572 )
			( 47 , 0.472431 )
			( 48 , 0.42887 )
			( 49 , 0.395904 )
			( 50 , 0.356457 )

		};
		\addlegendentry{\cite{Heusdens} mult}
		
		\addplot[color=red,dashed] plot coordinates {
			( 2 , 0.638717 )
			( 3 , 0.706456 )
			( 4 , 0.75264 )
			( 5 , 0.785535 )
			( 6 , 0.808088 )
			( 7 , 0.822234 )
			( 8 , 0.827112 )
			( 9 , 0.825888 )
			( 10 , 0.817674 )
			( 11 , 0.805186 )
			( 12 , 0.786477 )
			( 13 , 0.765616 )
			( 14 , 0.740995 )
			( 15 , 0.713741 )
			( 16 , 0.685113 )
			( 17 , 0.655269 )
			( 18 , 0.62492 )
			( 19 , 0.593429 )
			( 20 , 0.562031 )
			( 21 , 0.531299 )
			( 22 , 0.499851 )
			( 23 , 0.470491 )
			( 24 , 0.441664 )
			( 25 , 0.413227 )
			( 26 , 0.38559 )
			( 27 , 0.359625 )
			( 28 , 0.335093 )
			( 29 , 0.309973 )
			( 30 , 0.288419 )
			( 31 , 0.266412 )
			( 32 , 0.245675 )
			( 33 , 0.227242 )
			( 34 , 0.209501 )
			( 35 , 0.192578 )
			( 36 , 0.177422 )
			( 37 , 0.162922 )
			( 38 , 0.148786 )
			( 39 , 0.137333 )
			( 40 , 0.125116 )
			( 41 , 0.114032 )
			( 42 , 0.104512 )
			( 43 , 0.094727 )
			( 44 , 0.086792 )
			( 45 , 0.079016 )
			( 46 , 0.071804 )
			( 47 , 0.065558 )
			( 48 , 0.059175 )
			( 49 , 0.054355 )
			( 50 , 0.04881 )

		};
		\addlegendentry{\cite{ADMMSiegel} div}
		
		\addplot[color=blue,dashed] plot coordinates {
			( 2 , 0.638717 )
			( 3 , 0.706456 )
			( 4 , 0.75264 )
			( 5 , 0.785535 )
			( 6 , 0.808088 )
			( 7 , 0.822234 )
			( 8 , 0.827112 )
			( 9 , 0.825888 )
			( 10 , 0.817674 )
			( 11 , 0.805186 )
			( 12 , 0.786477 )
			( 13 , 0.765616 )
			( 14 , 0.740995 )
			( 15 , 0.713741 )
			( 16 , 0.685113 )
			( 17 , 0.655269 )
			( 18 , 0.62492 )
			( 19 , 0.593429 )
			( 20 , 0.562031 )
			( 21 , 0.531299 )
			( 22 , 0.499851 )
			( 23 , 0.470491 )
			( 24 , 0.441664 )
			( 25 , 0.413227 )
			( 26 , 0.38559 )
			( 27 , 0.359625 )
			( 28 , 0.335093 )
			( 29 , 0.309973 )
			( 30 , 0.288419 )
			( 31 , 0.266412 )
			( 32 , 0.245675 )
			( 33 , 0.227242 )
			( 34 , 0.209501 )
			( 35 , 0.192578 )
			( 36 , 0.177422 )
			( 37 , 0.162922 )
			( 38 , 0.148786 )
			( 39 , 0.137333 )
			( 40 , 0.125116 )
			( 41 , 0.114032 )
			( 42 , 0.104512 )
			( 43 , 0.094727 )
			( 44 , 0.086792 )
			( 45 , 0.079016 )
			( 46 , 0.071804 )
			( 47 , 0.065558 )
			( 48 , 0.059175 )
			( 49 , 0.054355 )
			( 50 , 0.04881 )

		};
		\addlegendentry{\cite{Wasson} div}
		
		\addplot[color=green,dashed] plot coordinates {
			( 2 , 5.325866 )
			( 3 , 7.677848 )
			( 4 , 9.380756 )
			( 5 , 10.702892 )
			( 6 , 11.724657 )
			( 7 , 12.496851 )
			( 8 , 13.016782 )
			( 9 , 13.376038 )
			( 10 , 13.535439 )
			( 11 , 13.59137 )
			( 12 , 13.469431 )
			( 13 , 13.270712 )
			( 14 , 12.994012 )
			( 15 , 12.633914 )
			( 16 , 12.204938 )
			( 17 , 11.746178 )
			( 18 , 11.273855 )
			( 19 , 10.778171 )
			( 20 , 10.250575 )
			( 21 , 9.746072 )
			( 22 , 9.202123 )
			( 23 , 8.706134 )
			( 24 , 8.205276 )
			( 25 , 7.720102 )
			( 26 , 7.241098 )
			( 27 , 6.797599 )
			( 28 , 6.36246 )
			( 29 , 5.951975 )
			( 30 , 5.576839 )
			( 31 , 5.199752 )
			( 32 , 4.839378 )
			( 33 , 4.53737 )
			( 34 , 4.227621 )
			( 35 , 3.959478 )
			( 36 , 3.705999 )
			( 37 , 3.45594 )
			( 38 , 3.224522 )
			( 39 , 3.038508 )
			( 40 , 2.8409 )
			( 41 , 2.661291 )
			( 42 , 2.50713 )
			( 43 , 2.358809 )
			( 44 , 2.228636 )
			( 45 , 2.10889 )
			( 46 , 2.006305 )
			( 47 , 1.902489 )
			( 48 , 1.811267 )
			( 49 , 1.736895 )
			( 50 , 1.658677 )

		};
		\addlegendentry{\cite{Wei} div}

		\addplot[color=cyan,dashed] plot coordinates {
			( 2 , 0.638717 )
			( 3 , 0.873572 )
			( 4 , 0.75264 )
			( 5 , 0.816461 )
			( 6 , 0.808088 )
			( 7 , 0.827061 )
			( 8 , 0.827112 )
			( 9 , 0.826547 )
			( 10 , 0.817674 )
			( 11 , 0.805284 )
			( 12 , 0.786477 )
			( 13 , 0.765628 )
			( 14 , 0.740995 )
			( 15 , 0.713744 )
			( 16 , 0.685113 )
			( 17 , 0.655269 )
			( 18 , 0.62492 )
			( 19 , 0.593429 )
			( 20 , 0.562031 )
			( 21 , 0.531299 )
			( 22 , 0.499851 )
			( 23 , 0.470491 )
			( 24 , 0.441664 )
			( 25 , 0.413227 )
			( 26 , 0.38559 )
			( 27 , 0.359625 )
			( 28 , 0.335093 )
			( 29 , 0.309973 )
			( 30 , 0.288419 )
			( 31 , 0.266412 )
			( 32 , 0.245675 )
			( 33 , 0.227242 )
			( 34 , 0.209501 )
			( 35 , 0.192578 )
			( 36 , 0.177422 )
			( 37 , 0.162922 )
			( 38 , 0.148786 )
			( 39 , 0.137333 )
			( 40 , 0.125116 )
			( 41 , 0.114032 )
			( 42 , 0.104512 )
			( 43 , 0.094727 )
			( 44 , 0.086792 )
			( 45 , 0.079016 )
			( 46 , 0.071804 )
			( 47 , 0.065558 )
			( 48 , 0.059175 )
			( 49 , 0.054355 )
			( 50 , 0.04881 )

		};
		\addlegendentry{\cite{Heusdens} div}
		
		\addplot[color=orange,dashed] plot coordinates {
			( 2 , 0.638717 )
			( 3 , 1.214353 )
			( 4 , 1.624121 )
			( 5 , 1.955617 )
			( 6 , 2.228282 )
			( 7 , 2.446594 )
			( 8 , 2.610512 )
			( 9 , 2.734561 )
			( 10 , 2.818242 )
			( 11 , 2.874108 )
			( 12 , 2.89446 )
			( 13 , 2.893932 )
			( 14 , 2.868662 )
			( 15 , 2.82304 )
			( 16 , 2.76184 )
			( 17 , 2.688547 )
			( 18 , 2.606008 )
			( 19 , 2.511603 )
			( 20 , 2.412027 )
			( 21 , 2.308596 )
			( 22 , 2.198168 )
			( 23 , 2.091957 )
			( 24 , 1.985453 )
			( 25 , 1.876078 )
			( 26 , 1.766669 )
			( 27 , 1.663133 )
			( 28 , 1.562938 )
			( 29 , 1.457829 )
			( 30 , 1.367507 )
			( 31 , 1.272237 )
			( 32 , 1.182285 )
			( 33 , 1.100413 )
			( 34 , 1.021721 )
			( 35 , 0.945373 )
			( 36 , 0.876547 )
			( 37 , 0.809609 )
			( 38 , 0.743775 )
			( 39 , 0.690395 )
			( 40 , 0.632182 )
			( 41 , 0.579176 )
			( 42 , 0.533479 )
			( 43 , 0.485888 )
			( 44 , 0.446802 )
			( 45 , 0.408967 )
			( 46 , 0.372845 )
			( 47 , 0.341906 )
			( 48 , 0.310268 )
			( 49 , 0.285971 )
			( 50 , 0.25797 )

		};
		\addlegendentry{Fix div}
		
		\end{axis}
		\end{tikzpicture}

%% file: low_complexity_5.tex
\begin{tikzpicture}[scale=0.8]
		\begin{axis}[
		xlabel=Check degree $d$,
		ylabel=Average number of operations,
		grid=both,
		legend style={legend pos=outer north east,font=\scriptsize}]
		\addplot[color=red] plot coordinates {
			( 2 , 23.611624 )
			( 3 , 37.992385 )
			( 4 , 53.466124 )
			( 5 , 70.125017 )
			( 6 , 87.729362 )
			( 7 , 106.201409 )
			( 8 , 125.206376 )
			( 9 , 144.680452 )
			( 10 , 164.513991 )
			( 11 , 184.59751 )
			( 12 , 204.392985 )
			( 13 , 224.319299 )
			( 14 , 243.976974 )
			( 15 , 263.192338 )
			( 16 , 282.118471 )
			( 17 , 300.589921 )
			( 18 , 318.319623 )
			( 19 , 335.7345 )
			( 20 , 351.890239 )
			( 21 , 368.007771 )
			( 22 , 382.404451 )
			( 23 , 397.152657 )
			( 24 , 410.370371 )
			( 25 , 423.258167 )
			( 26 , 435.126387 )
			( 27 , 445.739157 )
			( 28 , 456.366369 )
			( 29 , 465.197028 )
			( 30 , 474.888237 )
			( 31 , 482.198315 )
			( 32 , 489.319805 )
			( 33 , 496.258708 )
			( 34 , 502.090008 )
			( 35 , 507.121876 )
			( 36 , 511.978286 )
			( 37 , 515.808501 )
			( 38 , 519.195543 )
			( 39 , 522.003786 )
			( 40 , 523.756984 )
			( 41 , 525.65425 )
			( 42 , 527.773962 )
			( 43 , 528.699922 )
			( 44 , 529.272432 )
			( 45 , 530.183846 )
			( 46 , 530.755639 )
			( 47 , 530.215793 )
			( 48 , 529.454842 )
			( 49 , 529.281231 )
			( 50 , 527.740666 )

		};
		\addlegendentry{\cite{ADMMSiegel}}
		
		\addplot[color=blue] plot coordinates {
			( 2 , 23.798248 )
			( 3 , 37.686693 )
			( 4 , 52.869337 )
			( 5 , 69.153417 )
			( 6 , 86.293135 )
			( 7 , 104.182342 )
			( 8 , 122.581261 )
			( 9 , 141.387694 )
			( 10 , 160.531247 )
			( 11 , 179.889474 )
			( 12 , 199.031202 )
			( 13 , 218.320461 )
			( 14 , 237.354334 )
			( 15 , 256.004569 )
			( 16 , 274.504937 )
			( 17 , 292.567398 )
			( 18 , 309.969942 )
			( 19 , 327.167207 )
			( 20 , 343.290475 )
			( 21 , 359.382314 )
			( 22 , 374.062592 )
			( 23 , 388.924882 )
			( 24 , 402.530944 )
			( 25 , 415.889744 )
			( 26 , 428.341061 )
			( 27 , 439.672074 )
			( 28 , 451.112096 )
			( 29 , 460.965069 )
			( 30 , 471.471114 )
			( 31 , 480.064076 )
			( 32 , 488.4543 )
			( 33 , 496.657126 )
			( 34 , 503.862356 )
			( 35 , 510.511537 )
			( 36 , 516.857143 )
			( 37 , 522.400174 )
			( 38 , 527.594141 )
			( 39 , 532.212476 )
			( 40 , 535.895156 )
			( 41 , 539.797016 )
			( 42 , 543.821137 )
			( 43 , 546.840098 )
			( 44 , 549.523536 )
			( 45 , 552.628773 )
			( 46 , 555.229889 )
			( 47 , 556.960262 )
			( 48 , 558.477511 )
			( 49 , 560.577409 )
			( 50 , 561.358147 )

		};
		\addlegendentry{\cite{Wasson}}
		
		\addplot[color=green] plot coordinates {
			( 2 , 56.504597 )
			( 3 , 106.606058 )
			( 4 , 163.993257 )
			( 5 , 228.148278 )
			( 6 , 297.741689 )
			( 7 , 371.431245 )
			( 8 , 446.806079 )
			( 9 , 524.899266 )
			( 10 , 602.975597 )
			( 11 , 682.226476 )
			( 12 , 758.341869 )
			( 13 , 833.881916 )
			( 14 , 906.305049 )
			( 15 , 976.177961 )
			( 16 , 1042.174269 )
			( 17 , 1105.218637 )
			( 18 , 1164.684689 )
			( 19 , 1221.183513 )
			( 20 , 1269.655394 )
			( 21 , 1317.331045 )
			( 22 , 1356.108436 )
			( 23 , 1396.794943 )
			( 24 , 1428.097674 )
			( 25 , 1457.118702 )
			( 26 , 1482.379606 )
			( 27 , 1499.794708 )
			( 28 , 1516.244362 )
			( 29 , 1528.18866 )
			( 30 , 1541.489755 )
			( 31 , 1542.771944 )
			( 32 , 1545.110245 )
			( 33 , 1546.252419 )
			( 34 , 1542.146705 )
			( 35 , 1535.411592 )
			( 36 , 1528.890224 )
			( 37 , 1518.568798 )
			( 38 , 1502.564381 )
			( 39 , 1489.929771 )
			( 40 , 1470.630822 )
			( 41 , 1452.728136 )
			( 42 , 1432.420877 )
			( 43 , 1414.58185 )
			( 44 , 1391.910257 )
			( 45 , 1370.020266 )
			( 46 , 1351.908894 )
			( 47 , 1327.566842 )
			( 48 , 1300.607352 )
			( 49 , 1279.662071 )
			( 50 , 1253.699429 )

		};
		\addlegendentry{\cite{Wei}}

		\addplot[color=cyan] plot coordinates {
			( 2 , 34.661743 )
			( 3 , 58.990471 )
			( 4 , 75.284972 )
			( 5 , 98.346241 )
			( 6 , 117.753585 )
			( 7 , 139.211436 )
			( 8 , 159.117166 )
			( 9 , 179.440182 )
			( 10 , 198.82413 )
			( 11 , 218.157279 )
			( 12 , 236.399412 )
			( 13 , 254.789134 )
			( 14 , 272.353834 )
			( 15 , 289.406837 )
			( 16 , 305.930872 )
			( 17 , 322.062948 )
			( 18 , 337.259787 )
			( 19 , 352.395733 )
			( 20 , 366.22629 )
			( 21 , 380.164653 )
			( 22 , 392.871958 )
			( 23 , 405.968223 )
			( 24 , 417.762506 )
			( 25 , 429.463344 )
			( 26 , 440.135645 )
			( 27 , 450.348531 )
			( 28 , 460.482057 )
			( 29 , 469.735368 )
			( 30 , 479.477446 )
			( 31 , 487.781474 )
			( 32 , 495.824172 )
			( 33 , 504.078551 )
			( 34 , 511.450524 )
			( 35 , 518.420848 )
			( 36 , 525.545405 )
			( 37 , 531.987119 )
			( 38 , 538.018377 )
			( 39 , 544.140582 )
			( 40 , 549.587653 )
			( 41 , 555.204283 )
			( 42 , 560.887584 )
			( 43 , 566.016817 )
			( 44 , 571.359854 )
			( 45 , 576.70886 )
			( 46 , 581.722566 )
			( 47 , 586.287247 )
			( 48 , 590.731361 )
			( 49 , 596.035914 )
			( 50 , 599.768451 )

		};
		\addlegendentry{\cite{Heusdens}}
		
		\addplot[color=orange] plot coordinates {
			( 2 , 17.628465 )
			( 3 , 29.893039 )
			( 4 , 41.83592 )
			( 5 , 54.033233 )
			( 6 , 66.403412 )
			( 7 , 78.876214 )
			( 8 , 91.327879 )
			( 9 , 103.685317 )
			( 10 , 116.046813 )
			( 11 , 128.305867 )
			( 12 , 140.227816 )
			( 13 , 152.07639 )
			( 14 , 163.590233 )
			( 15 , 174.738458 )
			( 16 , 185.6468 )
			( 17 , 196.215626 )
			( 18 , 206.292346 )
			( 19 , 216.145906 )
			( 20 , 225.369303 )
			( 21 , 234.551107 )
			( 22 , 242.845046 )
			( 23 , 251.251268 )
			( 24 , 258.919897 )
			( 25 , 266.467611 )
			( 26 , 273.527332 )
			( 27 , 279.978136 )
			( 28 , 286.486299 )
			( 29 , 292.199612 )
			( 30 , 298.22284 )
			( 31 , 303.312453 )
			( 32 , 308.272 )
			( 33 , 313.177473 )
			( 34 , 317.600991 )
			( 35 , 321.697337 )
			( 36 , 325.781599 )
			( 37 , 329.489379 )
			( 38 , 332.959454 )
			( 39 , 336.195059 )
			( 40 , 339.081692 )
			( 41 , 342.055889 )
			( 42 , 345.156255 )
			( 43 , 347.792485 )
			( 44 , 350.163782 )
			( 45 , 352.94143 )
			( 46 , 355.474985 )
			( 47 , 357.555486 )
			( 48 , 359.642951 )
			( 49 , 362.028504 )
			( 50 , 363.813902 )

		};
		\addlegendentry{Fix}
		
		\end{axis}
		\end{tikzpicture}

%% file: mult_div_5.tex
\begin{tikzpicture}[scale=0.8]
		\begin{axis}[
		xlabel=Check degree $d$,
		ylabel=Average number of operations,
		grid=both,
		legend style={legend pos=outer north east,font=\scriptsize}]
		\addplot[color=red] plot coordinates {
			( 2 , 0.830233 )
			( 3 , 1.388578 )
			( 4 , 2.029294 )
			( 5 , 2.735235 )
			( 6 , 3.485377 )
			( 7 , 4.274867 )
			( 8 , 5.078641 )
			( 9 , 5.899749 )
			( 10 , 6.724199 )
			( 11 , 7.555417 )
			( 12 , 8.353565 )
			( 13 , 9.150445 )
			( 14 , 9.923418 )
			( 15 , 10.657712 )
			( 16 , 11.373422 )
			( 17 , 12.052848 )
			( 18 , 12.686012 )
			( 19 , 13.298036 )
			( 20 , 13.837455 )
			( 21 , 14.366629 )
			( 22 , 14.802482 )
			( 23 , 15.254292 )
			( 24 , 15.626452 )
			( 25 , 15.969578 )
			( 26 , 16.263736 )
			( 27 , 16.489544 )
			( 28 , 16.719757 )
			( 29 , 16.849058 )
			( 30 , 17.028111 )
			( 31 , 17.079234 )
			( 32 , 17.129326 )
			( 33 , 17.168455 )
			( 34 , 17.15068 )
			( 35 , 17.098476 )
			( 36 , 17.030993 )
			( 37 , 16.907836 )
			( 38 , 16.775323 )
			( 39 , 16.618939 )
			( 40 , 16.401896 )
			( 41 , 16.197508 )
			( 42 , 16.003488 )
			( 43 , 15.751409 )
			( 44 , 15.501043 )
			( 45 , 15.250603 )
			( 46 , 14.987929 )
			( 47 , 14.682186 )
			( 48 , 14.356381 )
			( 49 , 14.067949 )
			( 50 , 13.715495 )

		};
		\addlegendentry{\cite{ADMMSiegel} mult}
		
		\addplot[color=blue] plot coordinates {
			( 2 , 0.9997 )
			( 3 , 1.608269 )
			( 4 , 2.284243 )
			( 5 , 3.016331 )
			( 6 , 3.786581 )
			( 7 , 4.59146 )
			( 8 , 5.406677 )
			( 9 , 6.235745 )
			( 10 , 7.06743 )
			( 11 , 7.904699 )
			( 12 , 8.705535 )
			( 13 , 9.505481 )
			( 14 , 10.279929 )
			( 15 , 11.013475 )
			( 16 , 11.729533 )
			( 17 , 12.406756 )
			( 18 , 13.037529 )
			( 19 , 13.645915 )
			( 20 , 14.182269 )
			( 21 , 14.707739 )
			( 22 , 15.136955 )
			( 23 , 15.583716 )
			( 24 , 15.95043 )
			( 25 , 16.287789 )
			( 26 , 16.574602 )
			( 27 , 16.793198 )
			( 28 , 17.01751 )
			( 29 , 17.137955 )
			( 30 , 17.309431 )
			( 31 , 17.353384 )
			( 32 , 17.395191 )
			( 33 , 17.426982 )
			( 34 , 17.401363 )
			( 35 , 17.341812 )
			( 36 , 17.266208 )
			( 37 , 17.134561 )
			( 38 , 16.994263 )
			( 39 , 16.830108 )
			( 40 , 16.605587 )
			( 41 , 16.393863 )
			( 42 , 16.193142 )
			( 43 , 15.933758 )
			( 44 , 15.6763 )
			( 45 , 15.419724 )
			( 46 , 15.149992 )
			( 47 , 14.837334 )
			( 48 , 14.505029 )
			( 49 , 14.210709 )
			( 50 , 13.851374 )

		};
		\addlegendentry{\cite{Wasson} mult}
		
		\addplot[color=cyan] plot coordinates {
			( 2 , 1.084882 )
			( 3 , 2.213285 )
			( 4 , 2.041773 )
			( 5 , 2.705656 )
			( 6 , 2.875495 )
			( 7 , 3.272157 )
			( 8 , 3.490923 )
			( 9 , 3.744753 )
			( 10 , 3.922883 )
			( 11 , 4.100273 )
			( 12 , 4.216254 )
			( 13 , 4.339527 )
			( 14 , 4.423869 )
			( 15 , 4.496044 )
			( 16 , 4.544477 )
			( 17 , 4.587794 )
			( 18 , 4.600079 )
			( 19 , 4.618712 )
			( 20 , 4.598628 )
			( 21 , 4.591038 )
			( 22 , 4.547338 )
			( 23 , 4.525819 )
			( 24 , 4.471021 )
			( 25 , 4.418818 )
			( 26 , 4.352588 )
			( 27 , 4.280893 )
			( 28 , 4.207373 )
			( 29 , 4.125297 )
			( 30 , 4.050018 )
			( 31 , 3.956605 )
			( 32 , 3.861795 )
			( 33 , 3.775621 )
			( 34 , 3.676837 )
			( 35 , 3.58038 )
			( 36 , 3.481046 )
			( 37 , 3.381338 )
			( 38 , 3.278051 )
			( 39 , 3.180048 )
			( 40 , 3.071844 )
			( 41 , 2.973453 )
			( 42 , 2.877972 )
			( 43 , 2.777818 )
			( 44 , 2.681279 )
			( 45 , 2.590925 )
			( 46 , 2.498794 )
			( 47 , 2.404938 )
			( 48 , 2.310156 )
			( 49 , 2.228424 )
			( 50 , 2.132623 )

		};
		\addlegendentry{\cite{Heusdens} mult}
		
		\addplot[color=red,dashed] plot coordinates {
			( 2 , 0.589832 )
			( 3 , 0.634313 )
			( 4 , 0.66972 )
			( 5 , 0.701303 )
			( 6 , 0.728024 )
			( 7 , 0.751547 )
			( 8 , 0.769935 )
			( 9 , 0.785315 )
			( 10 , 0.797374 )
			( 11 , 0.807183 )
			( 12 , 0.811901 )
			( 13 , 0.815507 )
			( 14 , 0.816471 )
			( 15 , 0.814167 )
			( 16 , 0.810699 )
			( 17 , 0.805121 )
			( 18 , 0.797192 )
			( 19 , 0.788892 )
			( 20 , 0.777331 )
			( 21 , 0.766323 )
			( 22 , 0.751455 )
			( 23 , 0.738816 )
			( 24 , 0.72365 )
			( 25 , 0.70823 )
			( 26 , 0.692052 )
			( 27 , 0.674274 )
			( 28 , 0.658089 )
			( 29 , 0.639109 )
			( 30 , 0.623308 )
			( 31 , 0.604022 )
			( 32 , 0.585995 )
			( 33 , 0.568695 )
			( 34 , 0.550582 )
			( 35 , 0.53253 )
			( 36 , 0.514992 )
			( 37 , 0.496837 )
			( 38 , 0.479409 )
			( 39 , 0.462245 )
			( 40 , 0.444316 )
			( 41 , 0.427639 )
			( 42 , 0.412055 )
			( 43 , 0.395732 )
			( 44 , 0.380275 )
			( 45 , 0.36545 )
			( 46 , 0.351047 )
			( 47 , 0.336293 )
			( 48 , 0.321702 )
			( 49 , 0.308557 )
			( 50 , 0.294586 )

		};
		\addlegendentry{\cite{ADMMSiegel} div}
		
		\addplot[color=blue,dashed] plot coordinates {
			( 2 , 0.589832 )
			( 3 , 0.634313 )
			( 4 , 0.66972 )
			( 5 , 0.701303 )
			( 6 , 0.728024 )
			( 7 , 0.751547 )
			( 8 , 0.769935 )
			( 9 , 0.785315 )
			( 10 , 0.797374 )
			( 11 , 0.807183 )
			( 12 , 0.811901 )
			( 13 , 0.815507 )
			( 14 , 0.816471 )
			( 15 , 0.814167 )
			( 16 , 0.810699 )
			( 17 , 0.805121 )
			( 18 , 0.797192 )
			( 19 , 0.788892 )
			( 20 , 0.777331 )
			( 21 , 0.766323 )
			( 22 , 0.751455 )
			( 23 , 0.738816 )
			( 24 , 0.72365 )
			( 25 , 0.70823 )
			( 26 , 0.692052 )
			( 27 , 0.674274 )
			( 28 , 0.658089 )
			( 29 , 0.639109 )
			( 30 , 0.623308 )
			( 31 , 0.604022 )
			( 32 , 0.585995 )
			( 33 , 0.568695 )
			( 34 , 0.550582 )
			( 35 , 0.53253 )
			( 36 , 0.514992 )
			( 37 , 0.496837 )
			( 38 , 0.479409 )
			( 39 , 0.462245 )
			( 40 , 0.444316 )
			( 41 , 0.427639 )
			( 42 , 0.412055 )
			( 43 , 0.395732 )
			( 44 , 0.380275 )
			( 45 , 0.36545 )
			( 46 , 0.351047 )
			( 47 , 0.336293 )
			( 48 , 0.321702 )
			( 49 , 0.308557 )
			( 50 , 0.294586 )

		};
		\addlegendentry{\cite{Wasson} div}
		
		\addplot[color=green,dashed] plot coordinates {
			( 2 , 6.321751 )
			( 3 , 8.885718 )
			( 4 , 10.870073 )
			( 5 , 12.552678 )
			( 6 , 14.002162 )
			( 7 , 15.252175 )
			( 8 , 16.276788 )
			( 9 , 17.185266 )
			( 10 , 17.921038 )
			( 11 , 18.565792 )
			( 12 , 19.024705 )
			( 13 , 19.403436 )
			( 14 , 19.659815 )
			( 15 , 19.828647 )
			( 16 , 19.899464 )
			( 17 , 19.906669 )
			( 18 , 19.849605 )
			( 19 , 19.748365 )
			( 20 , 19.527365 )
			( 21 , 19.313507 )
			( 22 , 18.987676 )
			( 23 , 18.717337 )
			( 24 , 18.343083 )
			( 25 , 17.9657 )
			( 26 , 17.570735 )
			( 27 , 17.11006 )
			( 28 , 16.670483 )
			( 29 , 16.208992 )
			( 30 , 15.793709 )
			( 31 , 15.276466 )
			( 32 , 14.80335 )
			( 33 , 14.345685 )
			( 34 , 13.864252 )
			( 35 , 13.385648 )
			( 36 , 12.934047 )
			( 37 , 12.47216 )
			( 38 , 11.986538 )
			( 39 , 11.554231 )
			( 40 , 11.087233 )
			( 41 , 10.654324 )
			( 42 , 10.2226 )
			( 43 , 9.829278 )
			( 44 , 9.419808 )
			( 45 , 9.030984 )
			( 46 , 8.6871 )
			( 47 , 8.314748 )
			( 48 , 7.93849 )
			( 49 , 7.618071 )
			( 50 , 7.277846 )

		};
		\addlegendentry{\cite{Wei} div}

		\addplot[color=cyan,dashed] plot coordinates {
			( 2 , 0.589832 )
			( 3 , 0.874788 )
			( 4 , 0.66972 )
			( 5 , 0.765341 )
			( 6 , 0.728024 )
			( 7 , 0.766004 )
			( 8 , 0.769935 )
			( 9 , 0.788247 )
			( 10 , 0.797374 )
			( 11 , 0.807749 )
			( 12 , 0.811901 )
			( 13 , 0.815612 )
			( 14 , 0.816471 )
			( 15 , 0.814192 )
			( 16 , 0.810699 )
			( 17 , 0.805125 )
			( 18 , 0.797192 )
			( 19 , 0.788892 )
			( 20 , 0.777331 )
			( 21 , 0.766323 )
			( 22 , 0.751455 )
			( 23 , 0.738816 )
			( 24 , 0.72365 )
			( 25 , 0.70823 )
			( 26 , 0.692052 )
			( 27 , 0.674274 )
			( 28 , 0.658089 )
			( 29 , 0.639109 )
			( 30 , 0.623308 )
			( 31 , 0.604022 )
			( 32 , 0.585995 )
			( 33 , 0.568695 )
			( 34 , 0.550582 )
			( 35 , 0.53253 )
			( 36 , 0.514992 )
			( 37 , 0.496837 )
			( 38 , 0.479409 )
			( 39 , 0.462245 )
			( 40 , 0.444316 )
			( 41 , 0.427639 )
			( 42 , 0.412055 )
			( 43 , 0.395732 )
			( 44 , 0.380275 )
			( 45 , 0.36545 )
			( 46 , 0.351047 )
			( 47 , 0.336293 )
			( 48 , 0.321702 )
			( 49 , 0.308557 )
			( 50 , 0.294586 )

		};
		\addlegendentry{\cite{Heusdens} div}
		
		\addplot[color=orange,dashed] plot coordinates {
			( 2 , 0.589832 )
			( 3 , 1.065562 )
			( 4 , 1.410257 )
			( 5 , 1.712811 )
			( 6 , 1.979881 )
			( 7 , 2.215918 )
			( 8 , 2.419141 )
			( 9 , 2.597802 )
			( 10 , 2.757251 )
			( 11 , 2.900132 )
			( 12 , 3.015421 )
			( 13 , 3.120814 )
			( 14 , 3.205736 )
			( 15 , 3.27158 )
			( 16 , 3.325199 )
			( 17 , 3.366134 )
			( 18 , 3.390713 )
			( 19 , 3.408745 )
			( 20 , 3.408982 )
			( 21 , 3.408033 )
			( 22 , 3.385002 )
			( 23 , 3.367959 )
			( 24 , 3.337296 )
			( 25 , 3.301201 )
			( 26 , 3.25861 )
			( 27 , 3.206029 )
			( 28 , 3.15807 )
			( 29 , 3.094033 )
			( 30 , 3.042677 )
			( 31 , 2.972258 )
			( 32 , 2.906127 )
			( 33 , 2.839454 )
			( 34 , 2.768831 )
			( 35 , 2.695253 )
			( 36 , 2.624011 )
			( 37 , 2.54669 )
			( 38 , 2.470661 )
			( 39 , 2.396359 )
			( 40 , 2.315732 )
			( 41 , 2.240862 )
			( 42 , 2.170252 )
			( 43 , 2.093678 )
			( 44 , 2.021753 )
			( 45 , 1.952188 )
			( 46 , 1.883594 )
			( 47 , 1.812486 )
			( 48 , 1.741061 )
			( 49 , 1.676535 )
			( 50 , 1.607416 )

		};
		\addlegendentry{Fix div}
		\end{axis}
		\end{tikzpicture}

%% file: low_complexity_10.tex
\begin{tikzpicture}[scale=0.8]
		\begin{axis}[
		xlabel=Check degree $d$,
		ylabel=Average number of operations,
		grid=both,
		legend style={legend pos=outer north east,font=\scriptsize}]
		\addplot[color=red] plot coordinates {
			( 2 , 23.703125 )
			( 3 , 37.193932 )
			( 4 , 51.529258 )
			( 5 , 66.834501 )
			( 6 , 82.952819 )
			( 7 , 99.903196 )
			( 8 , 117.431577 )
			( 9 , 135.715977 )
			( 10 , 154.585732 )
			( 11 , 174.114521 )
			( 12 , 193.871494 )
			( 13 , 214.233345 )
			( 14 , 235.013219 )
			( 15 , 255.885241 )
			( 16 , 277.216411 )
			( 17 , 298.659986 )
			( 18 , 320.496973 )
			( 19 , 342.481207 )
			( 20 , 364.456107 )
			( 21 , 386.308756 )
			( 22 , 408.637841 )
			( 23 , 430.572639 )
			( 24 , 452.334174 )
			( 25 , 474.233367 )
			( 26 , 496.234084 )
			( 27 , 517.834089 )
			( 28 , 539.506183 )
			( 29 , 560.720454 )
			( 30 , 581.74014 )
			( 31 , 602.828752 )
			( 32 , 623.198921 )
			( 33 , 643.451508 )
			( 34 , 663.489285 )
			( 35 , 683.256448 )
			( 36 , 701.999285 )
			( 37 , 721.323829 )
			( 38 , 739.84335 )
			( 39 , 758.01413 )
			( 40 , 775.605947 )
			( 41 , 792.827587 )
			( 42 , 810.790142 )
			( 43 , 826.211439 )
			( 44 , 842.891421 )
			( 45 , 858.39637 )
			( 46 , 873.627355 )
			( 47 , 888.87555 )
			( 48 , 902.8892 )
			( 49 , 917.149659 )
			( 50 , 930.059785 )

		};
		\addlegendentry{\cite{ADMMSiegel}}
		
		\addplot[color=blue] plot coordinates {
			( 2 , 23.301216 )
			( 3 , 36.397606 )
			( 4 , 50.448382 )
			( 5 , 65.365604 )
			( 6 , 81.024864 )
			( 7 , 97.386079 )
			( 8 , 114.301123 )
			( 9 , 131.864345 )
			( 10 , 149.943176 )
			( 11 , 168.593982 )
			( 12 , 187.465755 )
			( 13 , 206.891273 )
			( 14 , 226.685361 )
			( 15 , 246.534841 )
			( 16 , 266.807994 )
			( 17 , 287.243577 )
			( 18 , 307.982715 )
			( 19 , 328.871188 )
			( 20 , 349.767395 )
			( 21 , 370.545105 )
			( 22 , 391.827896 )
			( 23 , 412.695183 )
			( 24 , 433.471537 )
			( 25 , 454.436279 )
			( 26 , 475.370986 )
			( 27 , 496.082825 )
			( 28 , 516.764745 )
			( 29 , 537.198062 )
			( 30 , 557.425141 )
			( 31 , 577.700511 )
			( 32 , 597.395331 )
			( 33 , 616.938805 )
			( 34 , 636.388323 )
			( 35 , 655.621743 )
			( 36 , 673.881389 )
			( 37 , 692.739616 )
			( 38 , 710.897686 )
			( 39 , 728.643741 )
			( 40 , 745.995647 )
			( 41 , 763.088429 )
			( 42 , 780.765179 )
			( 43 , 796.282668 )
			( 44 , 812.88771 )
			( 45 , 828.599882 )
			( 46 , 843.919292 )
			( 47 , 859.347893 )
			( 48 , 873.641703 )
			( 49 , 888.241189 )
			( 50 , 901.574846 )

		};
		\addlegendentry{\cite{Wasson}}
		
		\addplot[color=green] plot coordinates {
			( 2 , 72.143983 )
			( 3 , 130.976065 )
			( 4 , 198.965874 )
			( 5 , 276.233865 )
			( 6 , 360.908358 )
			( 7 , 452.933982 )
			( 8 , 549.519243 )
			( 9 , 652.753537 )
			( 10 , 760.211237 )
			( 11 , 871.690334 )
			( 12 , 985.430974 )
			( 13 , 1102.41966 )
			( 14 , 1220.570058 )
			( 15 , 1341.556218 )
			( 16 , 1461.547124 )
			( 17 , 1582.930287 )
			( 18 , 1705.050279 )
			( 19 , 1828.112096 )
			( 20 , 1948.860628 )
			( 21 , 2065.235878 )
			( 22 , 2184.583161 )
			( 23 , 2303.840186 )
			( 24 , 2414.857888 )
			( 25 , 2528.318395 )
			( 26 , 2639.071041 )
			( 27 , 2742.931389 )
			( 28 , 2849.701976 )
			( 29 , 2952.222399 )
			( 30 , 3051.652994 )
			( 31 , 3145.995208 )
			( 32 , 3235.254597 )
			( 33 , 3325.363427 )
			( 34 , 3409.952847 )
			( 35 , 3493.214257 )
			( 36 , 3568.305925 )
			( 37 , 3647.809776 )
			( 38 , 3713.219064 )
			( 39 , 3783.903906 )
			( 40 , 3838.520512 )
			( 41 , 3898.859998 )
			( 42 , 3958.113876 )
			( 43 , 4010.824706 )
			( 44 , 4051.565013 )
			( 45 , 4101.097453 )
			( 46 , 4141.461201 )
			( 47 , 4178.096648 )
			( 48 , 4216.205071 )
			( 49 , 4243.862274 )
			( 50 , 4262.384675 )

		};
		\addlegendentry{\cite{Wei}}

		\addplot[color=cyan] plot coordinates {
			( 2 , 33.298335 )
			( 3 , 58.124396 )
			( 4 , 70.653026 )
			( 5 , 93.291943 )
			( 6 , 109.205331 )
			( 7 , 129.48496 )
			( 8 , 147.077785 )
			( 9 , 166.361028 )
			( 10 , 184.604494 )
			( 11 , 203.491681 )
			( 12 , 221.671043 )
			( 13 , 240.491328 )
			( 14 , 258.914276 )
			( 15 , 277.347336 )
			( 16 , 295.662974 )
			( 17 , 314.009427 )
			( 18 , 332.186772 )
			( 19 , 350.501862 )
			( 20 , 368.353077 )
			( 21 , 386.021773 )
			( 22 , 404.05237 )
			( 23 , 421.833295 )
			( 24 , 438.974859 )
			( 25 , 456.432265 )
			( 26 , 473.273997 )
			( 27 , 490.247878 )
			( 28 , 506.748028 )
			( 29 , 523.482528 )
			( 30 , 540.015274 )
			( 31 , 556.215576 )
			( 32 , 571.815884 )
			( 33 , 587.578757 )
			( 34 , 602.922715 )
			( 35 , 617.955848 )
			( 36 , 632.806467 )
			( 37 , 647.96666 )
			( 38 , 661.922557 )
			( 39 , 676.030392 )
			( 40 , 689.843347 )
			( 41 , 703.526936 )
			( 42 , 717.424249 )
			( 43 , 729.841962 )
			( 44 , 743.463644 )
			( 45 , 755.983323 )
			( 46 , 768.458702 )
			( 47 , 780.435426 )
			( 48 , 792.538695 )
			( 49 , 804.616143 )
			( 50 , 815.569405 )

		};
		\addlegendentry{\cite{Heusdens}}
		
		\addplot[color=orange] plot coordinates {
			( 2 , 17.446305 )
			( 3 , 28.589279 )
			( 4 , 39.426108 )
			( 5 , 50.458599 )
			( 6 , 61.660798 )
			( 7 , 73.033099 )
			( 8 , 84.469809 )
			( 9 , 96.061784 )
			( 10 , 107.787326 )
			( 11 , 119.704535 )
			( 12 , 131.541936 )
			( 13 , 143.549081 )
			( 14 , 155.644656 )
			( 15 , 167.605739 )
			( 16 , 179.671265 )
			( 17 , 191.673536 )
			( 18 , 203.780401 )
			( 19 , 215.788038 )
			( 20 , 227.826622 )
			( 21 , 239.631928 )
			( 22 , 251.649338 )
			( 23 , 263.308906 )
			( 24 , 274.867694 )
			( 25 , 286.49362 )
			( 26 , 297.994894 )
			( 27 , 309.322195 )
			( 28 , 320.596748 )
			( 29 , 331.666462 )
			( 30 , 342.541388 )
			( 31 , 353.45647 )
			( 32 , 364.016306 )
			( 33 , 374.448685 )
			( 34 , 384.796978 )
			( 35 , 394.957575 )
			( 36 , 404.642029 )
			( 37 , 414.64136 )
			( 38 , 424.223507 )
			( 39 , 433.627881 )
			( 40 , 442.850904 )
			( 41 , 451.812946 )
			( 42 , 461.117602 )
			( 43 , 469.235202 )
			( 44 , 478.090872 )
			( 45 , 486.301261 )
			( 46 , 494.40419 )
			( 47 , 502.61045 )
			( 48 , 510.154528 )
			( 49 , 517.942875 )
			( 50 , 525.09414 )

		};
		\addlegendentry{Fix}
		
		\end{axis}
		\end{tikzpicture}

%% file: mult_div_10.tex
\begin{tikzpicture}[scale=0.8]
			\begin{axis}[
			xlabel=Check degree $d$,
			ylabel=Average number of operations,
			grid=both,
			legend style={legend pos=outer north east,font=\scriptsize}]
			\addplot[color=red] plot coordinates {
				( 2 , 0.907588 )
				( 3 , 1.433868 )
				( 4 , 2.007318 )
				( 5 , 2.628096 )
				( 6 , 3.282768 )
				( 7 , 3.975019 )
				( 8 , 4.687908 )
				( 9 , 5.434455 )
				( 10 , 6.203853 )
				( 11 , 7.000511 )
				( 12 , 7.799389 )
				( 13 , 8.623037 )
				( 14 , 9.462338 )
				( 15 , 10.29468 )
				( 16 , 11.14393 )
				( 17 , 11.994789 )
				( 18 , 12.855966 )
				( 19 , 13.719763 )
				( 20 , 14.570446 )
				( 21 , 15.408381 )
				( 22 , 16.267731 )
				( 23 , 17.105844 )
				( 24 , 17.920343 )
				( 25 , 18.735451 )
				( 26 , 19.556653 )
				( 27 , 20.347866 )
				( 28 , 21.13904 )
				( 29 , 21.902862 )
				( 30 , 22.652302 )
				( 31 , 23.403499 )
				( 32 , 24.108898 )
				( 33 , 24.812946 )
				( 34 , 25.496037 )
				( 35 , 26.166781 )
				( 36 , 26.778028 )
				( 37 , 27.417903 )
				( 38 , 28.014474 )
				( 39 , 28.592478 )
				( 40 , 29.138878 )
				( 41 , 29.661483 )
				( 42 , 30.230435 )
				( 43 , 30.662464 )
				( 44 , 31.154624 )
				( 45 , 31.596048 )
				( 46 , 32.017511 )
				( 47 , 32.43603 )
				( 48 , 32.799352 )
				( 49 , 33.175402 )
				( 50 , 33.480262 )

			};
			\addlegendentry{\cite{ADMMSiegel} mult}
			
			\addplot[color=blue] plot coordinates {
				( 2 , 1.000014 )
				( 3 , 1.56198 )
				( 4 , 2.166732 )
				( 5 , 2.813918 )
				( 6 , 3.491766 )
				( 7 , 4.204004 )
				( 8 , 4.933298 )
				( 9 , 5.694011 )
				( 10 , 6.476422 )
				( 11 , 7.285041 )
				( 12 , 8.093863 )
				( 13 , 8.925379 )
				( 14 , 9.772134 )
				( 15 , 10.611062 )
				( 16 , 11.46647 )
				( 17 , 12.321678 )
				( 18 , 13.187463 )
				( 19 , 14.055263 )
				( 20 , 14.909327 )
				( 21 , 15.749755 )
				( 22 , 16.612314 )
				( 23 , 17.451553 )
				( 24 , 18.268237 )
				( 25 , 19.085698 )
				( 26 , 19.908248 )
				( 27 , 20.699811 )
				( 28 , 21.491908 )
				( 29 , 22.255955 )
				( 30 , 23.004554 )
				( 31 , 23.756679 )
				( 32 , 24.461044 )
				( 33 , 25.165246 )
				( 34 , 25.846837 )
				( 35 , 26.516943 )
				( 36 , 27.125857 )
				( 37 , 27.764904 )
				( 38 , 28.359113 )
				( 39 , 28.935749 )
				( 40 , 29.480511 )
				( 41 , 30.000677 )
				( 42 , 30.568629 )
				( 43 , 30.996662 )
				( 44 , 31.48706 )
				( 45 , 31.926526 )
				( 46 , 32.34541 )
				( 47 , 32.761023 )
				( 48 , 33.120666 )
				( 49 , 33.494132 )
				( 50 , 33.795814 )

			};
			\addlegendentry{\cite{Wasson} mult}
			
			\addplot[color=cyan] plot coordinates {
				( 2 , 1.046106 )
				( 3 , 2.228243 )
				( 4 , 1.785499 )
				( 5 , 2.491755 )
				( 6 , 2.455314 )
				( 7 , 2.847167 )
				( 8 , 2.968677 )
				( 9 , 3.217002 )
				( 10 , 3.369873 )
				( 11 , 3.558093 )
				( 12 , 3.693505 )
				( 13 , 3.850527 )
				( 14 , 3.976796 )
				( 15 , 4.105293 )
				( 16 , 4.215411 )
				( 17 , 4.327095 )
				( 18 , 4.425583 )
				( 19 , 4.52649 )
				( 20 , 4.607504 )
				( 21 , 4.686078 )
				( 22 , 4.761151 )
				( 23 , 4.837071 )
				( 24 , 4.888448 )
				( 25 , 4.949197 )
				( 26 , 4.997072 )
				( 27 , 5.046158 )
				( 28 , 5.079869 )
				( 29 , 5.121234 )
				( 30 , 5.152589 )
				( 31 , 5.183999 )
				( 32 , 5.199312 )
				( 33 , 5.220412 )
				( 34 , 5.229236 )
				( 35 , 5.241826 )
				( 36 , 5.240679 )
				( 37 , 5.254444 )
				( 38 , 5.244684 )
				( 39 , 5.241171 )
				( 40 , 5.229327 )
				( 41 , 5.219705 )
				( 42 , 5.209963 )
				( 43 , 5.183441 )
				( 44 , 5.168077 )
				( 45 , 5.146147 )
				( 46 , 5.120244 )
				( 47 , 5.09233 )
				( 48 , 5.062362 )
				( 49 , 5.035956 )
				( 50 , 4.998021 )

			};
			\addlegendentry{\cite{Heusdens} mult}
			
			\addplot[color=red,dashed] plot coordinates {
				( 2 , 0.547398 )
				( 3 , 0.570422 )
				( 4 , 0.591518 )
				( 5 , 0.612486 )
				( 6 , 0.631458 )
				( 7 , 0.64997 )
				( 8 , 0.665841 )
				( 9 , 0.681614 )
				( 10 , 0.696315 )
				( 11 , 0.710683 )
				( 12 , 0.722398 )
				( 13 , 0.734213 )
				( 14 , 0.745384 )
				( 15 , 0.754285 )
				( 16 , 0.763187 )
				( 17 , 0.770878 )
				( 18 , 0.778305 )
				( 19 , 0.784979 )
				( 20 , 0.790141 )
				( 21 , 0.794194 )
				( 22 , 0.798822 )
				( 23 , 0.801931 )
				( 24 , 0.80383 )
				( 25 , 0.805503 )
				( 26 , 0.807225 )
				( 27 , 0.807663 )
				( 28 , 0.80796 )
				( 29 , 0.807305 )
				( 30 , 0.806084 )
				( 31 , 0.805098 )
				( 32 , 0.802603 )
				( 33 , 0.800151 )
				( 34 , 0.79723 )
				( 35 , 0.794067 )
				( 36 , 0.789337 )
				( 37 , 0.785665 )
				( 38 , 0.780987 )
				( 39 , 0.776075 )
				( 40 , 0.770591 )
				( 41 , 0.76465 )
				( 42 , 0.760248 )
				( 43 , 0.752651 )
				( 44 , 0.74688 )
				( 45 , 0.740159 )
				( 46 , 0.7333 )
				( 47 , 0.726629 )
				( 48 , 0.719042 )
				( 49 , 0.712053 )
				( 50 , 0.703862 )

			};
			\addlegendentry{\cite{ADMMSiegel} div}
			
			\addplot[color=blue,dashed] plot coordinates {
				( 2 , 0.547398 )
				( 3 , 0.570422 )
				( 4 , 0.591518 )
				( 5 , 0.612486 )
				( 6 , 0.631458 )
				( 7 , 0.64997 )
				( 8 , 0.665841 )
				( 9 , 0.681614 )
				( 10 , 0.696315 )
				( 11 , 0.710683 )
				( 12 , 0.722398 )
				( 13 , 0.734213 )
				( 14 , 0.745384 )
				( 15 , 0.754285 )
				( 16 , 0.763187 )
				( 17 , 0.770878 )
				( 18 , 0.778305 )
				( 19 , 0.784979 )
				( 20 , 0.790141 )
				( 21 , 0.794194 )
				( 22 , 0.798822 )
				( 23 , 0.801931 )
				( 24 , 0.80383 )
				( 25 , 0.805503 )
				( 26 , 0.807225 )
				( 27 , 0.807663 )
				( 28 , 0.80796 )
				( 29 , 0.807305 )
				( 30 , 0.806084 )
				( 31 , 0.805098 )
				( 32 , 0.802603 )
				( 33 , 0.800151 )
				( 34 , 0.79723 )
				( 35 , 0.794067 )
				( 36 , 0.789337 )
				( 37 , 0.785665 )
				( 38 , 0.780987 )
				( 39 , 0.776075 )
				( 40 , 0.770591 )
				( 41 , 0.76465 )
				( 42 , 0.760248 )
				( 43 , 0.752651 )
				( 44 , 0.74688 )
				( 45 , 0.740159 )
				( 46 , 0.7333 )
				( 47 , 0.726629 )
				( 48 , 0.719042 )
				( 49 , 0.712053 )
				( 50 , 0.703862 )

			};
			\addlegendentry{\cite{Wasson} div}
			
			\addplot[color=green,dashed] plot coordinates {
				( 2 , 8.27308 )
				( 3 , 11.094767 )
				( 4 , 13.360955 )
				( 5 , 15.374075 )
				( 6 , 17.152636 )
				( 7 , 18.787486 )
				( 8 , 20.219611 )
				( 9 , 21.585247 )
				( 10 , 22.826832 )
				( 11 , 23.970376 )
				( 12 , 24.992566 )
				( 13 , 25.944898 )
				( 14 , 26.793421 )
				( 15 , 27.594556 )
				( 16 , 28.279122 )
				( 17 , 28.912929 )
				( 18 , 29.491451 )
				( 19 , 30.028168 )
				( 20 , 30.473623 )
				( 21 , 30.811584 )
				( 22 , 31.162948 )
				( 23 , 31.485102 )
				( 24 , 31.668781 )
				( 25 , 31.868783 )
				( 26 , 32.021357 )
				( 27 , 32.079183 )
				( 28 , 32.166939 )
				( 29 , 32.201079 )
				( 30 , 32.199805 )
				( 31 , 32.144998 )
				( 32 , 32.040873 )
				( 33 , 31.952295 )
				( 34 , 31.815002 )
				( 35 , 31.673745 )
				( 36 , 31.465691 )
				( 37 , 31.307202 )
				( 38 , 31.035744 )
				( 39 , 30.822406 )
				( 40 , 30.487564 )
				( 41 , 30.214346 )
				( 42 , 29.946424 )
				( 43 , 29.640838 )
				( 44 , 29.258012 )
				( 45 , 28.957631 )
				( 46 , 28.604472 )
				( 47 , 28.239012 )
				( 48 , 27.900122 )
				( 49 , 27.503904 )
				( 50 , 27.063656 )

			};
			\addlegendentry{\cite{Wei} div}

			\addplot[color=cyan,dashed] plot coordinates {
				( 2 , 0.547398 )
				( 3 , 0.875066 )
				( 4 , 0.591518 )
				( 5 , 0.715175 )
				( 6 , 0.631458 )
				( 7 , 0.679103 )
				( 8 , 0.665841 )
				( 9 , 0.689055 )
				( 10 , 0.696315 )
				( 11 , 0.712485 )
				( 12 , 0.722398 )
				( 13 , 0.734659 )
				( 14 , 0.745384 )
				( 15 , 0.754405 )
				( 16 , 0.763187 )
				( 17 , 0.770906 )
				( 18 , 0.778305 )
				( 19 , 0.784986 )
				( 20 , 0.790141 )
				( 21 , 0.794195 )
				( 22 , 0.798822 )
				( 23 , 0.801931 )
				( 24 , 0.80383 )
				( 25 , 0.805503 )
				( 26 , 0.807225 )
				( 27 , 0.807663 )
				( 28 , 0.80796 )
				( 29 , 0.807305 )
				( 30 , 0.806084 )
				( 31 , 0.805098 )
				( 32 , 0.802603 )
				( 33 , 0.800151 )
				( 34 , 0.79723 )
				( 35 , 0.794067 )
				( 36 , 0.789337 )
				( 37 , 0.785665 )
				( 38 , 0.780987 )
				( 39 , 0.776075 )
				( 40 , 0.770591 )
				( 41 , 0.76465 )
				( 42 , 0.760248 )
				( 43 , 0.752651 )
				( 44 , 0.74688 )
				( 45 , 0.740159 )
				( 46 , 0.7333 )
				( 47 , 0.726629 )
				( 48 , 0.719042 )
				( 49 , 0.712053 )
				( 50 , 0.703862 )

			};
			\addlegendentry{\cite{Heusdens} div}
			
			\addplot[color=orange,dashed] plot coordinates {
				( 2 , 0.547398 )
				( 3 , 0.921154 )
				( 4 , 1.20174 )
				( 5 , 1.4467 )
				( 6 , 1.666029 )
				( 7 , 1.867447 )
				( 8 , 2.047208 )
				( 9 , 2.215832 )
				( 10 , 2.373774 )
				( 11 , 2.526258 )
				( 12 , 2.661885 )
				( 13 , 2.794162 )
				( 14 , 2.918786 )
				( 15 , 3.030656 )
				( 16 , 3.13737 )
				( 17 , 3.2367 )
				( 18 , 3.331309 )
				( 19 , 3.420911 )
				( 20 , 3.502003 )
				( 21 , 3.574263 )
				( 22 , 3.647982 )
				( 23 , 3.711666 )
				( 24 , 3.769061 )
				( 25 , 3.822891 )
				( 26 , 3.874715 )
				( 27 , 3.919354 )
				( 28 , 3.961094 )
				( 29 , 3.996387 )
				( 30 , 4.027426 )
				( 31 , 4.058503 )
				( 32 , 4.078638 )
				( 33 , 4.097811 )
				( 34 , 4.113638 )
				( 35 , 4.12702 )
				( 36 , 4.131427 )
				( 37 , 4.139754 )
				( 38 , 4.142382 )
				( 39 , 4.141537 )
				( 40 , 4.138103 )
				( 41 , 4.129178 )
				( 42 , 4.128788 )
				( 43 , 4.109312 )
				( 44 , 4.099683 )
				( 45 , 4.083039 )
				( 46 , 4.065368 )
				( 47 , 4.047616 )
				( 48 , 4.024309 )
				( 49 , 4.003172 )
				( 50 , 3.975189 )

			};
			\addlegendentry{Fix div}
			
			\end{axis}
			\end{tikzpicture}

%% file: probability.tex
\begin{tikzpicture}[scale=0.85]
		\begin{axis}[
		xlabel=Check degree $d$,
		ylabel=Probability,
		grid=both,
		legend style={legend pos=outer north east,font=\scriptsize}]

		\addplot[color=orange] plot coordinates {
			( 2 , 0.750115 )
			( 3 , 0.833351 )
			( 4 , 0.812386 )
			( 5 , 0.7322 )
			( 6 , 0.628253 )
			( 7 , 0.517251 )
			( 8 , 0.412501 )
			( 9 , 0.320767 )
			( 10 , 0.243626 )
			( 11 , 0.182253 )
			( 12 , 0.133928 )
			( 13 , 0.097109 )
			( 14 , 0.069566 )
			( 15 , 0.049109 )
			( 16 , 0.03455 )
			( 17 , 0.023817 )
			( 18 , 0.016641 )
			( 19 , 0.011274 )
			( 20 , 0.007614 )

		};
		\addlegendentry{$[-1,1)^d$}
		
%
%
%
%
%
		
		\addplot[color=red] plot coordinates {
			( 2 , 0.638717 )
			( 3 , 0.706456 )
			( 4 , 0.75264 )
			( 5 , 0.785535 )
			( 6 , 0.808088 )
			( 7 , 0.822234 )
			( 8 , 0.827112 )
			( 9 , 0.825888 )
			( 10 , 0.817674 )
			( 11 , 0.805186 )
			( 12 , 0.786477 )
			( 13 , 0.765616 )
			( 14 , 0.740995 )
			( 15 , 0.713741 )
			( 16 , 0.685113 )
			( 17 , 0.655269 )
			( 18 , 0.62492 )
			( 19 , 0.593429 )
			( 20 , 0.562031 )
			( 21 , 0.531299 )
			( 22 , 0.499851 )
			( 23 , 0.470491 )
			( 24 , 0.441664 )
			( 25 , 0.413227 )
			( 26 , 0.38559 )
			( 27 , 0.359625 )
			( 28 , 0.335093 )
			( 29 , 0.309973 )
			( 30 , 0.288419 )
			( 31 , 0.266412 )
			( 32 , 0.245675 )
			( 33 , 0.227242 )
			( 34 , 0.209501 )
			( 35 , 0.192578 )
			( 36 , 0.177422 )
			( 37 , 0.162922 )
			( 38 , 0.148786 )
			( 39 , 0.137333 )
			( 40 , 0.125116 )
			( 41 , 0.114032 )
			( 42 , 0.104512 )
			( 43 , 0.094727 )
			( 44 , 0.086792 )
			( 45 , 0.079016 )
			( 46 , 0.071804 )
			( 47 , 0.065558 )
			( 48 , 0.059175 )
			( 49 , 0.054355 )
			( 50 , 0.04881 )

		};
		\addlegendentry{$[-3,3)^d$}
		
		\addplot[color=green] plot coordinates {
			( 2 , 0.589832 )
			( 3 , 0.634313 )
			( 4 , 0.66972 )
			( 5 , 0.701303 )
			( 6 , 0.728024 )
			( 7 , 0.751547 )
			( 8 , 0.769935 )
			( 9 , 0.785315 )
			( 10 , 0.797374 )
			( 11 , 0.807183 )
			( 12 , 0.811901 )
			( 13 , 0.815507 )
			( 14 , 0.816471 )
			( 15 , 0.814167 )
			( 16 , 0.810699 )
			( 17 , 0.805121 )
			( 18 , 0.797192 )
			( 19 , 0.788892 )
			( 20 , 0.777331 )
			( 21 , 0.766323 )
			( 22 , 0.751455 )
			( 23 , 0.738816 )
			( 24 , 0.72365 )
			( 25 , 0.70823 )
			( 26 , 0.692052 )
			( 27 , 0.674274 )
			( 28 , 0.658089 )
			( 29 , 0.639109 )
			( 30 , 0.623308 )
			( 31 , 0.604022 )
			( 32 , 0.585995 )
			( 33 , 0.568695 )
			( 34 , 0.550582 )
			( 35 , 0.53253 )
			( 36 , 0.514992 )
			( 37 , 0.496837 )
			( 38 , 0.479409 )
			( 39 , 0.462245 )
			( 40 , 0.444316 )
			( 41 , 0.427639 )
			( 42 , 0.412055 )
			( 43 , 0.395732 )
			( 44 , 0.380275 )
			( 45 , 0.36545 )
			( 46 , 0.351047 )
			( 47 , 0.336293 )
			( 48 , 0.321702 )
			( 49 , 0.308557 )
			( 50 , 0.294586 )

		};
		\addlegendentry{$[-5,5)^d$}
		
		\addplot[color=cyan] plot coordinates {
			( 2 , 0.547398 )
			( 3 , 0.570422 )
			( 4 , 0.591518 )
			( 5 , 0.612486 )
			( 6 , 0.631458 )
			( 7 , 0.64997 )
			( 8 , 0.665841 )
			( 9 , 0.681614 )
			( 10 , 0.696315 )
			( 11 , 0.710683 )
			( 12 , 0.722398 )
			( 13 , 0.734213 )
			( 14 , 0.745384 )
			( 15 , 0.754285 )
			( 16 , 0.763187 )
			( 17 , 0.770878 )
			( 18 , 0.778305 )
			( 19 , 0.784979 )
			( 20 , 0.790141 )
			( 21 , 0.794194 )
			( 22 , 0.798822 )
			( 23 , 0.801931 )
			( 24 , 0.80383 )
			( 25 , 0.805503 )
			( 26 , 0.807225 )
			( 27 , 0.807663 )
			( 28 , 0.80796 )
			( 29 , 0.807305 )
			( 30 , 0.806084 )
			( 31 , 0.805098 )
			( 32 , 0.802603 )
			( 33 , 0.800151 )
			( 34 , 0.79723 )
			( 35 , 0.794067 )
			( 36 , 0.789337 )
			( 37 , 0.785665 )
			( 38 , 0.780987 )
			( 39 , 0.776075 )
			( 40 , 0.770591 )
			( 41 , 0.76465 )
			( 42 , 0.760248 )
			( 43 , 0.752651 )
			( 44 , 0.74688 )
			( 45 , 0.740159 )
			( 46 , 0.7333 )
			( 47 , 0.726629 )
			( 48 , 0.719042 )
			( 49 , 0.712053 )
			( 50 , 0.703862 )
		};
		\addlegendentry{$[-10,10)^d$}

		\end{axis}
		\end{tikzpicture}

%% file: paper.bbl
\begin{thebibliography}{10}
\providecommand{\url}[1]{#1}
\csname url@samestyle\endcsname
\providecommand{\newblock}{\relax}
\providecommand{\bibinfo}[2]{#2}
\providecommand{\BIBentrySTDinterwordspacing}{\spaceskip=0pt\relax}
\providecommand{\BIBentryALTinterwordstretchfactor}{4}
\providecommand{\BIBentryALTinterwordspacing}{\spaceskip=\fontdimen2\font plus
\BIBentryALTinterwordstretchfactor\fontdimen3\font minus
  \fontdimen4\font\relax}
\providecommand{\BIBforeignlanguage}[2]{{%
\expandafter\ifx\csname l@#1\endcsname\relax
\typeout{** WARNING: IEEEtran.bst: No hyphenation pattern has been}%
\typeout{** loaded for the language `#1'. Using the pattern for}%
\typeout{** the default language instead.}%
\else
\language=\csname l@#1\endcsname
\fi
#2}}
\providecommand{\BIBdecl}{\relax}
\BIBdecl

\bibitem{Feldman}
J.~Feldman, M.~J. Wainwright, and D.~R. Karger, ``Using linear programming to
  decode binary linear codes,'' \emph{IEEE Transactions on Information Theory},
  vol.~51, no.~3, pp. 954--972, March 2005.

\bibitem{csa}
X.~Zhang and P.~H. Siegel, ``Adaptive cut generation algorithm for improved
  linear programming decoding of binary linear codes,'' \emph{IEEE Transactions
  on Information Theory}, vol.~58, no.~10, pp. 6581--6594, October 2012.

\bibitem{Berlekamp}
E.~Berlekamp, R.~McEliece, and H.~van Tilborg, ``On the inherent intractability
  of certain coding problems (corresp.),'' \emph{IEEE Transactions on
  Information Theory}, vol.~24, no.~3, pp. 384--386, May 1978.

\bibitem{Boyd}
\BIBentryALTinterwordspacing
S.~Boyd, N.~Parikh, E.~Chu, B.~Peleato, and J.~Eckstein, ``Distributed
  optimization and statistical learning via the alternating direction method of
  multipliers,'' \emph{Found. Trends Mach. Learn.}, vol.~3, no.~1, pp. 1--122,
  January 2011. [Online]. Available: \url{http://dx.doi.org/10.1561/2200000016}
\BIBentrySTDinterwordspacing

\bibitem{Barman}
S.~Barman, X.~Liu, S.~C. Draper, and B.~Recht, ``Decomposition methods for
  large scale {L}{P} decoding,'' \emph{IEEE Transactions on Information
  Theory}, vol.~59, no.~12, pp. 7870--7886, December 2013.

\bibitem{ADMMSiegel}
X.~Zhang and P.~H. Siegel, ``Efficient iterative {L}{P} decoding of
  {L}{D}{P}{C} codes with alternating direction method of multipliers,'' in
  \emph{2013 IEEE International Symposium on Information Theory}, July 2013,
  pp. 1501--1505.

\bibitem{Wasson}
M.~Wasson and S.~C. Draper, ``Hardware based projection onto the parity
  polytope and probability simplex,'' in \emph{2015 49th Asilomar Conference on
  Signals, Systems and Computers}, November 2015, pp. 1015--1020.

\bibitem{Heusdens}
G.~Zhang, R.~Heusdens, and W.~B. Kleijn, ``Large scale {L}{P} decoding with low
  complexity,'' \emph{IEEE Communications Letters}, vol.~17, no.~11, pp.
  2152--2155, November 2013.

\bibitem{projectionl1ball}
\BIBentryALTinterwordspacing
J.~Duchi, S.~Shalev-Shwartz, Y.~Singer, and T.~Chandra, ``Efficient projections
  onto the l1-ball for learning in high dimensions,'' in \emph{Proceedings of
  the 25th International Conference on Machine Learning}, ser. ICML '08.\hskip
  1em plus 0.5em minus 0.4em\relax New York, NY, USA: ACM, 2008, pp. 272--279.
  [Online]. Available: \url{http://doi.acm.org/10.1145/1390156.1390191}
\BIBentrySTDinterwordspacing

\bibitem{Cormen}
T.~H. Cormen, C.~Stein, R.~L. Rivest, and C.~E. Leiserson, \emph{Introduction
  to Algorithms}, 2nd~ed.\hskip 1em plus 0.5em minus 0.4em\relax McGraw-Hill
  Higher Education, 2001.

\bibitem{Wei}
H.~Wei and A.~H. Banihashemi, ``An iterative check polytope projection
  algorithm for {A}{D}{M}{M}-based {L}{P} decoding of {L}{D}{P}{C} codes,''
  \emph{IEEE Communications Letters}, vol.~22, no.~1, pp. 29--32, January 2018.

\bibitem{ADMMLookup}
X.~Jiao, J.~Mu, Y.~C. He, and C.~Chen, ``Efficient {A}{D}{M}{M} decoding of
  {L}{D}{P}{C} codes using lookup tables,'' \emph{IEEE Transactions on
  Communications}, vol.~65, no.~4, pp. 1425--1437, April 2017.

\bibitem{ADMMLookup2}
X.~Jiao, Y.~C. He, and J.~Mu, ``Memory-reduced look-up tables for efficient
  {A}{D}{M}{M} decoding of {L}{D}{P}{C} codes,'' \emph{IEEE Signal Processing
  Letters}, vol.~25, no.~1, pp. 110--114, January 2018.

\bibitem{penalizedADMM}
X.~Liu and S.~C. Draper, ``The {A}{D}{M}{M} penalized decoder for {L}{D}{P}{C}
  codes,'' \emph{IEEE Transactions on Information Theory}, vol.~62, no.~6, pp.
  2966--2984, June 2016.

\bibitem{trappingsets}
------, ``{A}{D}{M}{M} decoding on trapping sets,'' in \emph{2015 IEEE
  International Symposium on Information Theory (ISIT)}, June 2015, pp.
  2663--2667.

\bibitem{ADMMirregular}
X.~Jiao, H.~Wei, J.~Mu, and C.~Chen, ``Improved {A}{D}{M}{M} penalized decoder
  for irregular low-density parity-check codes,'' \emph{IEEE Communications
  Letters}, vol.~19, no.~6, pp. 913--916, June 2015.

\bibitem{ReducedComplexityADMM}
H.~Wei, X.~Jiao, and J.~Mu, ``Reduced-complexity linear programming decoding
  based on {A}{D}{M}{M} for {L}{D}{P}{C} codes,'' \emph{IEEE Communications
  Letters}, vol.~19, no.~6, pp. 909--912, June 2015.

\bibitem{ImprovedPenaltyADMM}
B.~Wang, J.~Mu, X.~Jiao, and Z.~Wang, ``Improved penalty functions of
  {A}{D}{M}{M} penalized decoder for {L}{D}{P}{C} codes,'' \emph{IEEE
  Communications Letters}, vol.~21, no.~2, pp. 234--237, February 2017.

\bibitem{AcceleratedADMM}
X.~Jiao, J.~Mu, and J.~Guo, ``A comparison study of {LDPC} decoding using
  accelerated {ADMM} and over-relaxed {ADMM},'' in \emph{2016 2nd IEEE
  International Conference on Computer and Communications (ICCC)}, October
  2016, pp. 191--195.

\bibitem{fastADMM}
\BIBentryALTinterwordspacing
T.~Goldstein, B.~O'Donoghue, S.~Setzer, and R.~Baraniuk, ``Fast alternating
  direction optimization methods,'' \emph{SIAM Journal on Imaging Sciences},
  vol.~7, no.~3, pp. 1588--1623, 2014. [Online]. Available:
  \url{https://doi.org/10.1137/120896219}
\BIBentrySTDinterwordspacing

\bibitem{TwoStepADMM}
X.~Jiao and J.~Mu, ``Lowering the error floor of {A}{D}{M}{M} penalized decoder
  for {L}{D}{P}{C} codes,'' \emph{China Communications}, vol.~13, no.~8, pp.
  127--135, August 2016.

\bibitem{WassonHardware}
\BIBentryALTinterwordspacing
M.~Wasson, M.~Milicevic, S.~C. Draper, and P.~G. Gulak, ``Hardware-based linear
  program decoding with the alternating direction method of multipliers,''
  \emph{CoRR}, vol. abs/1611.05975, 2016. [Online]. Available:
  \url{http://arxiv.org/abs/1611.05975}
\BIBentrySTDinterwordspacing

\bibitem{DebbabiHardware}
I.~Debbabi, N.~Khouja, F.~Tlili, B.~Le~Gal, and C.~Jégo, ``Evaluation of the
  hardware complexity of the {ADMM} approach for {L}{D}{P}{C} decoding,'' in
  \emph{2016 IEEE Wireless Communications and Networking Conference}, April
  2016, pp. 1--6.

\bibitem{MulticoreADMM}
\BIBentryALTinterwordspacing
I.~Debbabi, B.~Le~Gal, N.~Khouja, F.~Tlili, and C.~J{\'e}go, ``Multicore and
  manycore implementations of {A}{D}{M}{M}-based decoders for {L}{D}{P}{C}
  decoding,'' \emph{Journal of Signal Processing Systems}, September 2017.
  [Online]. Available: \url{https://doi.org/10.1007/s11265-017-1284-0}
\BIBentrySTDinterwordspacing

\bibitem{ADMMScheduling}
------, ``Fast converging {ADMM}-penalized algorithm for {L}{D}{P}{C}
  decoding,'' \emph{IEEE Communications Letters}, vol.~20, no.~4, pp. 648--651,
  April 2016.

\bibitem{ADMMScheduling2}
X.~Jiao, J.~Mu, and H.~Wei, ``Reduced complexity node-wise scheduling of
  {A}{D}{M}{M} decoding for {L}{D}{P}{C} codes,'' \emph{IEEE Communications
  Letters}, vol.~21, no.~3, pp. 472--475, March 2017.

\bibitem{ADMMconvolutional}
H.~B. Thameur, B.~Le~Gal, N.~Khouja, F.~Tlili, and C.~Jégo, ``Low complexity
  {A}{D}{M}{M}-{L}{P} based decoding strategy for {L}{D}{P}{C} convolutional
  codes,'' in \emph{2017 25th International Conference on Software,
  Telecommunications and Computer Networks (SoftCOM)}, September 2017, pp.
  1--5.

\bibitem{ADMMequalization}
W.~Xu, X.~Jiao, and J.~Mu, ``Evaluating the performance of turbo equalization
  with {A}{D}{M}{M} decoding,'' in \emph{2017 IEEE International Conference on
  Signal Processing, Communications and Computing (ICSPCC)}, October 2017, pp.
  1--5.

\bibitem{Feldman03PhD}
J.~Feldman, ``{D}ecoding {E}rror-{C}orrecting {C}odes via {L}inear
  {P}rogramming,'' Ph.D. dissertation, Massachusetts Institute of Technology,
  2003.

\bibitem{Taghavi}
M.~H. Taghavi~N. and P.~H. Siegel, ``Adaptive methods for linear programming
  decoding,'' \emph{IEEE Transactions on Information Theory}, vol.~54, no.~12,
  pp. 5396--5410, December 2008.

\bibitem{Jeroslow}
\BIBentryALTinterwordspacing
R.~Jeroslow, ``On defining sets of vertices of the hypercube by linear
  inequalities,'' \emph{Discrete Mathematics}, vol.~11, no.~2, pp. 119 -- 124,
  1975. [Online]. Available:
  \url{http://www.sciencedirect.com/science/article/pii/0012365X75900035}
\BIBentrySTDinterwordspacing

\end{thebibliography}
